\documentclass[a4paper,11pt]{article}
\pdfoutput=1
\usepackage{jheppub} % for details on the use of the package, please
\usepackage{booktabs}
\usepackage[utf8]{inputenc} %Support for additional character in the bibliography
\usepackage{url}
\usepackage{color}
\usepackage{amsmath,amsfonts,amssymb,mathrsfs,verbatim,epsfig, bbm, wasysym}
\usepackage{slashed, cancel}
\usepackage{bbold}%For identity-matrix symbol
\usepackage{graphicx}
\allowdisplaybreaks
\usepackage{latexsym}
\usepackage[dvipsnames]{xcolor}

\graphicspath{{./figs/}{./Figures/}}

\definecolor{rossos}{rgb}{0.8,0.2,0.3}
\definecolor{bluscuro}{rgb}{0.15, 0.2, .85}
\definecolor{bluchiaro}{cmyk}{1,.3,0.,0.1}
\definecolor{orange}{rgb}{1,0.5,0}
\definecolor{blue}{rgb}{0,0,1}
\hypersetup{colorlinks, citecolor=bluscuro, linkcolor=bluscuro, urlcolor=bluscuro}

\def\beq{\begin{equation}}
\def\eeq{\end{equation}}
\def\beqa{\begin{eqnarray}}
\def\eeqa{\end{eqnarray}}

\def\ltap{\ \raise.3ex\hbox{$<$\kern-.75em\lower1ex\hbox{$\sim$}}\ }
\def\gtap{\ \raise.3ex\hbox{$>$\kern-.75em\lower1ex\hbox{$\sim$}}\ }
\def\ba{\begin{array}}
\def\ea{\end{array}}
\def\bea{\begin{eqnarray}}
\def\eea{\end{eqnarray}}
\def\bean{\begin{eqnarray*}}
\def\eean{\end{eqnarray*}}

\def\mx{m_{\rm DM}}
\def\mf{m_{f}}

\def\vDM{v_{\rm DM}}
\def\mX{m_{\chi}}
\def\gX{g_{\chi}}

\def\d{{\rm d}}
\def\cO{{\cal O}}
\newcommand{\GeV}{{\rm \,GeV}}
\newcommand{\TeV}{{\rm \,TeV}}

\newcommand{\fb}{{\rm fb}}
\newcommand{\cm}{{\rm \,cm}}
\newcommand{\met}{\slashed{E}_T}
\newcommand{\sSD}{\sigma^\text{SD}_{\chi p}}
\newcommand{\GZp}{\Gamma_{Z'}}
\newcommand{\mZp}{m_{Z'}}
\newcommand{\mZ}{m_{Z}}
\newcommand{\GZ}{\Gamma_{Z}}
\newcommand{\gZ}{g_Z}

\newcommand{\spsi}{s_\psi}
\newcommand{\UY}{U(1)_Y}
\newcommand{\UBL}{U(1)_{B-L}}
\newcommand{\Up}{U(1)'}
\newcommand{\cth}{\cos\theta}
\newcommand{\sth}{\sin\theta}
\newcommand{\cL}{{\cal L}} 
\newcommand{\mF}{\mathcal F}
\newcommand{\gZp}{g_{Z'}}
\newcommand{\nDM}{n_{\rm DM}}
\newcommand{\Gcap}{\Gamma_{\rm cap}}
\newcommand{\Gann}{\Gamma_{\rm ann}}
\newcommand{\asv}{\langle \sigma v\rangle}
\newcommand{\bes}[1]{\mathscr K_{#1}}
\newcommand{\aW}{\alpha_W}
\newcommand{\aem}{\alpha_{\rm em}}
\newcommand{\tW}{\theta_W}
\newcommand{\sv}{\langle \sigma v\rangle}

\newcommand{\mU}{m_{u^i}}
\newcommand{\mD}{m_{d^i}}
\newcommand{\cZp}{g}
\newcommand{\cLu}{\cZp^L_{u^i}}
\newcommand{\cRu}{\cZp^R_{u^i}}
\newcommand{\cLd}{\cZp^L_{d^i}}
\newcommand{\cRd}{\cZp^R_{d^i}}
\newcommand{\cVu}{\cZp^V_{u^i}}
\newcommand{\cAu}{\cZp^A_{u^i}}
\newcommand{\cVd}{\cZp^V_{d^i}}
\newcommand{\cAd}{\cZp^A_{d^i}}
\newcommand{\cVf}{\cZp^V_f}
\newcommand{\cAf}{\cZp^A_f}
\newcommand{\cVq}{\cZp^V_q}
\newcommand{\cAq}{\cZp^A_q}
\newcommand{\Nc}{N_c^i}
\newcommand{\cAZp}{c_{A, f}^{Z'}}
\newcommand{\cVZp}{c_{V, f}^{Z'}}
\newcommand{\cAZ}{c_{A, f}^{Z}}
\newcommand{\cVZ}{c_{V, f}^{Z}}
\newcommand{\cZ}{c}
\newcommand{\cZLf}{\cZ^L_f}
\newcommand{\cZRf}{\cZ^R_f}
\newcommand{\new}[1]{#1}
\newcommand{\replace}[1]{(\textit{Replaces Fig.~#1 of the original paper.})}

\subheader{\footnotesize
\vspace*{-3.7em}
\begin{flushright}
CERN-TH-2016-099\\
SISSA 22/2016/FISI
\end{flushright}
}

\title{\boldmath
Complementarity of DM Searches in a Consistent Simplified Model: the Case of $Z'$
}

\author[a]{Thomas~Jacques,}
\author[b,c]{Andrey~Katz,}
\author[c]{Enrico~Morgante,}
\author[c]{Davide~Racco,}
\author[d]{Mohamed~Rameez,}
\author[c]{and Antonio~Riotto}

\affiliation[a]{
SISSA and INFN, via Bonomea 265, 34136 Trieste, Italy}

\affiliation[b]{Theory Division, CERN, CH-1211 Geneva 23, Switzerland}

\affiliation[c]{
D\'epartement de Physique Th\'eorique and Center for Astroparticle Physics (CAP), \\ 
Universit\'e de Gen\`eve, 24 quai Ansermet, CH-1211 Gen\`eve 4, Switzerland}

\affiliation[d]{D\'epartement de Physique Nucl\'eaire et Corpusculaire, \\ 
Universit\'e de Gen\`eve, 24 quai Ansermet, CH-1211 Gen\`eve 4, Switzerland}

\emailAdd{thomas.jacques@sissa.it}
\emailAdd{andrey.katz@cern.ch}
\emailAdd{enrico.morgante@unige.ch}
\emailAdd{davide.racco@unige.ch}
\emailAdd{mohamed.rameez@unige.ch}
\emailAdd{antonio.riotto@unige.ch}

\abstract{ \noindent 
We analyze the constraints from direct and indirect detection on fermionic  Majorana Dark Matter (DM).
Because the interaction with the Standard Model (SM) particles is 
spin-dependent, \emph{a priori} the constraints that one gets from  neutrino telescopes, the LHC, direct \new{and indirect} detection experiments
are comparable. We study the complementarity of these searches in a particular example, 
in which a heavy $Z'$ mediates the interactions between the SM and the DM.  
\new{We find that for heavy dark matter indirect detection provides the strongest bounds on this scenario, while IceCube bounds are typically stronger than those from direct detection. The LHC constraints are dominant for smaller dark matter masses.}
These light masses  are less motivated by thermal relic abundance considerations. 
We show that the dominant annihilation channels of the light DM in the Sun \new{and the Galactic Center} are either $b \bar b$ or $t \bar t$, 
while the heavy DM annihilation is completely dominated by $Zh$ channel. The latter produces a hard neutrino
spectrum which has not been previously analyzed.
We study the neutrino spectrum yielded by DM and recast IceCube constraints to allow proper comparison with constraints from direct \new{and indirect} detection experiments and LHC exclusions. \\
\textit{Note that the original version of the paper contains an important error: the contribution of the $Z$-mediated processes was overlooked. This changes some of the results of the paper. For the details and the correct results please see the erratum attached to this file as an appendix.}}

\begin{document}

\maketitle

%%%%%%%%%%%%%%%%%%%%%%%%%%%%%%%%
\section{Introduction}
\label{sec:intro}
% !TEX root = main_draft.tex
Although there is overwhelming evidence that  Dark Matter (DM) 
comprises nearly a quarter of the energy budget of the Universe \cite{Ade:2015xua}, the precise
nature of the DM still remains unknown. Until now all the evidence
for  DM has come from gravitational interactions, while 
searches for other DM interactions have yielded negative results. 
However, nowadays there is a well-developed program of searches for DM via its presumed non-gravitational
interactions. This program includes direct detection experiments,
a plethora of various LHC searches, indirect searches for the annihilation of DM
captured  by the Earth and the Sun, as well as searches
for  DM annihilations in the Milky Way Galactic Center and in
nearby dwarf Galaxies. 

One of the main motivations for thinking about  DM interactions beyond
the gravitational is the idea of thermal freezeout. This idea finds its natural home in the so-called WIMP-miracle
scenario: a particle $\chi$ with a
mass of order $\sim $TeV, charged under a weak force, yields the measured relic abundance. 
Direct detection experiments have mostly been aiming to discover this
particular DM candidate, which also looked very attractive as part of a 
solution to the SM naturalness problem, in particular via
supersymmetry and extra-dimensional scenarios.

The strongest  bounds on the spin-independent DM-proton scattering
cross-section in most cases are coming from
LUX~\cite{Akerib:2013tjd} that constrains this 
scattering at the level of $\sigma_{\chi p} \sim 10^{-44}- 10^{-45}$~cm$^2$. 
These cross sections are much smaller than the
natural cross sections one would expect from the WIMP miracle, and, in
fact, they  are more  
characteristic of ``higgs-portal" DM~\cite{Patt:2006fw,Kim:2006af}.
These direct 
detection  results are
very suggestive and put
the entire idea of WIMPs under pressure. 

However, there are of course
several noticeable caveats in this logic, with the most important
being the possibility of spin-dependent (SD) DM. 
If the nucleon-DM scattering rate depends on the spin of the nucleus,
the \emph{total} cross section in direct
detection experiments is reduced by
orders of magnitude.
The strongest direct detection bounds on 
SD DM, coming from PICO~\cite{Amole:2015pla}, constrain
$\sSD 
\lesssim 10^{-38} - 10^{-39}\cm^2$, perfectly within the naive
ballpark of  the WIMP DM. LUX constraints on the SD DM are
comparable to those of PICO~\cite{Akerib:2016lao}. 

Interestingly, at least naively, the direct detection bounds are not
the strongest bounds one can put on the SD nucleon-DM scattering. Even
stronger exclusions have been claimed both by \new{Fermi-LAT,} IceCube and by 
collider searches at the LHC. However, these constraints are more
model dependent and the comparison of these constraints to the direct
detection operators is not straightforward. 

The problem with comparing these results is twofold. Let us first
consider the signal in neutrino 
telescopes. Assuming that the DM capture and annihilation rates are already in equilibrium in the Sun's
core, the total neutrino flux is only proportional to the capture
rate, and independent of the DM annihilation rate~\cite{Gould:1991hx}.
However, in order to know the predicted neutrino flux, one 
should also know the annihilation channels. 
%The total annihilation on the Sun is only proportional to the
%scattering rate (assuming that the DM is already in equilibrium in
%the Sun core), but in order to know neutrino fluxes, one should also
%know partucular annihilation channels. 
As we will discuss later, these
are highly model dependent. The IceCube collaboration reports its
results~\cite{Aartsen:2012kia,Aartsen:2016exj} 
considering 
 annihilations into $WW$, $\tau \tau$ (optimistic scenarios with hard
 neutrinos) 
and $b \bar b$ (pessimistic scenario with soft neutrinos). However,
there is no guarantee that in a full UV complete picture any of these
channels is dominant. In fact, if the DM is a Dirac fermion, it will
most likely annihilate into a pair of SM fermions, and assuming flavor
universality, these will most likely be light flavors, leaving no
distinctive signature at IceCube. On the other hand, if the DM is
Majorana and the fermion current preserves chirality, the light fermion channels are velocity-suppressed,
giving way to other channels: $WW$, $ZZ$, $hh$, $Zh$ and $t \bar
t$. \emph{A priori}, there is 
no way to know which of these channels dominates the annihilation
process and sets the neutrino spectrum. 
\new{The same problem holds for the bounds obtained by indirect detection of the diffuse $\gamma$-rays from the Galactic 
Center and the dwarf satellite galaxies.
These bounds also depend on the choice of the primary DM annihilation channels
that give rise to the final spectrum of diffuse $\gamma$-rays.}

The second problem has to do with  the comparison of the direct detection exclusions 
to the LHC searches that nominally claim the
strongest bounds on  SD DM interactions with 
nucleons. 
However, this claim is strongly model-dependent. 
The LHC searches are extremely sensitive to the ratio between the mediator
couplings to the DM particles and to the SM particles. 
Moreover, the collider exclusions quickly weaken in the high DM mass region. 

In this paper we analyze the complementarity of all these
searches in one particular, but highly motivated scenario. 
Of course the question of complementarity is  meaningless if done purely within the effective field theory (EFT) of DM. 
%Firstly, the EFT DM does not allow us to calculate the annihilation channels of the DM into the weak gauge bosons and the Higgs, unless the DM is directly charged under the SM $SU(2) \times U(1)$. 
Firstly, the EFT may not include some relevant annihilation channels of DM into weak gauge bosons and the Higgs, if the particles integrated out directly interact with them.
Secondly, the EFT of DM is inadequate for analyzing the results of LHC
searches unless the mediator is very heavy, 
otherwise the typical momentum running in the
diagram is at least comparable to the mass of the 
mediator~\cite{Busoni:2013lha,Busoni:2014sya,Busoni:2014haa,
Buchmueller:2013dya,Racco:2015dxa,DeSimone:2016fbz}.
This leads to the inevitable conclusion that the question of complementarity can
only be addressed in the context of full models, even if these models
are simplified. 

In this particular paper we will address the question of the
complementarity between searches in the case that the interaction between
the DM and the baryonic matter is mediated by a heavy vector boson
($Z'$). This is probably the easiest scenario one can consider beyond
the proper WIMP (namely, the DM charged under the SM $SU(2)
\times U(1)$). It can also also be part of more complicated scenarios,
e.g.\ non-minimal SUSY or strongly coupled physics at the TeV scale. We
explicitly calculate the annihilation rate into the 
electroweak (EW) bosons and $Zh$ and show that, if the
annihilations into SM fermions are helicity suppressed, the
dominant channels (depending on the DM mass) are $b\bar b$, $t \bar t$ and $Zh$. 
The latter have been neglected in previous studies, however we show that for the heavy DM 
it is a dominant source of hard neutrinos, which can be detected by IceCube. 
Because even relatively small BRs into the gauge bosons that can potentially be important 
sources of the energetic neutrinos that can be detected by  IceCube, we calculate  
the rates into the EW gauge bosons at the NLO level. We further comment on this issue 
in Sec.~\ref{sec:IC}, while the details of the calculations are relegated to Appendix \ref{sec:annihilation}.
We show that the one-loop corrections to the $WW$ annihilation channel are indeed important
and noticeably harden the neutrino spectrum that one would naively estimate at the tree level.  
We also reanalyze the collider constraints, both on the direct production of $Z'$ and on the production of DM associated 
with a jet \new{and review the situation with the indirect detection from dwarf satelite galaxies.}

Our paper is structured as follows. In Sec.~\ref{sec:SD} we review the
basic features of SD DM, we describe in detail our setup and we review the constraints from direct searches for a 
$Z'$ at colliders and from the calculation of the relic density of DM.
\new{In Sec.~\ref{sec:IC} we illustrate the constraints on our model from the LHC, direct and indirect DM searches and from the 
IceCube experiment, and we discuss the branching ratios for DM pair annihilation into the 
SM channels (which are required in order to derive bounds from IceCube and Fermi).}
In Sec.~\ref{sec:results} we illustrate our results, and in Sec.~\ref{sec:conclusions} we summarize our conclusions. 
Appendix~\ref{sec:annihilation} contains further details about the formul\ae\ for the annihilation of DM.
 
%%%%%%%%%%%%%%%%%%%%%%%%%%%%%%%%

%%%%%%%%%%%%%%%%%%%%%%%%%%%%%%%%
\section{$Z'$-Mediated Spin-Dependent Interactions}
\label{sec:SD}
% !TEX root = main_draft.tex
Spin dependent DM--baryon scattering is a fairly generic
phenomenon. The non-relativistic (NR) operators that describe 
this
scattering~\cite{Fan:2010gt,Fitzpatrick:2012ix,Fitzpatrick:2012ib}
have explicit dependence on the nucleus spin.    
These interactions can 
arise 
from different types of  UV theories and 
studying all possible SD interactions case by case would a tedious and
unilluminating 
exercise. However, an   $\vec
S_\chi \cdot \vec S_N$ interaction, where $\vec S_{\chi}, \vec S_{N}$ are the spin of the DM particle and the nucleon 
respectively,  is in many senses unique in the long list of
NR SD operators. 
It is the only operator arising from an
interaction mediated by a heavy particle, for which the 
NR limit of the DM-nucleon interaction is not also suppressed  by halo velocity; that is, either powers of $\vec q$ or $\vec v$ in the language of the NR EFT. 
These extra suppressions often render the direct detection and  solar capture rates too small for most practical purposes. 

In the high-energy EFT this operator descends from the axial
vector - axial vector interaction~\cite{Anand:2013yka}, namely 
\beq\label{eq:AxialAxial}
\cL = \big( \overline \chi \gamma_\mu \gamma^5 \chi \big)
\, \big(\overline N \gamma^\mu \gamma^5 N \big) \,. 
\eeq
As we have described in the introduction, this relativistic EFT description is not very useful, if we would like to compare  direct detection searches for
 DM with the constraints from the LHC and IceCube. 
 The UV completion for this operator is needed. 
There are several options for how this interaction can be completed at the renormalizable level. 
The simplest and most economical possibility, from the point of view of model building, is to mediate the interaction via a  massive neutral vector boson, which couples to the SM axial current and to the DM axial current. This might either be a $Z$-boson of the SM (corresponding to a standard WIMP scenario) or a new $Z'$ boson.  

The simplest realization of this scenario would be a ``classical''
WIMP, namely a Majorana fermion that is charged under the SM $SU(2) \times U(1)$ (but not under the electromagnetism)
and couples to the heavy weak gauge bosons at the tree level. One can view a pure
supersymmetric electroweakino as a prototype of this kind of DM -- either a doublet or a triplet of  $SU(2)_L$, 
possibly mixed with a singlet
(bino).\footnote{For a generalization of this scenario to 
arbitrary representations see~\cite{Cirelli:2005uq,Cirelli:2009uv}.}
SUSY dark matter also satisfies the criteria that we have listed above: the direct detection cross-section is 
spin-dependent\footnote{
\new{Of course we assume very small mixings between the electroweakinos and  suppressed squark interactions. 
This can naturally happen in various scenarios, e.g.\ split SUSY~\cite{ArkaniHamed:2004fb,Giudice:2004tc} or 
recently proposed mini-split, motivated by the 125~GeV higgs~\cite{Arvanitaki:2012ps,ArkaniHamed:2012gw}}.}
and therefore the bounds from direct detection experiments are relatively weak. 
Because the generic SUSY DM candidate is a Majorana fermion (rather than Dirac), annihilations into the light quarks are helicity suppressed. 
In the neutralino case, the annihilation into the EW gauge bosons proceeds at the leading order and in this case 
the $WW$, $ZZ$ and $t \bar t$ channels dominate the annihilation
branching ratios. The only relevant NLO annihilation channels for the neutralino spin-dependent DM are $\gamma \gamma$ and $\gamma Z$, which have been fully analyzed in Refs.~\cite{Bern:1997ng,Bergstrom:1997fh}.

Another straightforward yet less explored UV-completion is
mediation of interaction~\eqref{eq:AxialAxial} by a heavy $Z'$,
remnant of a new gauge symmetry at the TeV scale. In particular, we
assume the following renormalizable interaction between the DM, $Z'$
mediator and the SM fermions:
\begin{equation}
\label{eq:L}
\cL = i \gZp \gX \bar \chi \gamma^\mu \gamma^5 \chi Z'_{\mu} + i \gZp \cAf \bar
\psi \gamma^\mu \gamma^5 \psi Z'_{\mu} + i \gZp \cVf  \bar \psi
\gamma^{\mu} \psi Z'_{\mu}~,
\end{equation}
where $\chi$ is a Majorana DM particle and $\psi$ stands for the SM
fermions. Note that we have not yet specified the couplings of the
$Z'$ to the SM Higgs (if they exist at all).
We expect a generic $Z'$ to couple  to both a vector and 
an axial current, and therefore we write down both these couplings for
the SM fermions. 
For the DM fermions we have two choices: they might be either Dirac or Majorana fermions. 
If they are Dirac, it would be difficult to explain why the $Z'$ does not couple to the vector current, 
opening up spin-independent scattering with baryons. 
On the other hand, if the DM is \emph{Majorana}, Eq.~\eqref{eq:L}
presents the most general couplings of the $Z'$ to the DM and the SM
fermions. Hence, we will further consider only  Majorana DM (see also~\cite{Matsumoto:2016hbs}). 

The scenario of  DM which couples to the SM via a TeV-scale $Z'$ has been addressed in a number of references 
e.g.~\cite{Lebedev:2014bba,
Kahlhoefer:2015bea,
Bell:2015rdw,
Buchmueller:2014yoa,
Blennow:2015gta,
Alves:2013tqa,
Alves:2015pea,
Alves:2015mua,
An:2012va,
An:2012ue,
Frandsen:2012rk,
Arcadi:2013qia,
Shoemaker:2011vi,
Frandsen:2011cg,
Gondolo:2011eq,
Fairbairn:2014aqa,
Harris:2014hga,
Chala:2015ama,
Jacques:2015zha,
Brennan:2016xjh,
Dreiner:2013vla,
Ghorbani:2015baa},  
however it is still unclear how various direct and indirect searches compare in this case. 
The most important caveat in this comparison has to do with the signal from neutrino telescopes, because it is very sensitive to the annihilation
branching ratios of the DM into the EW bosons. 
The annihilation cross sections of DM into the SM fermions are helicity suppressed (except for $\mX \sim m_f$, with $f$ a SM fermion, where there is no suppression into $f \overline f $), and therefore bosonic channels are expected to be the dominant source of neutrinos that can be detected by IceCube from DM annihilation in the Sun.  

Parenthetically we note that another option for mediating the axial vector is via diagrams  with new scalars in the $t$- and $u$-channels~\cite{Agrawal:2010fh} (see also~\cite{Abdallah:2014hon}). 
In this work we will put these examples aside, because of the generic problem of flavor changing neutral currents that these scalars can mediate.
However it would also be interesting to see what the dominant annihilation channels are in this case.\footnote{An obvious way to avoid dangerous FCNC would be to impose some flavor symmetries, similar to those one usually assumes in flavor-blind SUSY models.} 

In this study we exclusively concentrate on the $Z'$ mediator in the $s$-channel, as it appears in Eq.~\eqref{eq:L}. 
Of course not every $Z'$ with arbitrary charges would be a consistent UV-completion of the effective contact term~\eqref{eq:AxialAxial}. 
If the $Z'$ is a gauge boson of a broken symmetry at multi-TeV scale, we should make sure that we are analyzing an anomaly free theory. 
This is a crucial demand, because without specifying the full matter content of the anomaly free theory, the annihilation rates of  DM into the gauge bosons  remain, strictly speaking,  incalculable. Indeed, some of them could occur via loop diagrams, whose value is only guaranteed to be finite  if the theory is renormalizable.
 
Based on the demands of renormalizability and anomaly cancellation,
one can relatively easily parametrize all possible flavor-blind
$U(1)'$ models. Any such model will be a linear combination of the SM
hypercharge and $U(1)_{B-L}$ (see Sec.~22.4 of~\cite{Weinberg:1996kr}). 
In  extreme cases we either get a $Y$-sequential theory or a pure $U(1)_{B-L}$. As we will shortly see, the latter is quite a boring case for DM direct and indirect
searches because it is halo-momentum suppressed.\footnote{So-called
  ``anomalous'' $U(1)$ has also been 
  discussed in literature, e.g.\ $U(1)_\psi$ \cite{Robinett:1982tq}. Note that these models demand the introduction of 
  spectator fermions to cancel the
  anomalies. One should necessarily take into account the
  contribution of spectators when calculating the DM annihilation branching ratios to avoid non-physical results.}

Because the charges of the SM fermions under the new $U(1)$ are a two-parameter family, we will conveniently parametrize the generator of
the new symmetry as 
\beq\label{eq:thetadef}
\cos \theta\ t_Y + \sin \theta\ t_{B-L}
\eeq
where $t_Y$  and $t_{B-L}$ stand  for the generators of the hypercharge and $B-L$ symmetries respectively.  
To ensure that the fermion masses are gauge invariant, the SM fermion
charges under the $U(1)'$ unambiguously determine the SM Higgs charge
under the $U(1)'$. For completeness we list all the charges under the
gauge symmetries, including the $U(1)'$, in
Table~\ref{tab:U1charges}. 

\begin{table}[t]
\centering
\begin{tabular}{cccccc}
\toprule
 & $SU(3)$ & $SU(2)$ & $\UY$  & $\UBL$  & $\Up$ \\
\midrule
$\begin{pmatrix} \nu^{\ell_i}_L \\ \ell^i_L \end{pmatrix}$ &
$\textbf{1}$ & $\textbf{2}$ & $-\frac 12$ & $-1$ & $-\frac 12 \cth
-\sth$ \\ 
$\left(\ell^i_R\right)^\text{C}$ & $\textbf{1}$ & $\textbf{1}$ & $1$ &
$+1$ & $\cth+\sth$ \\ 
$\begin{pmatrix} u^i_L\\ d^i_L\end{pmatrix}$ & $\textbf{3}$ &
$\textbf{2}$ & $\frac 16$ & $+\frac 13$ & $\frac 16 \cth + \frac 13
\sth$\\ 
$\left( u^i_R\right)^\text{C}$ & $\overline{\textbf{3}}$ &
$\textbf{1}$ & $-\frac 23$ & $-\frac 13$ & $-\frac 23 \cth -\frac
13\sth$ \\ 
$\left( d^i_R\right)^\text{C}$ & $\overline{\textbf{3}}$ & 
$\textbf{1}$ & $\frac 13$ & $-\frac 13$ & $\frac 13 \cth - \frac 13
\sth$ \\ 
$\Phi= \begin{pmatrix} \phi^+ \\ \phi^0\end{pmatrix}$ & $\textbf{1}$ &
$\textbf{2}$ & $\frac 12$ & $0$ & $\frac 12 \cth$ \\ 
\bottomrule
\end{tabular}
\caption{Charges of the SM matter content under the gauge symmetries of the SM and the gauge $U(1)'$ with the generator~\eqref{eq:thetadef}. 
  $i$  stands for the  family index.} 
\label{tab:U1charges}
\end{table}

These charges have a strong impact on the DM phenomenology in this scenario. 
Because the SM Higgs couples to the $Z'$, tree level couplings between the $Z'$ and the EW gauge bosons are induced after EW symmetry breaking. 
In this case $Z'$ mixes with the $Z$. This allows annihilations of the DM to  EW gauge bosons at the tree level.   

\subsection{Direct constraints on $Z'$ from LHC searches}
Here we review  direct constraints on this $Z'$ from the LHC.
In addition, we will consider monojet constraints on DM production in Sec.~\ref{sec:monojet}.
The easiest way to
spot a $Z'$ at a collider is via an analysis of the leptonic modes, unless
they are highly suppressed. For these purposes we recast a CMS
search for a narrow $Z'$ in the leptonic
channel~\cite{Khachatryan:2014fba}, which conveniently phrases the
constraints in terms of
\begin{equation}
R_\sigma \equiv\frac{ \sigma(pp \to Z')\times
BR(Z' \to l^+ l^-)}{\sigma (pp \to Z) \times BR(Z \to l^+ l^-)}
\end{equation}
and for the reference point we take $\sigma (pp \to Z ) \times BR(Z \to
l^+ l^- ) = 1.15$~nb at $\sqrt{s} = 8$~TeV~\cite{Chatrchyan:2014mua}.

\begin{figure}[t]
\centering 
\includegraphics[width=0.48\textwidth]{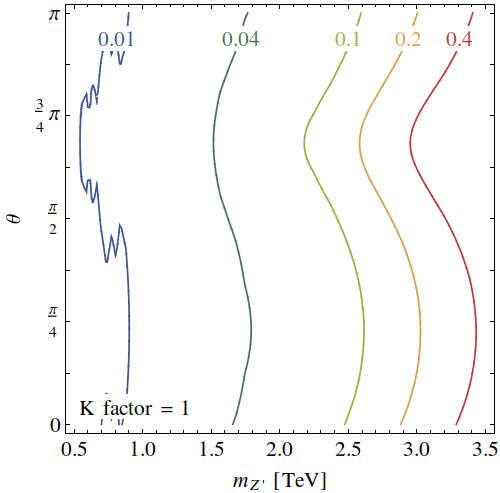} \hfill
\includegraphics[width=0.48\textwidth]{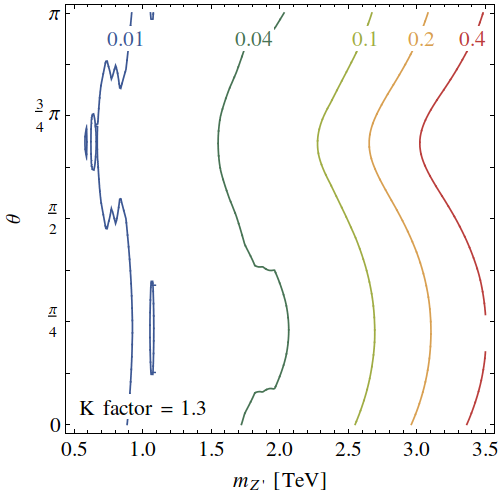}
\caption{Maximal allowed gauge couplings $\gZp$ of the hidden $U(1)'$ as a function of the angle $\theta$ (Eq.~\eqref{eq:thetadef})
  and of $\mZp$. On the LH side we assume the nominal LO
  cross sections, on the RH side we apply a flat $k = 1.3$ factor.}
\label{fig:Z'constraints}
\end{figure}

We show the results of our recast in Fig.~\ref{fig:Z'constraints} as a
function of the mass $\mZp$ of the $Z'$ and the angle $\theta$ as
defined in Eq.~\eqref{eq:thetadef}.
Note that, for a given $\gZp$ and $\theta$, all  couplings to the SM are fixed, therefore both production cross sections and  
BRs are unambiguously determined by these values.
In Fig.~\ref{fig:Z'constraints} we show the exclusions both for nominal
cross sections, as we get from {\tt MadGraph5}~\cite{Alwall:2014hca},
as well as for cross 
sections one gets by applying the flat $k$-factor $k =1.3$ (similar
to the suggestion in~\cite{CMS-PAS-EXO-15-005}).
It is apparent from these lines that, in order to have a $Z'$ with $\mathcal{O}(1)$ couplings, its mass must be $\mZp\gtrsim 3\TeV$.
Note that the
constraints from LEP, coming from the mixing 
between the $Z$ and $Z'$, which further affect the $T$-parameter, are
much weaker than the direct LHC constraints. 

\subsection{DM annihilation to SM particles}
The annihilation proceeds via $Z'$ into the SM fermions and the EW gauge bosons. 
The couplings of the $Z'$ to the EW gauge bosons, in particular to $Zh$ and $W^+W^-$, arise at the \emph{tree level} after  EW symmetry breaking, when the Higgs Goldstone modes are ``eaten'' by the massive gauge bosons. 
Before  EWSB one can write down the couplings of the
$Z'$ to the SM as follows:
\begin{equation}
\label{eq:vpa}
\begin{aligned}
\cL_{Z'-\text{SM}} = & \gZp Z'_\mu \Big( \overline e^i (c_e^V \gamma^\mu+c_e^A \gamma^\mu \gamma_5) e^i 
+ \overline \nu^i (c_\nu^V \gamma^\mu+c_\nu^A \gamma^\mu \gamma_5)
\nu^i \\ 
& + \overline u^i (c_u^V \gamma^\mu+c_u^A \gamma^\mu \gamma_5) u^i 
+ \overline d^i (c_d^V \gamma^\mu+c_d^A \gamma^\mu \gamma_5) d^i 
\Big) + (D_\mu H)^\dagger (D^\mu H)
\end{aligned}
\end{equation}
with $D_\mu = D_\mu^{\rm SM}  - \frac{i}{2} \cth\, \gZp Z'_\mu$ and
the vector/axial vector couplings of the $Z'$ to the SM fermions
given in Table~\ref{tab:U1coefficients}. 
The Lagrangian describing DM is 
\begin{equation}
\label{eq:L DM}
\cL_{\rm DM} = \frac 12 \overline \chi \Big(i\slashed \partial -\mX\Big) \chi 
+\frac 12 \gZp\gX  Z'_\mu \Big( \overline \chi \gamma^\mu\gamma_5 \chi \Big) \,,
\end{equation}
where $\gX$ is the coupling of $\chi$ to the $Z'$.
In this paper we do not consider kinetic mixing, and we neglect the effects of renormalization group equations 
that could mix $Z$ and $Z'$ via loop effects. We also do not take into account the running of the operator coefficients 
due to the RG flow because the quantitative effect is expected to be mild (see recent Ref.~\cite{D'Eramo:2016atc}
for the details). 
%\footnote{\new{We also do not take into account the running of the operator coefficients due to the renormalization group flow, because the quantitative impact would be very mild: the leading correction to the coefficients $\cAq$, $\cVq$ would come from the Yukawa contribution involving the top quark, and amount to about the 10-20\% \cite{D'Eramo:2016atc}. \comm{Understand better this point!}
%}}
The latter is indeed a minor effect, if the model is such that the mixing is zero at a scale close to the electroweak one.
We do not specify the dynamics of the spontaneous symmetry breaking sector of $\Up$, and for our purposes we just assume 
that it provides a mass term $\tfrac 12 \mZp^2 Z^{\prime\, 2}$.

As we expand Eq.~\eqref{eq:vpa} we find that the $Z'$ mixes with $Z$, with
a mixing angle $\psi$ fully determined by the mass $\mZp$ of the physical mass eigenstate, the value of the coupling $\gZp$ and the angle $\theta$.
If we denote the mass of the lighter mass eigenstate by $m_Z=91.2$ GeV, then $\psi$  turns out to be of order $\psi\sim \gZp \cth \big( \frac{m_Z}{\mZp}
\big)^2 \ll 1$ in the regime $\mZp\gg m_Z$ that we consider in this work. 
For this reason, in the remainder of the paper  we ignore the mixing for simplicity of notation, and denote both the interaction or mass eigenstate equivalently by $Z'$, and the lighter mass eigenstate identifiable with the SM vector boson by $Z$ .

\begin{table}[h!]
\centering
\begin{tabular}{ccc}
\toprule
SM fermion $f$ & $\cVf$: coeff.\ of $\gZp\overline f \slashed{Z'} f$ & $\cAf$: coeff.\ of $\gZp \overline f \slashed{Z'}\gamma_5 f$ \\
\midrule
leptons & $- \frac 34 \cth -\sth$ & $-\frac 14 \cth$ \\
neutrinos & $- \frac 14 \cth -\frac 12 \sth$ & $\frac 14 \cth +\frac
12 \sth$ \\ 
up quarks & $\frac{5}{12} \cth +\frac 13 \sth$ & $\frac 14 \cth$ \\
down quarks & $-\frac{1}{12} \cth +\frac 13 \sth$ & $-\frac 14 \cth$
\\ 
\bottomrule
\end{tabular}
\caption{Coefficients of the vector and axial vector bilinear currents
  for the SM fermions ({\it cf} Eq.~\eqref{eq:vpa}). With obvious meaning of the notation, the
  coefficients $g^V$ and $g^A$ are obtained from $g^L,\, g^R$ of Tab.~\ref{tab:U1charges} via $g^L
  P_L +g^R P_R = \tfrac{g^L+g^R}{2} + \tfrac{-g^L+g^R}{2}\gamma_5$, so that $g^V= \tfrac{g^L+g^R}{2},\, g^A=\tfrac{-g^L+g^R}{2}$.} 
\label{tab:U1coefficients}
\end{table}

Due to the $Z$-$Z'$ mixing, the heavy $Z'$ couples at tree level to $Zh$ with a vertex 
\begin{equation}
\label{eq:vertex Z'Zh}
\frac{1}{2}\cth \gZp g_Z v \, \Big( Z' Z h \Big) \,, 
\end{equation}
where $g_Z=\sqrt{g^2+g^{\prime\, 2}} = 2 m_Z /v$ with $v=246$ GeV,
and with $W^+ W^-$ with a vertex equal to $\sin \psi$ times the SM vertex for $ZW^+W^-$,
\begin{equation}
\label{eq:vertex Z'WW}
\sin\psi \cdot  ig \cos \theta_\text{W} \Bigl[ \Bigl(W^+_\mu W^-_\nu - W^+_\nu W^-_\mu \Bigr) \partial^\mu Z^{\prime\,\nu} + W^+_{\mu\nu} W^{-\, \mu}Z^{\prime\, \nu} - W^-_{\mu\nu} W^{+\, \mu}Z^{\prime\, \nu} \Bigr] \,,
\end{equation}
where $W^\pm_{\mu\nu}=\partial_\mu W^\pm_\nu-\partial_\nu W^\pm_\mu$. 
Notice that both \eqref{eq:vertex Z'Zh} and \eqref{eq:vertex Z'WW} are proportional to $\cth$, thus they vanish in the pure $\UBL$ limit.
Interaction~\eqref{eq:vertex Z'WW} turns out to be velocity suppressed, 
% is proportional to the momentum flowing along the vertex,
thus it gives a small cross section in the low DM velocity regime of DD and IC, and both \eqref{eq:vertex Z'Zh} and \eqref{eq:vertex Z'WW} are proportional to $\cth$, thus vanish in the pure $\UBL$ limit.
Because of these suppressions, loop channels are also relevant at low kinetic energy, as discussed in 
Sec.~\ref{sec:BR}.
In particular, the annihilation of $\chi\chi$ to $Zh$ occurs at tree level, except for the case in which the $U(1)'$ 
extension is a pure $\UBL$ gauge symmetry. At low velocity, annihilation into $W^+W^-$ is instead dominantly 
driven by diagrams with a fermionic loop  (see Appendix~\ref{sec:annihilation}).

\subsection{Calculation of DM relic density}
Now that we have all the tree level couplings of the $Z'$, we are ready to
calculate the thermal relic abundance of the Majorana DM. 
We calculate the annihilation rates and perform the thermal average using the procedure
of~\cite{Gondolo:1990dk}. The result for the thermally averaged self-annihilation cross section as a function of temperature $T$ is 
\begin{equation}
\label{eq:asv}
\asv = \frac{x}{8\,\mX^5}\frac{1}{\big(\bes{2}(x)\big)^2} \int_{4\mX^2}^\infty \sigma_\text{ann} \sqrt s \, \left(s-4 \mX^2 \right) \,\bes{1}\left(\frac{x\sqrt s}{\mX} \right)\, \d s\,,
\end{equation}
where $x=\mX/T$, and $\bes{i}$ is the modified Bessel function of order $i$.

%The final DM abundance is then computed following \cite{Kolb:1990vq}. 
We do not approximate the thermally averaged cross section with a low DM velocity expansion, 
since close to the resonance $\mX \lesssim \mZp/2$, terms of higher order can also yield important contributions  
to the relic density~\cite{Griest:1990kh}.

\begin{figure}[h!]
\centering
  \includegraphics[width=0.7\textwidth]{gZpRelicDensity}
  \includegraphics[width=0.7\textwidth]{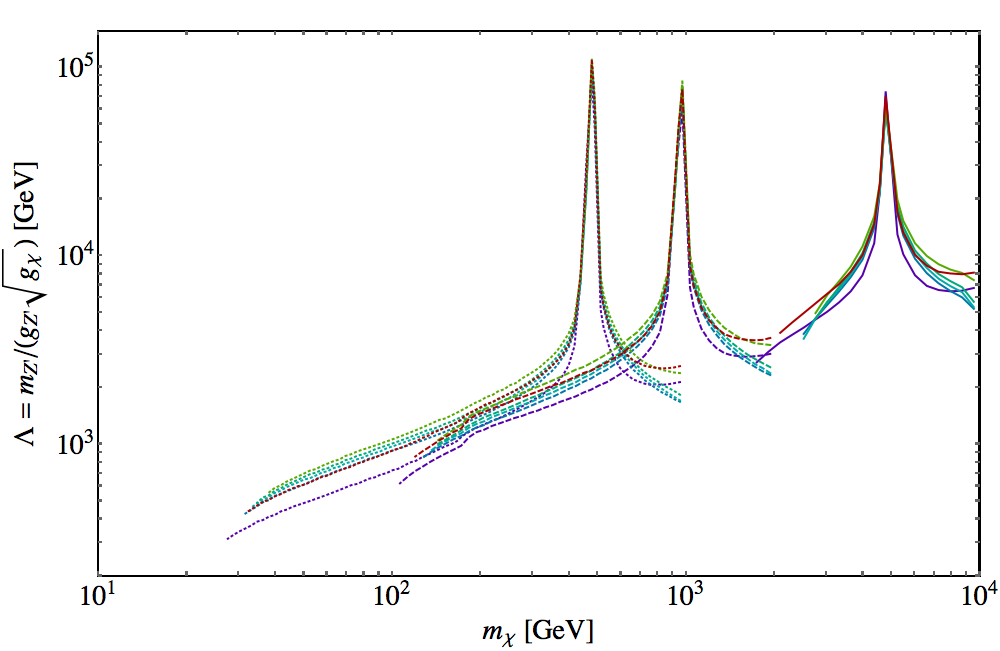}
  \caption{Top: Value of $\gZp \sqrt{\gX}$ that yields to the correct relic density, for three masses of the $Z'$ and different values of $\theta$. The regions colored with multiple shades of gray in the upper part of the plot (one for each $\theta$, $\mZp$) show the regions where $\GZp$ becomes of the order of $\mZp$, signalling the transition to the non-perturbative regime.
  Bottom: Same data as left, in the plane $\Lambda$ {\it vs.} $\mX$.
  }
  \label{fig:gZp relic}
\end{figure}

Once we fix the values of the angle $\theta$ and $\mZp$, we are left with $\gZp$ as the only free parameter. 
In Fig.~\ref{fig:gZp relic} we show the value of $\gZp$ that yields the correct relic density as measured 
by~\cite{Ade:2015xua}. The areas above (below) the lines correspond to points of the parameter space where the DM 
is under- (over-) abundant in the thermal 
scenario.

For the calculation of the relic density we relied only on tree level cross sections, which we confirmed are dominant at  typical freeze-out energies.
Varying the value of $\gX$ (which for this computation was set to 1) while keeping  the product $\gZp \sqrt{\gX}$ fixed only very mildly affects the decay width $\GZp$, and would have practically no effect on Fig.~\ref{fig:gZp relic}.

The lines in Fig.~\ref{fig:gZp relic} stop at $\mX=\mZp$: above that threshold, annihilation into $Z'Z'$ opens up, and in principle one would need to specify the details of the spontaneous symmetry breaking sector of $U(1)'$ in order 
to compute  the relic abundance precisely. This is not required for our purpose 
of understanding the plausible range of values for $\gZp$ that yield a relic abundance close to the observed one.

%%%%%%%%%%%%%%%%%%%%%%%%%%%%%%%% 

%%%%%%%%%%%%%%%%%%%%%%%%%%%%%%%%
\section{Overview of direct and indirect bounds}
\label{sec:IC}
% !TEX root = main_draft.tex
The constraints on the scenario that we have described above come from
three different primary sources: direct detection, neutrino telescopes
and the LHC. 
Since we assume a Majorana DM particle, all interactions that we get between the nuclei and the DM are either spin dependent or halo velocity suppressed, or both. 
We will comment on these interactions in detail in Subsec.~\ref{sec:DD}. 
This particular feature renders the direct detection results much less efficient than in the case of spin-independent interactions, while other experiments,
notably neutrino telescopes and the LHC, become competitive. 
In the following section we will carefully go through all of these different
experiments and discuss the bounds that they produce.     

\subsection{Direct Detection Experiments}
\label{sec:DD}
In this type of experiment, in a model with a $Z'$ mediator the effective DM theory is always valid, because the transferred momentum never exceeds hundreds of MeV, well below the mediator scale.
The effective contact terms that one gets between the DM and SM quarks are 
\begin{equation}
\cL_{eff} = \frac{\cVq}{\Lambda^2} \bar \chi \gamma^\mu \gamma^5 \chi \, \bar q \gamma_\mu
q + \frac{\cAq}{\Lambda^2} \bar \chi \gamma^\mu \gamma^5 \chi \, \bar q \gamma_\mu
\gamma^5  q 
\label{eq:efflagrangianscattering}
\end{equation}
%\beq
%\cL_{eff} = c_{AV} \bar \chi \gamma^\mu \gamma^5 \chi \, \bar q \gamma_\mu
%q + c_{AA} \bar \chi \gamma^\mu \gamma^5 \chi \, \bar q \gamma_\mu
%\gamma^5  q 
%\label{eq:efflagrangianscattering}
%\eeq
with coefficients $g$ given %(up to the overall coefficient $\gZp^2 \gX/\mZp^2$)
in Table~\ref{tab:U1coefficients}, and $1/\Lambda^2 = \gZp^2 \gX/\mZp^2$.       

It is straightforward to translate these interactions to the more
intuitive language of the NR effective theory. 
Using the dictionary of Ref.~\cite{Anand:2013yka} and considering a nucleus $N$ instead of the partons $q$ we get 
\beqa
\bar \chi \gamma^\mu \gamma^5 \chi \bar N \gamma_\mu \gamma^5 N & \to &
-4 \vec S_\chi \cdot \vec S_N = 4 \cO_4 \\
\bar \chi \gamma^\mu \gamma^5 \chi N \gamma_\mu N & \to & 2 \vec v
\cdot \vec S_\chi + 2i \vec S_\chi \cdot \left(\vec S_N \times
  \frac{\vec q}{m_N} \right) = 2\cO_8 + 2\cO_9.
\label{eq:NRops}
\eeqa 
For a generic $Z'$ (arbitrary $\theta$ angle in~\eqref{eq:thetadef})
we get both axial current - axial current (AA) and axial current - vector
current (AV)
interactions with the coefficients being roughly of the same order of
magnitude. However, the latter induces interactions that are halo
velocity suppressed ($\cO_8$) and both nuclear spin and halo velocity
suppressed ($\cO_9$). 
This usually renders the AV interactions smaller than 
the AA  interactions for  direct detection
and  solar capture, although we do include AV interactions when we derive our bounds. 
In fact, the halo-velocity
suppression in $\cO_8$ is sometimes comparable to the suppression due to the
spin-dependence in $\cO_4$. The operator $\cO_9$, which is both
spin-dependent and halo velocity dependent via the exchange momentum
$\vec q$ is completely negligible. The one-loop contributions may induce a spin-independent scattering cross section of the DM with the nucleons, but their quantitative impact is negligible in our model \cite{Haisch:2013uaa}.

For direct detection, there is a special point in parameter space where the usual
spin-dependent operator completely shuts down. This happens when our
new gauge symmetry is exactly $U(1)_{B-L}$, or, in our language,
$\theta = \frac{\pi}{2}$. In this case the $Z'$ couples only to the
vector current of the  SM, and therefore the NR interaction operators
proceed only via $\cO_8$ and subdominantly via $\cO_9$. 

We derive the exclusion bounds obtained by the experiments LUX, XENON100, CDMS-Ge, COUPP, PICASSO,  SuperCDMS with the help of the tables made available by~\cite{DelNobile:2013sia}, and the bounds by PICO from \cite{Amole:2015pla}. The strongest constraints among them are shown in Figs.~\ref{fig:results theta=0} and \ref{fig:results other thetas}, together with the constraints from IceCube and monojet searches.

\subsection{LHC monojet constraints}
\label{sec:monojet}
In this section, we explore the bounds on our $\Up$ model coming from LHC searches for DM, analyzing events with one hard jet plus missing transverse energy ($\met$).

Currently, the strongest exclusion limits are  from the CMS analysis \cite{Khachatryan:2014rra}, which analyzes $19.7\, \fb^{-1}$ at a collision energy of 8 TeV.

Despite the fact that the use of EFT to investigate dark matter signatures through missing energy advocated in Refs.~\cite{Goodman:2010ku,Goodman:2010yf,Bai:2010hh} is severely   limited at the LHC energies \cite{Busoni:2013lha,Busoni:2014sya,Racco:2015dxa,Buchmueller:2013dya},  we can nevertheless use the EFT interpretation of the exclusion bounds, as it is consistent for $\mZp\gtrsim 2$ TeV \cite{Busoni:2013lha,Busoni:2014sya,Racco:2015dxa,Buchmueller:2013dya}. The effective operators describing the interaction between DM and quarks, in the high $\mZp$ limit, are
\begin{equation}
\label{eq:operator D6}
\frac{1}{\Lambda^2} \sum_q \cVq \, \big( \overline \chi \gamma^\mu \gamma_5 \chi\big) \,\big( \
\overline q \gamma_\mu q\big)
\end{equation}
and 
\begin{equation}
\label{eq:operator D8}
\frac{1}{\Lambda^2} \sum_q \cAq \, \big( \overline \chi \gamma^\mu \gamma_5 \chi\big) \,\big( \overline q \gamma_\mu \gamma_5 q\big) \,,
\end{equation}
where $\Lambda\equiv \mZp/(\gZp \sqrt{\gX})$ and the coefficients $\cVq,\, \cAq$ are given in Table~\ref{tab:U1coefficients}.

At  LHC energy scales, the occurrence of a vector or an axial vector current in the fermion bilinears in~\eqref{eq:operator D6} 
and~\eqref{eq:operator D8} does not affect the cross section for the production of DM.
This is also apparent from the experimental exclusion limits reported in~\cite{Khachatryan:2014rra}.

The CMS analysis recasts the exclusion bound as a function of the coefficient $\Lambda$ of Eq.~\eqref{eq:operator D8}, assuming that {\bf 1)} all the $\cAq$ coefficients are equal to 1, and {\bf 2)} that $\chi$ is a Dirac fermion (with canonically normalized kinetic term). 
These two assumptions are not true for our analysis. The second assumption gives an overall factor of 2 in the cross section for the Majorana case relative to the Dirac case.
%\footnote{The amplitude derived from \eqref{eq:operator D8}, where $\chi$ is Majorana fermion with canonically normalized kinetic term, is larger of a factor 2 with respect to the same operator written for a Dirac fermion. This results in a factor 4 in the modulus squared. When integrating over the momenta of the final state, another factor of $\tfrac 12$ has to be included in the Majorana case, in order not to double count the indistinguishable configurations. }
We take into account both the first assumption and the convolution with the parton distribution functions (PDFs) of quarks, by means of a parton level simulation performed with \texttt{MadGraph 5}~\cite{Alwall:2014hca}. 
We simulate the signal both with the EFT defined by CMS and with our model, and we compute for each value of $\theta$ (which determines $\cVq,\, \cAq$) and $\mX$ a rescaling factor that we use to rescale the nominal limit reported by the CMS analysis. 

\begin{figure}[t]
\centering
  \includegraphics[width=0.6\textwidth]{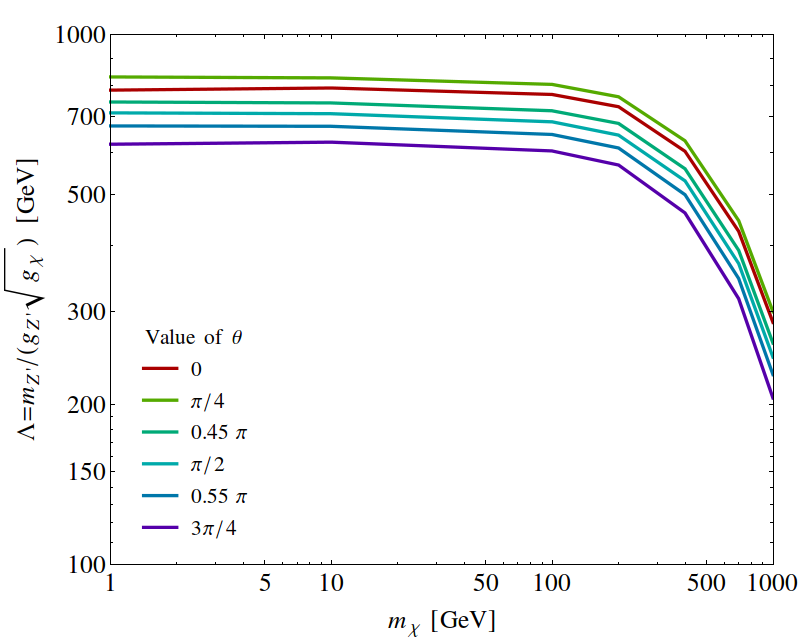}
  \caption{Bound on $\Lambda=\mZp/(\gZp \sqrt{\gX})$ as a function of $\mX$, for the six values of $\theta$ shown in the legend, from the CMS monojet search \cite{Khachatryan:2014rra}.}
  \label{fig:bound monojet}
\end{figure}

The final result\footnote{\new{
We remark the following. If the $Z'$ mass $\mZp$ is larger than a few TeV, the bounds shown in Fig.~\ref{fig:bound monojet} and the following fall in a region in which the product $\gZp g_\chi^{1/2}$ is necessarily $\gtrsim 1$. This is in contrast with the fact that our $Z'$ model has a rather large mediator width, and it must be $\gZp g_\chi^{1/2} \lesssim 1$ in order to have $\GZp\lesssim\mZp$. For this reason, with the present experimental sensitivity the lines of Fig.~\ref{fig:results other thetas} correspond to a realistic physical situation only for $\mZp\sim\,{\rm few\ TeV}$.
}}
 for the bounds on $\Lambda$ from monojet searches are reported in Fig.~\ref{fig:bound monojet}.

%%%%%%%%%%%%%%%%%%%%%%%%%%%%%%%
%%%%%%%%%%%%%%%%%%%%%%%%%%%%%%%
%%%%%%%%%%%%%%%%%%%%%%%%%%%%%%%

\new{
\subsection{Constraints from observations of $\gamma$-ray spectrum}

We now examine the exclusion bounds that can be obtained from the analysis of the $\gamma$-ray continuum spectrum. 
Limits coming from $\gamma$-ray lines are irrelevant for our model because the 
$\gamma\gamma$, $Z\gamma$ and $h\gamma$ channels are strongly suppressed.
The most stringent and robust bounds on the $\gamma$-ray continuum spectrum come from the observation of a set of 15 Dwarf Spheroidal Galaxies (dSph) performed by Fermi-LAT~\cite{Drlica-Wagner:2015xua,Ackermann:2015zua}.
The robustness of these bounds against astrophysical uncertainties comes mostly from the fact that the photon flux is integrated over the whole volume of the dwarf galaxy, and in this way the inherent uncertainty due to the choice of the DM profile is largely diluted (as a reference, results are presented for the Navarro-Frenk-White profile~\cite{Navarro:1996gj}). However, the bounds are practically 
independent on the profile choice and the variation of the bounds due to J-factor uncertainties typically does not exceed 
30\%~\cite{Ackermann:2015zua}. 

Here we should also briefly comment on HESS searches for DM using diffuse $\gamma$-rays from the Galactic Center. 
These bounds, claimed by the HESS collaboration~\cite{HESS-2016}, are nominally much stronger than Fermi dSph bounds for heavy 
DM, $\mX \gtrsim 1$~TeV. 
However one should also consider the uncertainties on these bounds. Unlike Fermi-LAT searches for emission from the Galactic Center, which mask a large region around the Galactic Center,\footnote{For instance, Ref.~\cite{Ackermann:2012rg} practically exclude an area of $10^\circ$ from consideration. 
Similarly, theoretical studies (see {\it e.g.}~\cite{Daylan:2014rsa}) mask out $1^\circ$ to $2^\circ$ around the Galactic Center. 
We do not show the Fermi-LAT Galactic Center bounds 
on our plots because they are inferior to the Fermi-LAT dSph bounds. 
It is also worth mentioning that measurements of the dSphs are essentially foregrounds-free, which renders them extremely robust.} 
%HESS merely masks a tiny region of $0.3^\circ$ around the Galactic Center, mainly because of the backgrounds for the cosmic rays, which is of course not present for the Fermi-LAT. 
HESS merely masks a tiny region of $0.3^\circ$ around the Galactic Center, mainly to avoid the cosmic ray photon background, which is of course not present for the space-based Fermi-LAT.
This makes the search much more vulnerable both 
to the astrophysical uncertainties and to the choice of the DM profile. HESS assumes cuspy DM profiles in its search (NFW
and Einasto), and if the profile is cored, the bounds can be attenuated by orders of magnitude. This point was nicely 
illustrated in the context of a different (wino) DM candidate in Refs.~\cite{Cohen:2013ama,Fan:2013faa}.
Therefore we decided  not to show HESS' bounds.

\begin{figure}[h!]
\begin{center}
\includegraphics[width=0.49\textwidth]{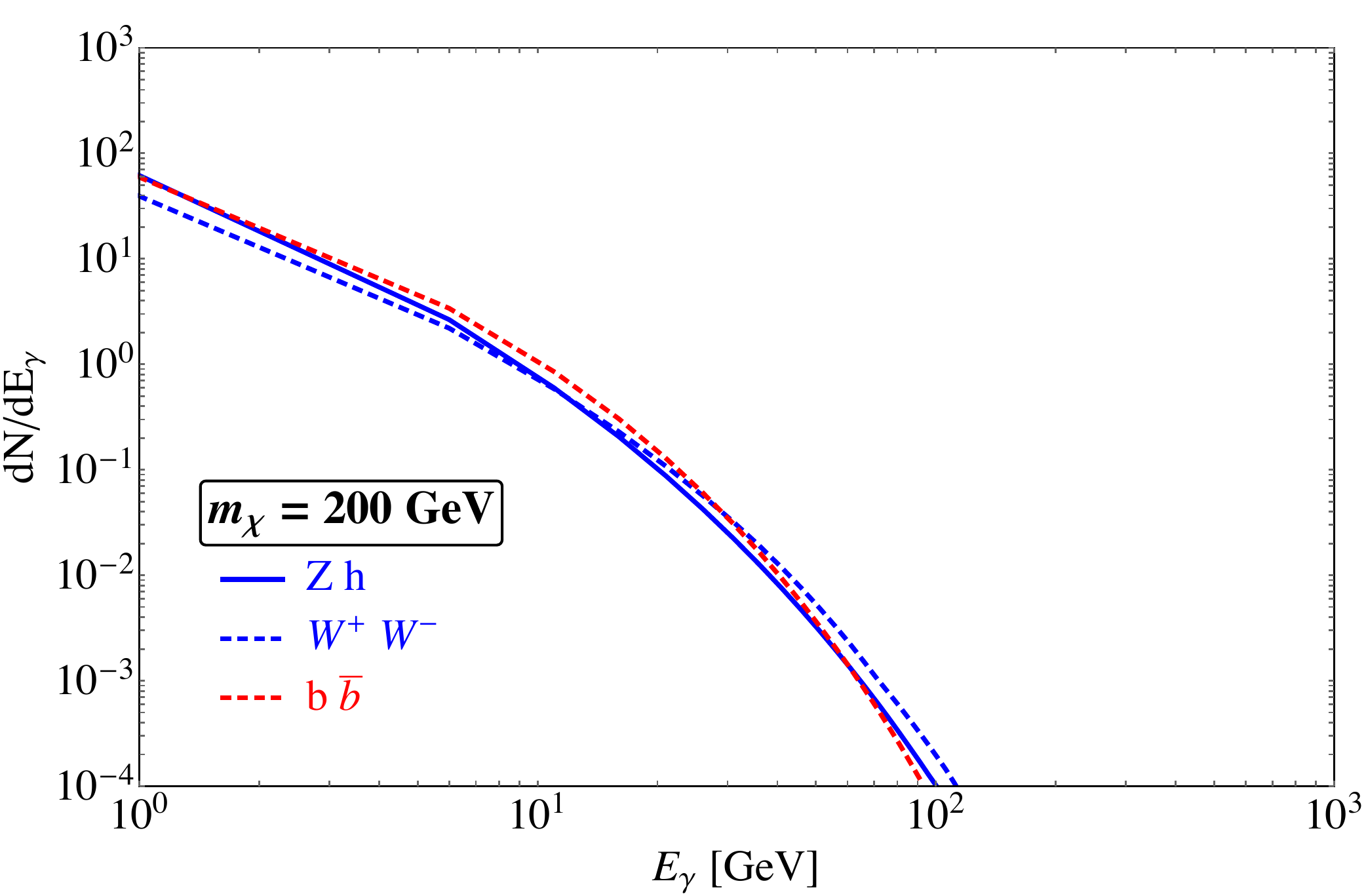}
\includegraphics[width=0.49\textwidth]{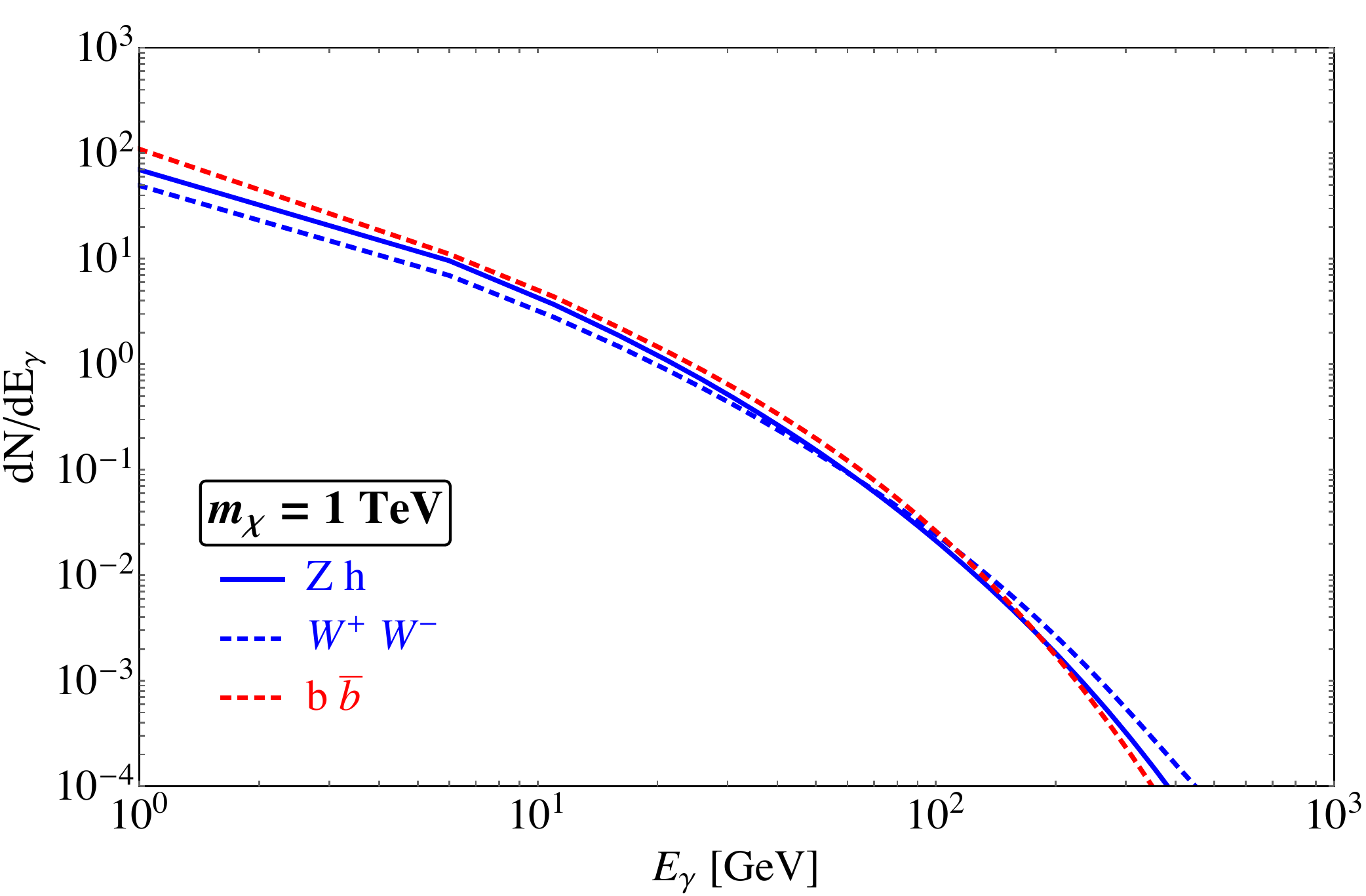}
\includegraphics[width=0.49\textwidth]{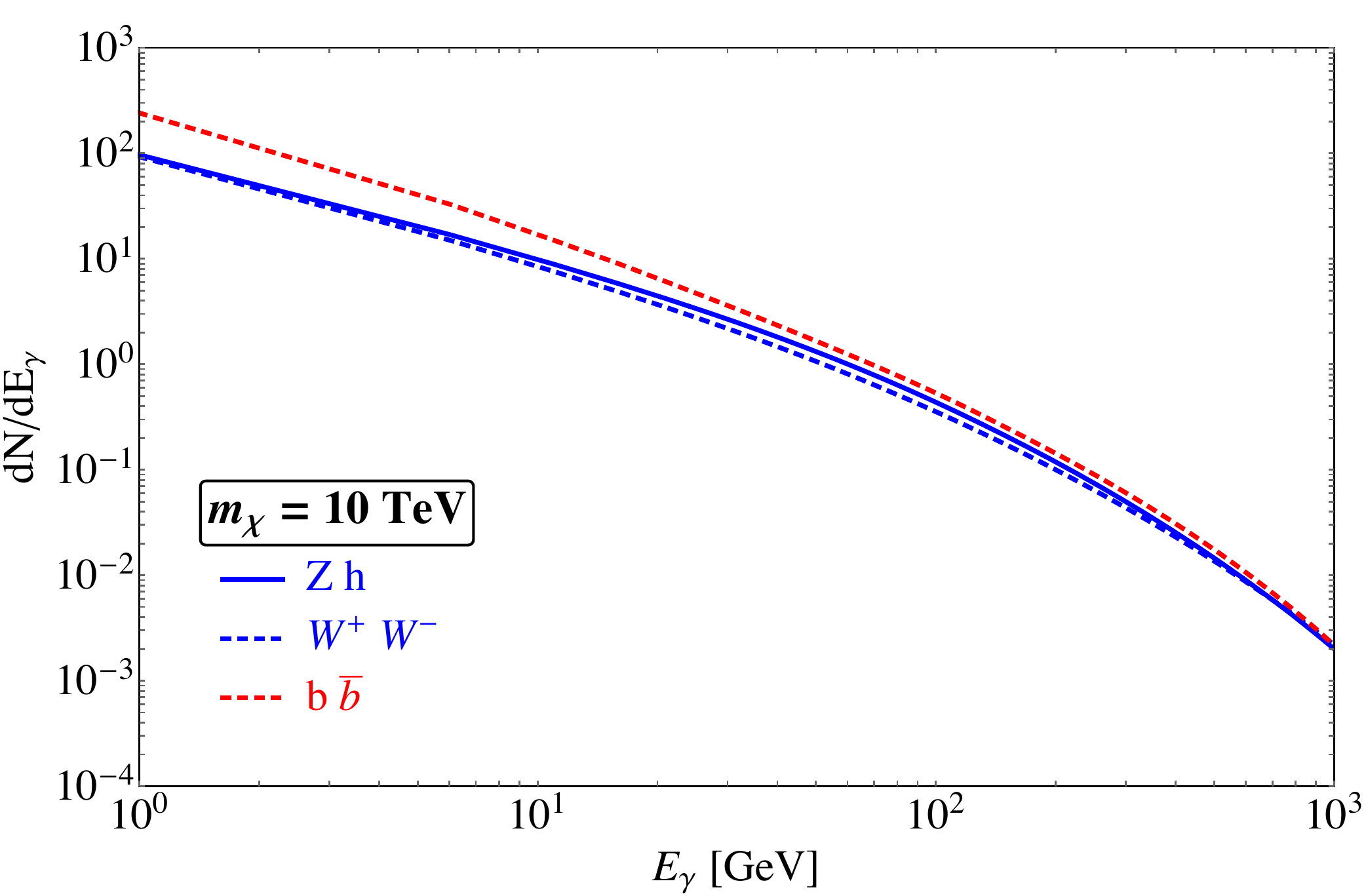}
\end{center}
\caption{Secondary photon spectra for different primary annihilation channels, for three reference DM masses.}
\label{fig:gammaspectrum}
\end{figure}

In order to properly recast the results of~\cite{Ackermann:2015zua} for our model two ingredients are necessary. 
The first one is a knowledge of the spectrum of $\gamma$-rays from DM annihilations, 
which can be computed using the tables provided in Ref.~\cite{Cirelli:2010xx}. Results of this calculation are shown in 
Fig.~\ref{fig:gammaspectrum}, for three reference values of the DM mass.
The second ingredient would be the exclusion limits on the flux of $\gamma$ rays, information which is not provided by 
the Fermi-LAT collaboration.
For this reason, we adopted a simplified recasting procedure. Firstly, we identified in each interval in $\mX$ the leading annihilation channel providing secondary photons, and approximated the total annihilation cross section with the one into that particular channel (or, in the case of multiple relevant primary channels with a similar $\gamma$-ray spectrum, we considered their sum).
Secondly, we used the results of~\cite{Cirelli:2010xx} to compare the photon flux from our dominant primary channel to the benchmark fluxes, namely $e^+e^-$, $\mu^+\mu^-$, $\tau^+\tau^-$, $u \bar u$, $b \bar b$ and $W^+W^-$, which are shown in the left panel of Fig.~\ref{fig:gamma rays constraints}.
% for Fermi and $\mu^+\mu^-$, $\tau^+\tau^-$, $b \bar b$, $t \bar t$ and $W^+W^-$ for HESS.
We used the limit on $\asv$ from the channel with the most similar photon flux as the limit on our channel.
Finally, the limit on $\asv$ is converted into a limit on $\Lambda$.
Though rough, we expect our procedure to provide bounds with at least an order of magnitude accuracy on $\sSD$ (which translates into a factor of $\lesssim 2$ on $\Lambda$). %Moreover, our bounds are always conservative, since we include only the main annihilation channels.

As we will discuss in Sec.~\ref{fig:BR}, there are two regions of interest, the first for $\mX<m_W$ and the second for $\mX>m_W$.
Leaving aside for the moment the peculiar case $\theta = \pi/2$, for $\mX<m_W$ the dominant channels are $b\bar b$ and $c \bar c$, which give a similar $\gamma$ spectrum, so the Fermi-LAT limit on $b \bar b$ can be assumed.
On the other hand, for $\mX>m_W$ the dominant annihilation channel is $Zh$, complemented by $W^+W^-$, $ZZ$ and $t \bar t$, all of which give a similar photon flux. Since the flux of photons in the $Zh$ channel is similar (up to a factor $\lesssim 2$) to that in the $b\bar b$ channel, we again picked the Fermi-LAT limit on the $b\bar b$ channel.
%HESS limits are presented only for $\mX\gtrsim 200\GeV$, and we picked the limit on the $W^+W^-$ channel.
In the peculiar case $\theta=\pi/2$, the dominant channels are leptonic for $\mX<m_W$ and $W^+W^-$, $ZZ$ for $\mX>m_W$. Therefore, for $\mX<m_W$ we picked the $\tau^+\tau^-$ channel (which, among leptons, gives the strongest bounds), while for $\mX>m_W$ we summed the $W^+W^-$ and $ZZ$ contributions and compared with the limits on the $W^+W^-$ channel.
Fig.~\ref{fig:gamma rays constraints} shows the result of this recast in the right panel.
\begin{figure}[h!]
\includegraphics[width=0.49\textwidth]{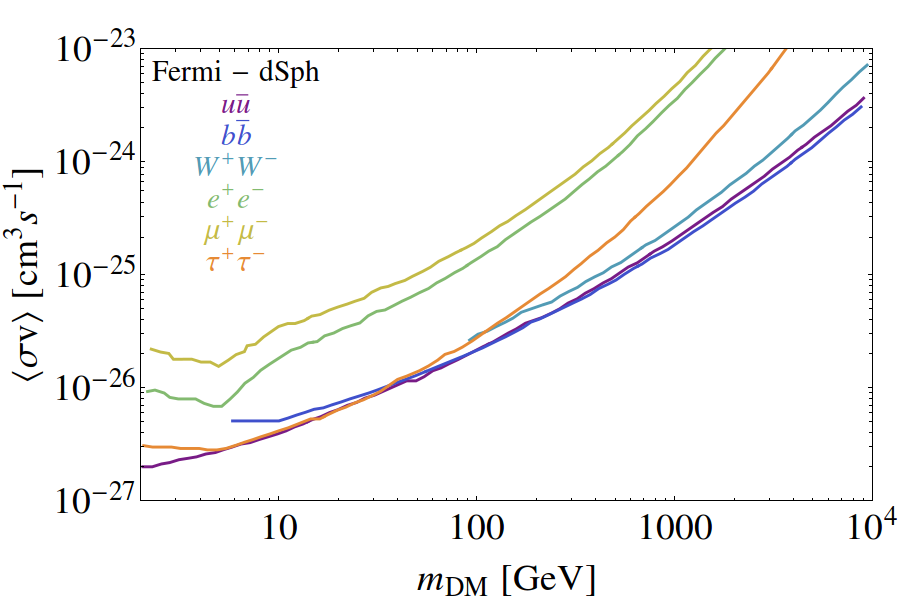} \hfill
\includegraphics[width=0.49\textwidth]{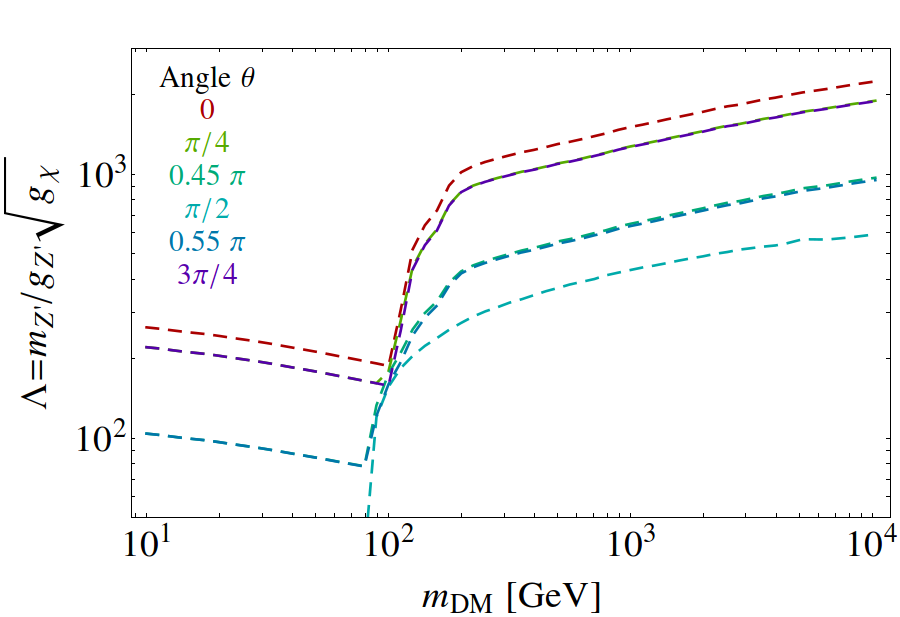}
\caption{Left: bounds reported by the Fermi-LAT~\cite{Ackermann:2015zua} 
%and HESS~\cite{HESS-2016}
collaboration assuming that DM annihilates in the specified channel. Right: recast of these limits in our model, for the different 
values of~$\theta$.}
\label{fig:gamma rays constraints}
\end{figure}

}

%%%%%%%%%%%%%%%%%%%%%%%%%%%%%%%
%%%%%%%%%%%%%%%%%%%%%%%%%%%%%%%
%%%%%%%%%%%%%%%%%%%%%%%%%%%%%%%

\subsection{Neutrino Telescopes --  IceCube}
\label{sec:IC recast}
A stringent constraint on the spin dependent WIMP-Nucleon cross section comes from IceCube, a Cherenkov neutrino detector in the deep glacial ice at the South Pole \cite{Achterberg:2006md}, through a search  for neutrinos from WIMP annihilations in the Sun~\cite{Montaruli:2015six}. 
Let us briefly review the main points of the physics related to DM annihilations in the Sun. We refer to Ref.~\cite{Jungman:1995df} for a more detailed discussion.

DM particles in the Galactic halo can scatter with atomic nuclei inside the Sun and lose some of their energy. 
If the final velocity of the DM particle is low enough, it can remain gravitationally bound in the Sun. 
This accumulation process is counterbalanced by the evaporation process, i.e.\ the escape of DM particles from the Sun, and by the annihilation of DM pairs\footnote{It is worth noting that the annihilation of \textit{pairs} of DM pairs is implied by a possible $\mathbb Z_2$ symmetry that makes the DM candidate stable. 
Besides this, we comment that annihilation could be very suppressed in an asymmetric DM scenario, in which the conservation of the DM quantum number would require the annihilation of a DM particle and antiparticle.} 
into SM particles. 
The evaporation process can be safely neglected, if the DM particle mass is above $\sim 10$ GeV. 
The interplay between DM capture and annihilation drives the DM number density $\nDM$ towards an equilibrium. 
The final value of $\nDM$, and the equilibrium time by which this value is attained, depend on the capture rate $\Gcap$ and on the annihilation rate $\Gann$, the latter being proportional to $\nDM^2$. 
When equilibrium is attained, then $\Gcap = 2\Gann$, and the annihilation cross section can be directly related to the cross section of elastic scattering between DM and proton. 

Neutrinos are among the final annihilation products of DM, even more so when including the electroweak 
(EW) corrections~\cite{Ciafaloni:2010ti}. 
If the energy of the final states is above the weak scale, then electroweak interactions imply a considerable emission 
of EW bosons in the final state, further amplifying the neutrino flux. 
%All SM particles are indeed charged under $SU(2)_L\times U(1)_Y$, and the emission of $W$ and $Z$ bosons 
%increases the final flux of hard neutrinos. % \AK{I am not sure what do we try to say in this paragraph.}

We include the effects of EW corrections and of the propagation of neutrinos by means of the PPPC 4 DM ID 
code~\cite{Cirelli:2010xx,Baratella:2013fya}.
This code provides a set of interpolation functions for the neutrino flux at Earth due to annihilation in the Sun, for a range of DM annihilation channels, and includes the effects of electroweak corrections. 
The code takes into account the cascade of the primary annihilation products into neutrinos within the Sun, as well as the subsequent neutrino oscillation in the Sun, in vacuum, and within the Earth. 

What matters, in particular, are the branching ratios (BR) of DM pairs into SM final states: once equilibrium is achieved, the neutrino flux at Earth depends only on the relative annihilation cross sections to SM particles. 
This property is particularly interesting from the phenomenological point of view, because in the computation of the branching ratios, interesting simplifications occur and even the excitement of a resonance may have only a modest impact on the neutrino flux (see Sections~\ref{sec:BR}, \ref{sec:results} and Appendix~\ref{sec:annihilation}).

Since the flux and spectrum of neutrinos from WIMP annihilations depend upon the preferred annihilation channels of the WIMPs, these constraints have traditionally been provided for extreme scenarios of WIMPs annihilating 100\% into $W^+W^-$ or $\tau^+\tau^-$, `hard' channels corresponding to many high energy neutrinos and consequently a very stringent bound on the $\sSD$, or $b\bar{b}$, a `soft' channel producing a very weak bound on  $\sSD$ \cite{Aartsen:2012kia}.
 These results can be recast for the scenario of known annihilation channels by the following method. 
 The search utilizes the Unbinned Maximum Likelihood ratio method~\cite{IceCubeSearchMethodPaper, ICRCSolarWIMP}, 
 for which the sensitivity improves as ${\rm Signal}/\sqrt{\rm Background}$.  
For an unbinned maximum likelihood search of variable resolution, the background level varies as $\Psi^2$ where $\Psi$ is the median angular resolution \cite{PunziPaper}.
For a given  differential (anti)-muon neutrino flux $\mF (E)$, the total number of signal events expected within a sample can be calculated as
\begin{equation}
n_s\left(\mF \right) = \int_{E_{\rm threshold}}^{\mx} \mF (E) \times A_{\rm eff}(E)\, \d E,
\label{eq:nsintegral}
\end{equation}
where $A_{\rm eff}$ is the effective area from Fig. 3 of Ref.~\cite{ICRCSolarWIMP}. Since the fluxes and effective areas of $\nu_{\mu}$ and $\bar{\nu}_{\mu}$ are different, Eq.~\ref{eq:nsintegral} has to be evaluated separately for $\nu_{\mu}$ and $\bar{\nu}_{\mu}$.
The median energy $E_{\rm med}(\mF)$ is then defined through
\begin{equation}
\int_{E_{\rm threshold}}^{E_{\rm med}}  \mF(E) \cdot A_{\rm eff}(E)\, \d E = \int_{E_{\rm med}}^{\mx}  \mF (E) \cdot A_{\rm eff}(E)\, \d E.
\label{eq:E_med}
\end{equation}
If the capture and annihilation processes in the Sun are in equilibrium, the neutrino flux, the capture/annihilation rate, as well as WIMP-nucleon scattering cross section all scale linearly with respect to each other.
Thus the theoretical limit on the WIMP-nucleon scattering cross-section for a given flux prediction $\mF_{\rm theory}$ can be derived as
\begin{equation}
 \sigma_{\rm theory} = \sigma_{\rm benchmark} \cdot \frac{n_s(\mF_{\rm benchmark} )}{n_s(\mF_{\rm theory})} \cdot \frac{\Psi(E_{\rm med}(\mF_{\rm theory}))}{\Psi(E_{\rm med}(\mF_{\rm benchmark}))}
\label{eq:scaling1}
\end{equation}
where the first term in the RHS accounts for the variation in the level of signal events while the second term accounts for the variation in background due to the shift in median angular resolution. An analogous scaling relation can also be used to obtain theoretical limits on the annihilation rate $\Gann$.
%
%\begin{equation}
% \Gamma^{ann}_{\rm theory} = \Gamma^{ann}_{\rm benchmark} \times \frac{n_s(\frac{\d \phi_{\rm benchmark}}{\d E})}{n_s(\frac{\d \phi_{\rm theory}}{\d E})} \times \frac{\Psi(E_{\rm med}(\frac{\d \phi_{\rm theory}}{\d E}))}{\Psi(E_{\rm med}(\frac{\d \phi_{\rm benchmark}}{\d E}))}
%\label{eq:scaling2}
%\end{equation}

The bounds on $\Gann$ for the IceCube benchmark channels can be derived from the limits on $\sigma$ by mean of the tools provided by WimpSim and DarkSUSY.

%For a given $\d \phi_{\rm limit}/\d E$, $\sigma_{\rm limit}$ can be calculated w.r.t any of the three benchmark IceCube channels. The different calculations are consistent to within $\sim 30\%$ and are thus averaged. The results are shown in Figures \ref{fig:ICLimits_AllMediators} and \ref{fig:ICLimits_Noresonances}.

The IceCube limit on the neutrino flux $\mF_{\rm limit}$ requires knowledge of the neutrino spectrum per annihilation as it would be observed at Earth.
The first step is to calculate the branching ratio to all relevant
final states. Results are discussed in Sec.~\ref{sec:BR}. 

In order to convert these branching ratios into the required neutrino spectrum, we use the PPPC 4 DM ID 
code. %\footnote{%\comm{READ CAREFULLY!!!} Previous studies of the constraints coming from the IceCube experiment were done in 
%The IceCube constraints have been previously considered in Refs.~\cite{Blumenthal:2014cwa,Heisig:2015ira}, however 
%the BRs and the dominant sources of neutrinos have not been properly identified and the mixing effects between 
%$Z$ and $Z'$ were not considered. } 
%In these works, the authors examine simplified models without considering the mixing of $Z$ and $Z'$ imposed by the Higgs boson charge under $\Up$, and therefore they do not include annihilation channels which turn out to be the dominant ones.
%Moreover, they weight the nominal IceCube benchmarks (which assume 100\% annihilation into one channel) by the annihilation cross sections into different channels computed in their model.
%Moreover, they do not take into account EW corrections.
%consider the two benchmark cases $b\overline b$ and $W^+W^-$, assuming a 100\% BR into one of the two channels and without including EW corrections.
%In our work, we compute the BR's into various SM channels in a complete and consistent model, and we compute the neutrino fluxes including the EW corrections with the help of PPPC4DM ID.
%We infer the exclusion bounds with a proper recast of IceCube limits, as explained in this section: in this way we use the full shape of the neutrino fluxes to obtain the new bound.} 
This is combined with the results for the branching ratios to determine the
final spectrum of muon neutrinos and antineutrinos per DM annihilation
event. The $Zh$, $\gamma h$ and $\gamma Z$ final states are not
available in the code, and so we use the average of the two
pair-production spectra for each of these final states. 
We assume that the differences in the kinematical distributions, due to the different masses of $Z$, $h$ and $\gamma$, have a minor impact on the shape of the final neutrino flux. 

For the theoretical flux prediction thus obtained, the number of
expected signal events as well as the median energy can be obtained
from the expressions in Eqs.~(\ref{eq:nsintegral}) and~(\ref{eq:E_med})
for each of the three IceCube samples described in
Ref.~\cite{ICRCSolarWIMP}. The integrals are evaluated separately for
$\nu_{\mu}$ and $\bar{\nu}_{\mu}$. These quantities can also be
evaluated for the IceCube benchmark channel flux predictions ($\mF_{\rm benchmark}$) obtained from WimpSim 3.03 and nusigma
1.17. Subsequently, theoretical limits on $\sigma$ and $\Gann$
can be obtained using the scaling relation~\eqref{eq:scaling1} and the analogous one for $\Gann$. 

For a given $\mF_{\rm theory}$, $\sigma_{\rm theory}$ can be calculated w.r.t any of the three benchmark IceCube channels. The different calculations are consistent to within $\sim 30\%$ and are thus averaged.

\subsection{Results for the Branching Ratios}
\label{sec:BR}
As discussed in the previous section, in order to extract bounds from IceCube observations, 
the branching ratios for the annihilations of DM pairs into SM final states must be computed. 
In this section we present the final results for the BR's, based on the cross sections that 
are computed in detail in Appendix~\ref{sec:annihilation}.

Fig.~\ref{fig:BR} shows the BR's into SM  two-body final states as a function 
of $\mX$, for $\mZp=10$ TeV and for six different values of $\theta$ (defined in Eq.~\eqref{eq:thetadef}). 
The BR's  for the final states shown on the plots do not depend at all on $\gZp$ and $\gX$, 
because they cancel in the ratios of cross sections, as  can be seen from the formul\ae\ in Appendix~\ref{sec:annihilation}.
Leaving aside for a moment  the pure $\UBL$, the main annihilation channels are the heavy fermion 
pairs ($t\overline t$, $b\overline b$ and $\tau^+ \tau^-$) and $Zh$. These are indeed the only tree level channels 
for which $a\not\approx 0$ in the low velocity expansion $\sigma \vDM \sim a+b\,\vDM^2$.

The main annihilation channel  below the kinematic threshold $\mX=108$ GeV for  $Zh$ production is 
$b\overline b$, while the BR's into $c\overline c$ and $\tau^+ \tau^-$ are less than 10\% each.
Even if its branching ratio is not the dominant one, the $\tau^+ \tau^-$ channel dominates the 
IC bound below the $Zh$ threshold because it yields more energetic neutrinos than the $b\overline b$ one.

In the region $\mX \sim 80 \GeV - 108 \GeV$ annihilation into $W^+W^-$ may overcome the one into $b\overline b$, depending on the value of $\theta$.
Notice that, as will be explained at the end of this section, the one loop contribution to the $W^+W^-$ cross section dominates over the tree level one, which is suppressed by the small $Z-Z'$ mixing angle and by the fact that, in the low velocity expansion, it has $a=0$.

When the $Zh$ channel opens up it overcomes all  others and remains the only relevant channel 
unless $\mX \gtrsim m_\text{top}$, where the cross section into $t\overline t$ is comparable to the one to $Zh$. 
At higher DM masses, the cross section to $Zh$ in the low $\vDM$ limit is proportional  to $\mX^2$, 
while $\sigma(\chi \chi \to t\overline t )$ is basically constant in $\mX$. 
The former proportionality comes from the final state with a longitudinally polarised $Z$ boson and a Higgs boson, and is ultimately due  to the derivative coupling of would-be Goldstone bosons. 
This  explains why $Zh$ is the only relevant channel at large $\mX$.

Around the resonance $\sigma(\chi \chi \to Zh)$ goes to 0 because the coefficient $a$ in the low velocity 
expansion $\sigma \vDM \sim a +b\,\vDM^2$ vanishes. The reason is explained in Appendix~\ref{sec:annihilation}. 
Therefore, in a small window around the resonance, other channels dominate. The position of this window is
 basically the only way in which the branching ratios depend on $\mZp$, as can be seen from 
 Figs.~\ref{fig:results theta=0} and~\ref{fig:results other thetas}.

The previous picture applies for all values of $\theta$, except $\theta =\pi /2$ which corresponds to the pure $\UBL$ case. In that case, there is no mixing between $Z$ and $Z'$, and the channel $Zh$ disappears at tree level.
Below the $W^+W^-$ threshold $\mx=m_W$, annihilation predominantly happens at tree level into fermionic channels.
For $\mX>m_W$  the dominant channel is instead $W^+W^-$, with a $\mathcal{O}(10\%)$ contribution from $ZZ$. Annihilation into these channels is due to a diagram with a triangular fermionic loop, as discussed in Appendix~\ref{sec:annihilation}.
The fermion channels, in the zero velocity limit, have a cross section proportional to $\mf^2 c_f^{A\,2}$, where $\mf$ is the fermion mass and $c_f^A$ is the coupling of $Z'$ to the axial vector fermion bilinear. 
When $\Up$ is reduced to $\UBL$ the $Z'$ couples to the vector current only. Thus in this limit the $\sigma(\chi \chi \to f\overline f)$ has $a=0$.
The coefficient $b$ is not proportional to $\mf^2$ (as it is $a$ because of the helicity suppression), thus in the fourth plot of Fig.~\ref{fig:BR} the fermions  contribute equally to the annihilation cross section (apart from a factor $B^2\times(\# {\rm \,colors})=1/3$ which penalises quarks with respect to leptons), unlike what happens for $\theta \neq \pi /2$.

%Leaving aside for a moment  the pure $\UBL$, the main annihilation channels are the heavy fermion 
%pairs ($t\overline t$, $b\overline b$ and $\tau^+ \tau^-$) and $Zh$. These are indeed the only tree level channels for which $a\not\approx 0$ in the low velocity expansion $\sigma v \sim a+b\,\vDM^2$.
%Even if the branching ratio to $\tau^+ \tau^-$ is smaller than that to $b\overline b$ by one order of magnitude, it is the $\tau^+ \tau^-$ channel that dominates the IC bound below the $Zh$ threshold because it yields more energetic neutrinos.
%A fundamental point for our discussion is as follows: what matters for IC is the energy spectrum of the final neutrinos, thus in order to determine the most important channel for IC we must take into account not only the BR's into the various channels, but also their impact on IC bounds. From the plots shown in Fig.~\ref{fig:BR}, we see that below the $Zh$ threshold the $\tau^+ \tau^-$ channel is smaller with respect to $b\overline b$ by just one order of magnitude. The IC bound assuming a 100\% BR into $\tau^+ \tau^-$ is stronger of two to three orders of magnitude with respect to the bound obtained assuming that DM annihilates only to $b\overline b$. Therefore the most important channel for IC, below the $Zh$ threshold, is $\tau^+ \tau^-$, even if it is not the main annihilation channel.

Let us conclude this section by explaining why the tree level contribution for $\chi\chi \to W^+W^-$ has $a=0$, which, together with the additional  suppression by $\sin\psi$, selects $Zh$ as the main channel at low velocities.
The initial state $\chi\chi$, in the limit $\vDM\to 0$, has total angular momentum $J=0$ and CP eigenvalue $-1$.
The final state $Zh$ is not a CP eigenstate, unlike $W^+ W^-$. Now, a pair of vector bosons can have a CP eigenvalue $-1$ and a total angular momentum $J=0$ only if they are both transversally polarized \cite{Drees:1992am}. 
In this case, the tree level interaction $Z'WW$ \eqref{eq:vertex Z'WW} turns out to give a vanishing cross section.
% WRONG: In this case, the derivative coupling of the $Z'WW$ interaction \eqref{eq:vertex Z'WW} turns out to give leads to a contraction of the $W$ bosons momenta with the $W$ polarisation vectors, which turns out to vanish for a transverse polarization.\\ 
%\DR{This is the calculation: from \eqref{eq:vertex Z'WW} one sees that, by denoting with $p_{3,4}$ the momenta of the $W$'s, the vertex is proportional to 
%\begin{equation}
%\begin{aligned}
%p_3^\mu \ \varepsilon_\mu^{(\lambda)}(p_4)  &=
%p_3^0 \ \varepsilon_0^{(\lambda)}(p_4) -p_3^i \ \varepsilon^{i\, (\lambda)}(p_4)
% \overset{\lambda = T \Rightarrow \varepsilon_0^{(T)}=0}{=}
% -p_3^i \ \varepsilon^{i\, (\lambda)}(p_4)=\\
%& \overset{\vec p_3=-\vec p_4}{=} 
% +p_4^i \ \varepsilon^{i\, (\lambda)}(p_4) 
%  \overset{\lambda = T \Rightarrow \varepsilon_0^{(T)}=0}{=}
%p_4^\mu \ \varepsilon_\mu^{(\lambda)}(p_4) = 0
%\end{aligned}
%\end{equation}
%}
%We notice that this argument does not apply to the final states $ZZ$ or $WW$ at one loop, which do not have a derivative coupling. In those cases, it turns out from the calculations that $a\neq 0$, and the interaction with $\chi\chi$ is suppressed only by the loop factors. \DR{The leading operator that we expect from the fermion loop is 
%$Z'_{\mu \nu} W^{+\,\mu \rho} W^{-\,\nu}_{\ \ \rho}$. This is the operator of lowest dimension allowed by $SU(2)_L$ and $CP$ invariance. This operator 
%seems to include contractions as $W^+\cdot W^-$, which would not fall in the case considered in the argument.}
We notice that this argument does not apply to the $W^+W^-$ and $ZZ$ 
amplitudes when the triangular fermion loop is included (see Fig.~\ref{fig:feynman}).
In those cases, the effective $Z'W^+W^-$, $Z'ZZ$ vertices contain the terms
\begin{equation}
f_5^{Z'WW} \epsilon^{\mu\nu\rho\sigma}(k_1-k_2)_\sigma Z'_\mu W^+_\nu W^-_\rho,
\qquad
f_5^{Z'ZZ} \epsilon^{\mu\nu\rho\sigma}(k_1-k_2)_\sigma Z'_\mu Z_\nu Z_\rho,
\end{equation}
where $k_1,k_2$ are the four-momenta of the outgoing bosons (see also~\cite{Hagiwara:1986vm,Gounaris:2000tb}).
These terms lead to $a\neq 0$ in the cross section, making the $W^+W^-$ and $ZZ$ channels relevant despite of the loop suppression.
%As a check of our results, we assigned to the $Z'$ the SM $Z$ couplings to fermions and we compared with the $ZZZ$ effective vertex computed in~\cite{Gounaris:2000tb}.

%\DR{Here is an \textbf{issue} that would be nice to understand, even if not compulsory.
%{\begin{small}
%$Zh$ has $a\neq 0$, while for $WW$ we get $a=0$. By exclusion, this is due to the extra momentum in the $WWZ'$ vertex: indeed, in the old model, $WW$, $ZZ$, $\gamma \gamma$ had $a\neq 0$ only for $TT$ polarizations, and $WW$ was coming from the fermion vertex. Now, there is no reason why $a_{WW}=0$ now, unless this is due to the trilinear interaction vertex.\\
%Why does this imply $a=0$? It is not clear. Two suggestions:
%\begin{itemize}
%\vspace{-.8em}
%\item The proportionality to momentum of the trilinear gauge vertex seems suspicious (gives $a=0$ both in the tree level and in the bosonic loop channel), but by itself is not enough to conclude that $\sigma(\chi\chi \to W^+ W^-)\propto \vDM^2$.
%\vspace{-.8em}
%\item CP arguments imply that $S$ in the final state is odd (initial eigenvalue $-1$, final eigenvalue $(-1)^{2L+S}=(-1)^{S}$ or $(-1)^{L}$ since $J=0$ by total angular momentum conservation).
%\end{itemize}
%\end{small}
%\vspace{-.8em}
%}}

\begin{figure}[h!]
\centering 
\includegraphics[width=0.49\textwidth]{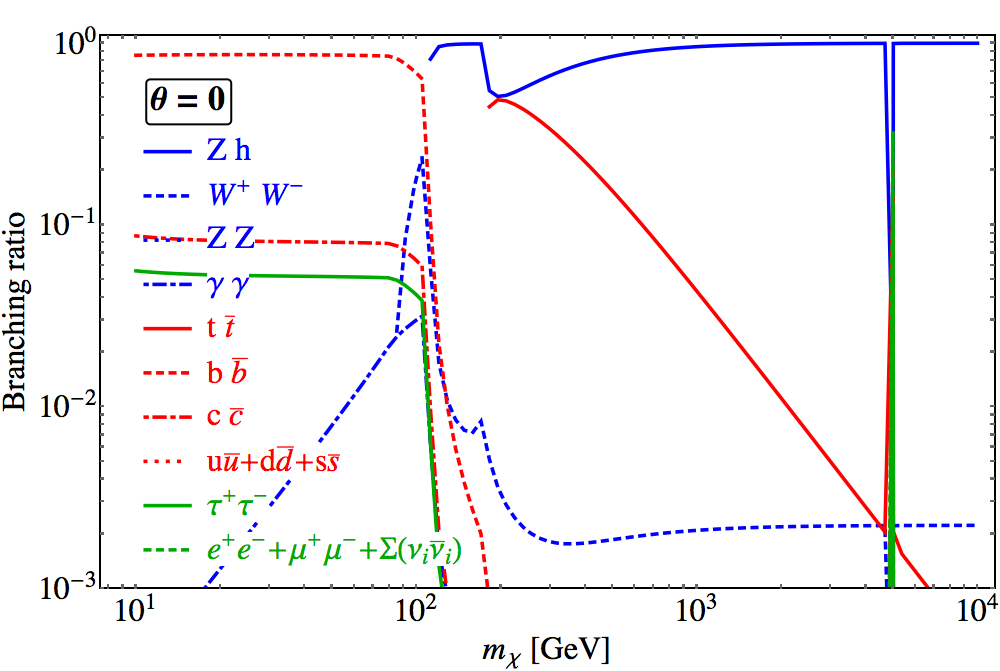} \hfill
\includegraphics[width=0.49\textwidth]{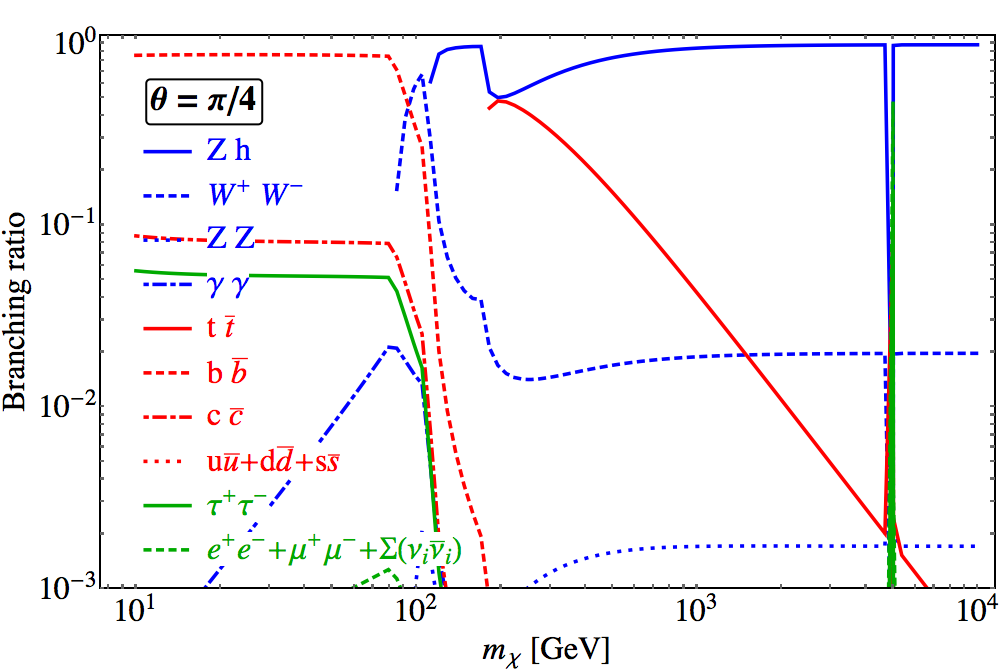}\\
\includegraphics[width=0.49\textwidth]{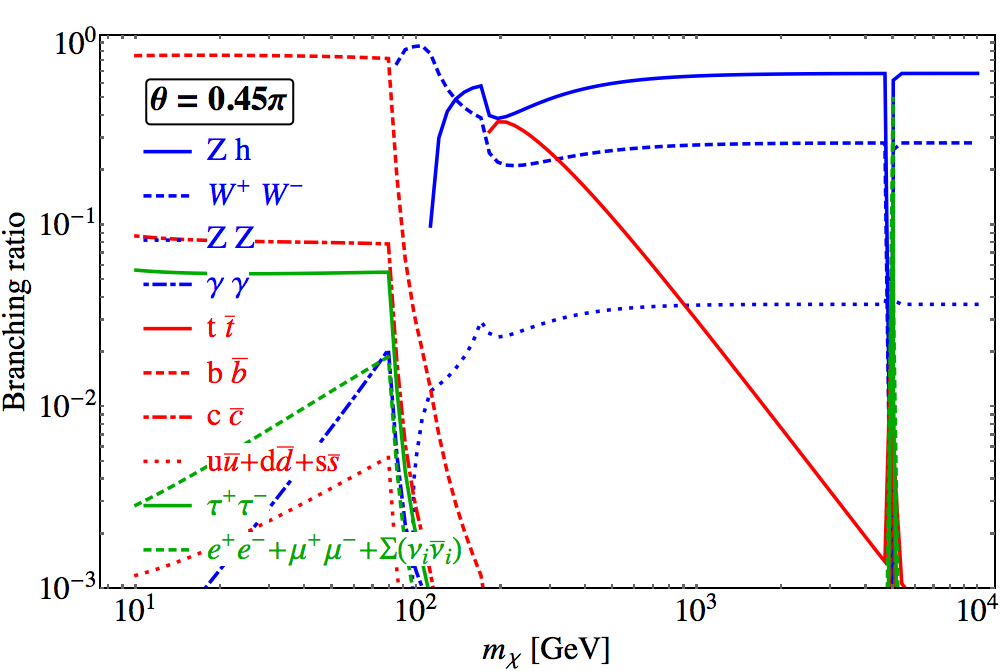} \hfill
\includegraphics[width=0.49\textwidth]{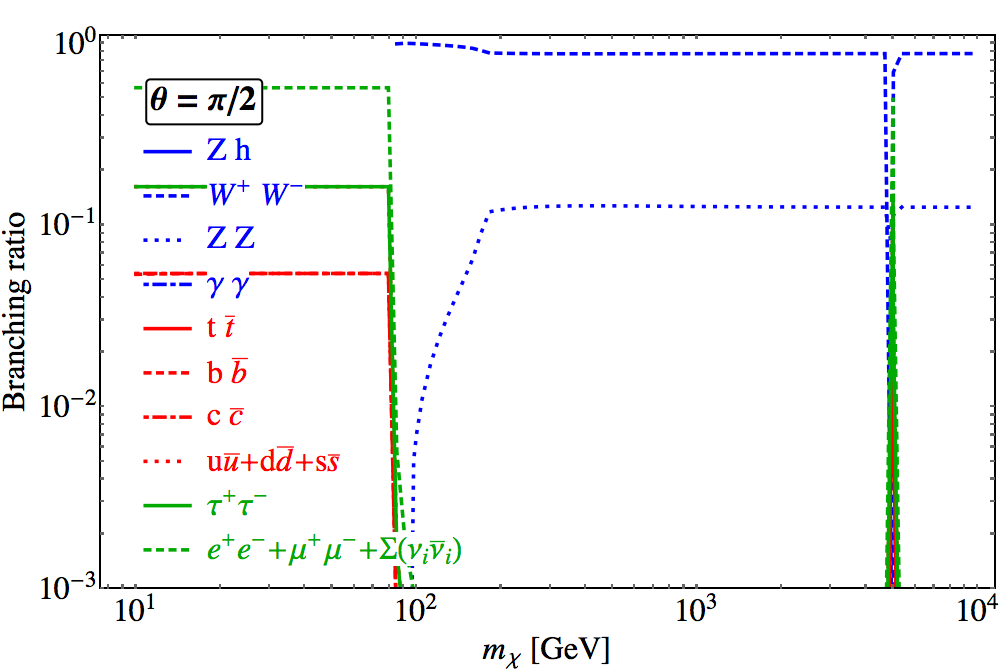} \\
\includegraphics[width=0.49\textwidth]{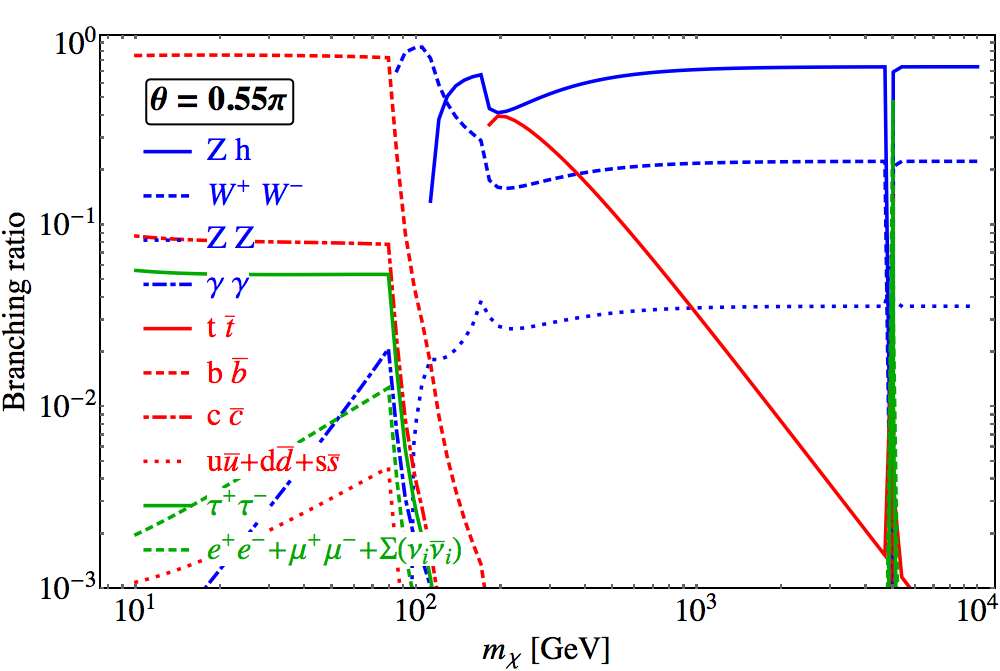} \hfill
\includegraphics[width=0.49\textwidth]{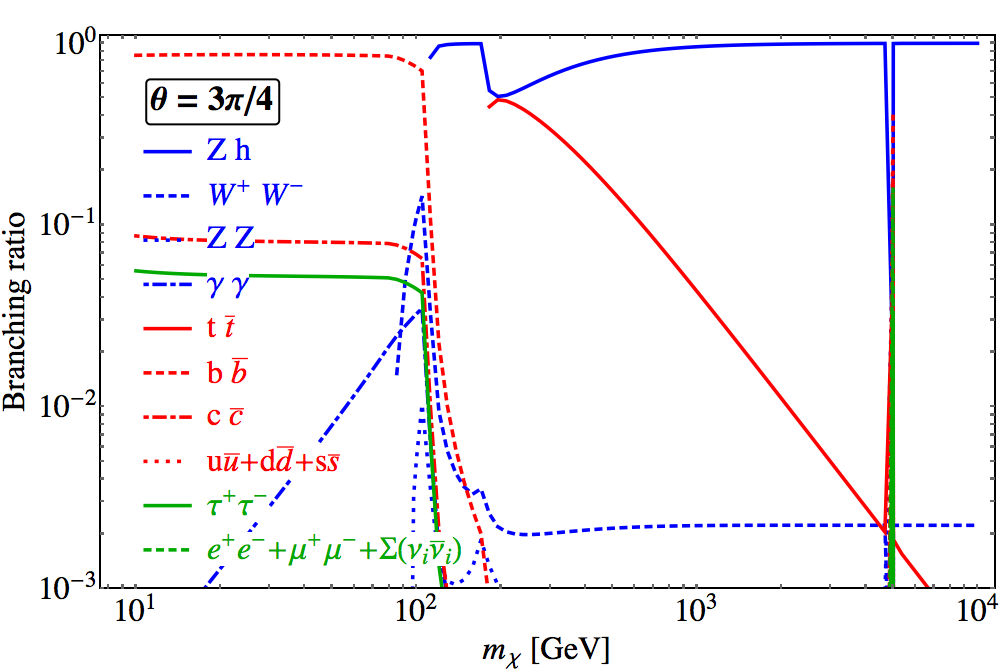}
\caption{Branching ratios for the annihilation of $\chi\chi$ into pairs of SM particles, for $\mZp=10$ TeV and a kinetic energy of the DM particles equal to the thermal one in the Sun core. There is no dependence on  $\gZp$ and $\gX$. The six plots, from left to right and top to bottom, correspond to $\theta=0,\, \pi/4,\, 0.45 \pi, \pi/2,\, 0.55\pi,\, 3\pi/4$. Only the channels with a BR greater than $10^{-3}$ are shown.}
\label{fig:BR}
\end{figure}
%%%%%%%%%%%%%%%%%%%%%%%%%%%%%%%% 

%%%%%%%%%%%%%%%%%%%%%%%%%%%%%%%%
\section{Summary of results}
\label{sec:results}
\begin{figure}[h!]
\centering 
\includegraphics[width=0.7\textwidth]{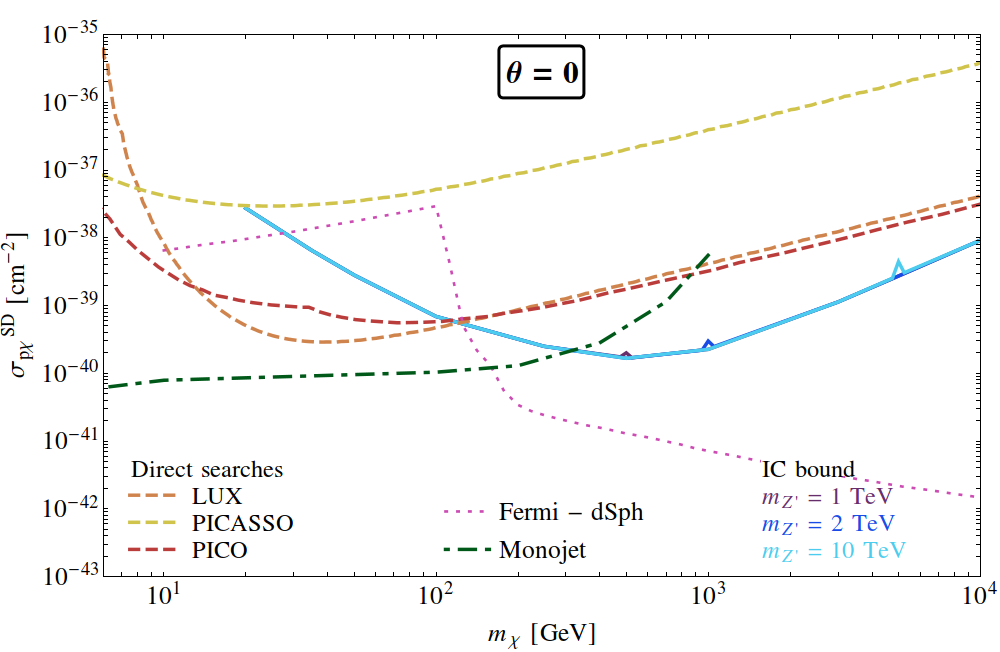}
\caption{Bound on $\sSD$ from direct detection, LHC's monojet analysis, IceCube and Fermi-LAT, for $\theta = 0$.}
\label{fig:results theta=0}
\end{figure}

\begin{figure}[h!]
\centering 
\includegraphics[width=0.49\textwidth]{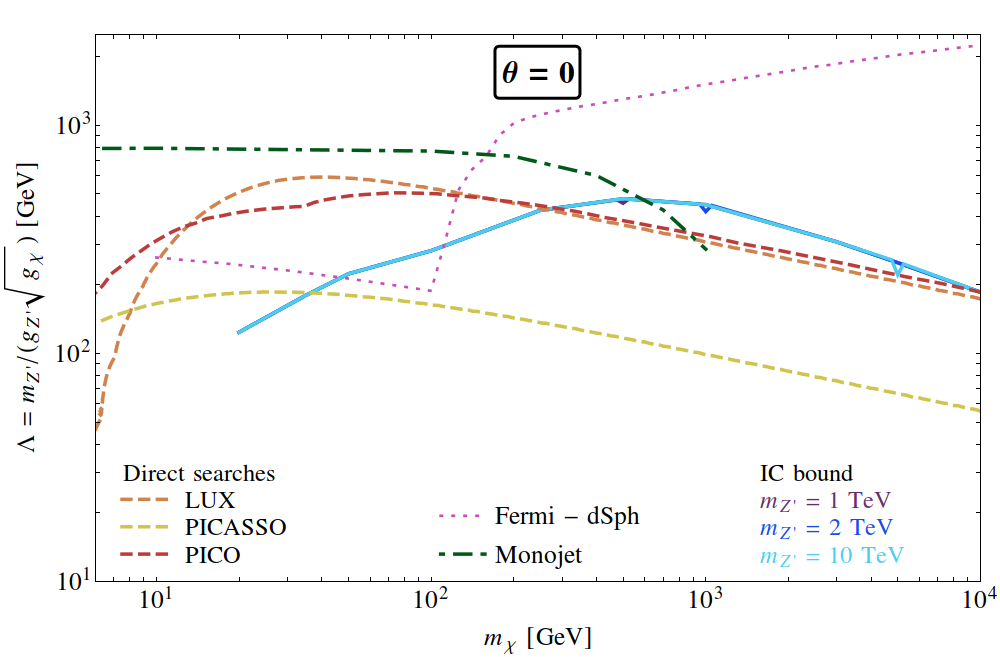}\hfill
\includegraphics[width=0.49\textwidth]{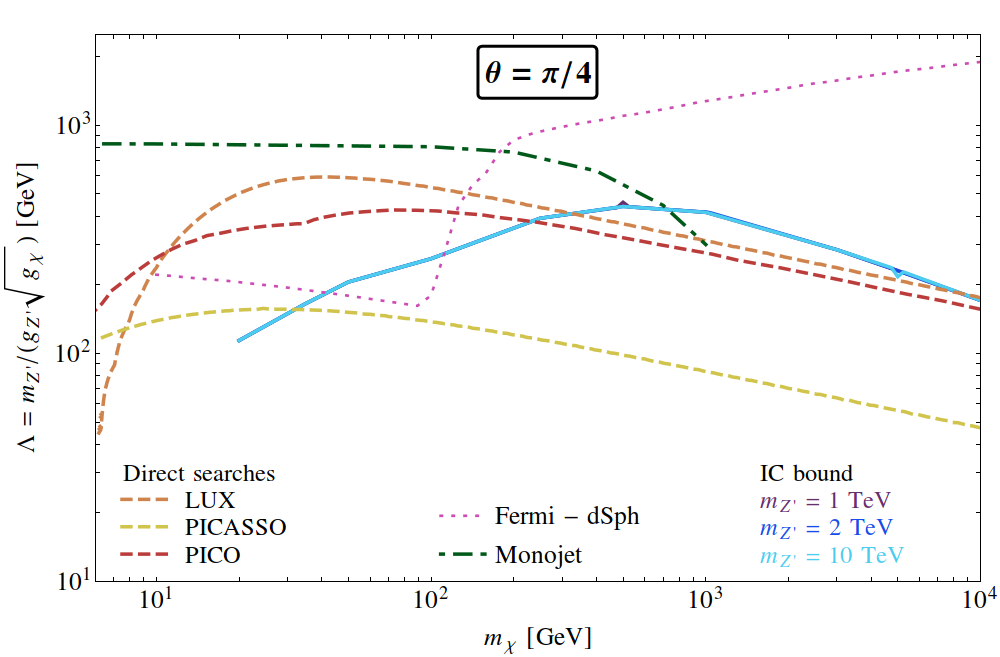}\\
\includegraphics[width=0.49\textwidth]{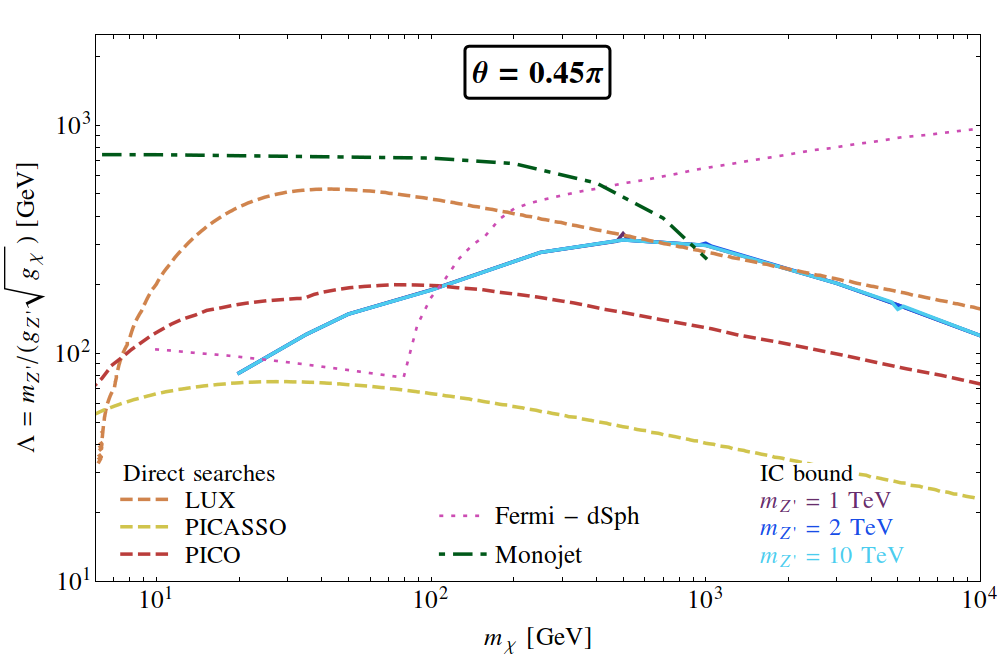} \hfill
\includegraphics[width=0.49\textwidth]{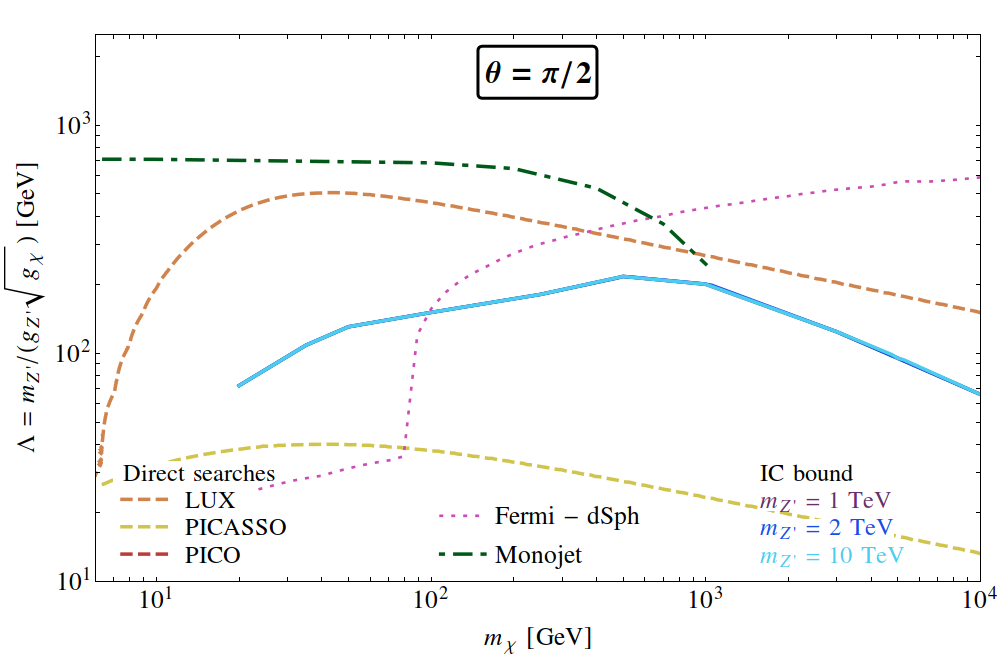} \\
\includegraphics[width=0.49\textwidth]{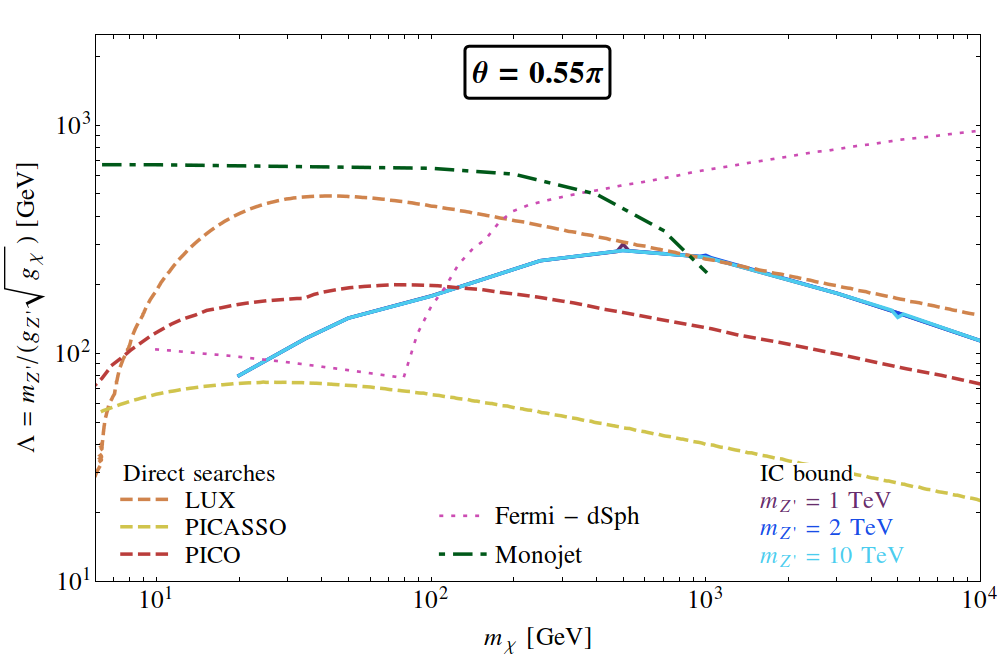} \hfill
\includegraphics[width=0.49\textwidth]{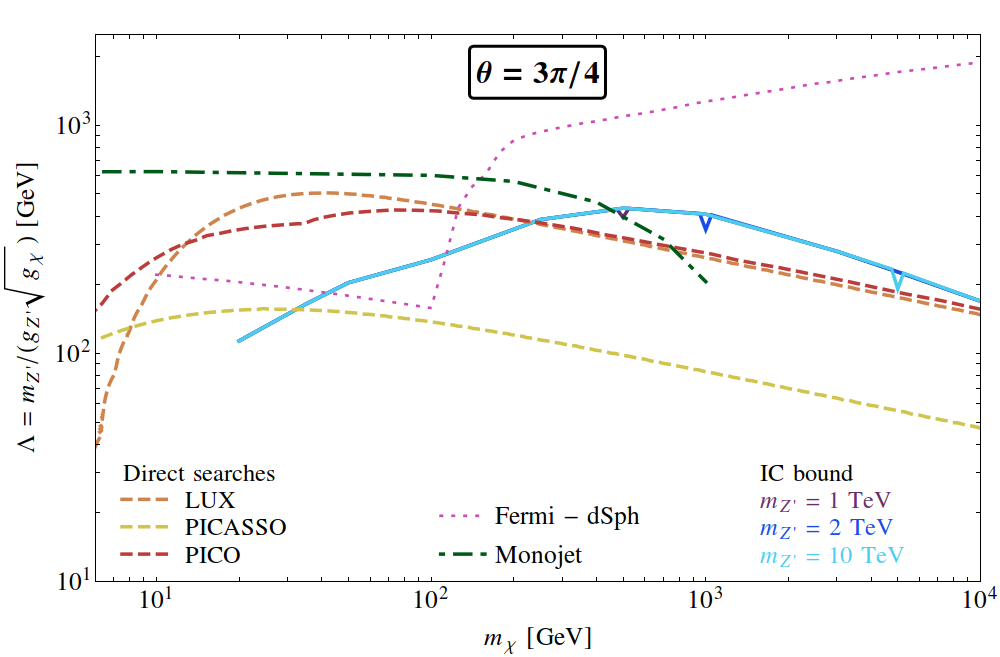}
\caption{Bound on $\Lambda = \mZp/(\gZp\sqrt{\gX})$ from direct detection, LHC's monojet analysis, IceCube and Fermi-LAT, for different values of $\theta$.}
\label{fig:results other thetas}
\end{figure}

% !TEX root = main_draft.tex
We show in Fig.~\ref{fig:results theta=0} bounds on $\sSD$ for the case $\theta =0$, comparing direct detection results with those coming from IceCube, LHC's mono-jet searches \new{and $\gamma$-ray searches}.
Analogous results for different values of  $\theta$ are presented in Fig.~\ref{fig:results other thetas}, in the plane $\Lambda$ {\it vs.} $\mX$.
As representative values of $\theta$ we choose $\theta = 0$, $\pi/4$, $\pi/2$ and $3\pi/4$. We did not consider $\theta=\pi$ because it is exactly equivalent to $\theta = 0$.
Since $\theta = \pi/2$ is quite a peculiar point, we added two values of $\theta$ in its vicinity, namely $0.45\pi$ and $0.55\pi$.

As is clear from Eq.~\ref{eq:NRops} and Table~\ref{tab:U1coefficients}, for a generic angle $\theta$ the DM-nucleon scattering is mediated by a linear combination of $\cO_4, \cO_8$ and $\cO_9$, with coefficients similar in magnitude.
The contributions from $\cO_8$ and $\cO_9$ can be safely ignored for IceCube: given the composition of the solar environment, their nuclear form factors  are between 100 and 1000 times smaller than the one for $\cO_4$ \cite{Catena:2015uha}.
This is not the case for DD experiments, where the three operators give a similar contribution, and the scattering cross section is not exactly $\sSD$, which is defined as being given by the operator $\cO_4$ alone.

To obtain a bound in the usual $\sSD$ {\it vs.} $\mX$ plane, one needs to first obtain a bound on the total scattering cross section, which is easily done using the results of \cite{DelNobile:2013sia}.
Then, this bound is translated into a bound on $\Lambda = \mZp/(\gZp \sqrt{\gX})$, using the expression for the cross section obtained from Eq.~\ref{eq:efflagrangianscattering}.
Finally, the bound on $\Lambda$ is translated into a bound on $\sSD$ using the expression
\begin{equation}
\sSD = \frac{3}{\pi}\left(-\frac{c_4}{\,\Lambda^2}\right)^2 \mu_p^2
\label{eq:sigmaSD-Lambda}
\end{equation}
which gives the scattering cross section when only the $\cO_4$ operator is involved, where $\mu_p$ is the reduced mass of the proton--DM system, and the dimensionless coefficient is given by $c_4=\tfrac 14 \cos\theta (\Delta_u-\Delta_d-\Delta_s)\simeq \tfrac 14 \cth \cdot 1.35$, with $\Delta_q$ parametrizing the quark spin content of each nucleon, and is assumed to be equal for protons and neutrons \cite{DelNobile:2013sia}.

For $\theta\neq0$, we do not  show bounds on $\sSD$ but only on $\Lambda$.
The reason is that, since $\sSD$ is defined as the contribution to the scattering cross section due to the operator $\cO_4$ only, given the limit on $\Lambda$ we have $\sSD\propto\cos^2\theta/\Lambda^4_{\rm lim}$.
When $\theta$ gets close to $\pi/2$, the computed value of $\sSD$ goes to $0$ independently of $\Lambda_{\rm lim}$, resulting in a spuriously strong bound.

The situation is slightly different when $\theta=\pi/2$.
In this case, the coupling of the $Z'$ to the vectorial current of the quark fields is identically $0$, and therefore the coefficient of the $\cO_4$ operator in the NR expansion vanishes.
While for IceCube the contribution of $\cO_9$ is subdominant with respect to that of $\cO_8$ and can be ignored, both of them have to be taken into account to obtain DD bounds.
A recast of the bound on $\Lambda$ in terms of the usual $\sSD$ would make no sense in this scenario.

The PICO experiment gives interesting direct detection bounds.
For $\theta=0$ ({\it i.e.} for the usual spin-dependent operator $\cO_4$), bounds on $\sSD$ can be read directly from Ref.~\cite{Amole:2015pla}, and translated into a bound on $\Lambda$.
Bounds on $\Lambda$ for other values of $\theta$ can not be obtained following the procedure we adopted for the other experiments, because PICO is not yet included between the Test Statistic functions given in \cite{DelNobile:2013sia}.
Therefore, we obtained a conservative bound on $\Lambda$ by rescaling the PICO limit on $\sSD$ as $\cos^2\theta$ (since this enters the coefficient of the $\cO_4$ operator) and then applying Eq.~\eqref{eq:sigmaSD-Lambda}.

Let us now comment briefly on the results shown in Figs.~\ref{fig:results theta=0} and \ref{fig:results other thetas}.
Bounds coming from LHC searches are typically the strongest ones in the low mass region, up to $\mX\sim400 - 700\GeV$ for large $\theta$.
IceCube searches have their maximal sensitivity in the region between a few hundred GeV and a few TeV.
When $\theta$ is small, in this mass region they give \new{a constraint on $\Lambda$ which is stronger than the direct detection one}.
In particular, for $\theta=0$ the constraint from IceCube is the dominant one from $\mX\sim 100\GeV$ up to $10\TeV$ and beyond.
\new{The Fermi-LAT bound from dwarf spheroidal galaxies appears to be the dominant one for $\mX$ mass above $200 - 300\GeV$. This is mainly due to the fact that, in our model, the dominant annihilation channel is $Zh$, which produces a large photon flux thanks to EW corrections. We expect that, in a leptophilic model in which DM annihilates copiously into $\tau$ and $\mu$ pairs, energetic neutrinos produced in their decay would make IceCube bounds the dominant ones.}

We notice that the bounds shown in Fig.~\ref{fig:results other thetas} fall in the region where the DM is underabundantly produced via the freeze-out mechanism (compare with Fig.~\ref{fig:gZp relic}).  The difference between the two values of $\Lambda$ goes up to one order of magnitude for large $\mX$.

An important remark about IceCube  results is that they are almost independent of $\mZp$ and $\GZp$ (together with $\gZp$, $\gX$, as stressed in Sec.~\ref{sec:BR}).
There are two reasons for this: first, as explained previously, when the equilibrium between annihilation and capture in the Sun is reached, neutrino fluxes at Earth only depend on branching ratios, which are not modified dramatically by the resonance except for a  very narrow region around it.
Electroweak corrections further dilute the difference, and as a result IceCube bounds for different values of $\mZp$ are almost superimposed (except for a small bump around the resonant point, whose width is $\sim\GZp$).

Previous studies of the constraints coming from the IceCube experiment were done in Refs.~\cite{Blumenthal:2014cwa,Heisig:2015ira}.
In these works, the authors examined simplified models without considering the mixing of $Z$ and $Z'$ imposed by the Higgs boson charge under $\Up$, and therefore they do not include annihilation channels which turn out to be the dominant ones.
Moreover, they weight the nominal IceCube benchmarks (which assume 100\% annihilation into one channel) by the annihilation cross sections into different channels computed in their model, and they do not take into account EW corrections.
In our work, we compute the BR's into various SM channels in a complete and consistent model, and we compute the neutrino fluxes including the EW corrections with the help of PPPC4DM ID.
We infer the exclusion bounds with a recast of IceCube limits, as explained in Sec.~\ref{sec:IC recast}, using the full shape of the neutrino fluxes to obtain the new bound.

%%%%%%%%%%%%%%%%%%%%%%%%%%%%%%%% 

%%%%%%%%%%%%%%%%%%%%%%%%%%%%%%%%
\section{Conclusions}
\label{sec:conclusions}
% !TEX root = main_draft.tex
While the spin-independent WIMP scenario has been probed experimentally to very high precision 
and  direct detection experiments basically disfavor this possibility, the bounds on  SD DM are much milder.
WIMP-strength interactions between  spin-dependent 
DM and  baryonic matter are still perfectly allowed. 
Moreover, often the strongest bounds on  SD DM do not come from direct detection experiments, but rather from  LHC searches (direct or indirect) \new{and Fermi-LAT. The IceCube experiment also produces interesting bounds, which are typically stronger than the direct detection ones above a few hundred GeV.}

In this work we  analyzed and compared these constraints, coming from different experiments. 
In order to be concrete, we concentrated on a particular set of spin-dependent DM models, in which the DM-baryon interaction is mediated by a heavy gauge $Z'$.  
If we restrict ourself to the models without spectators at the EW scale, the parameter space of  gauge $Z'$ models can be conveniently parametrized by three quantities: the mass of the heavy $Z'$, the gauge coupling and the mixing angle $\theta$ between the hypercharge and $\UBL$ generators. 
 
Although this study is not completely generic, as it does not cover all possible consistent models of  SD DM, there are good reasons to believe that these models capture important phenomenological features that are generic to  WIMP-like SD DM. 
We identified the region of parameter space favored by the observed thermal relic abundance and we have reanalyzed the existing
LHC constraints, both direct (from the monojet searches) and indirect, on the $Z'$ mass and coupling. 

More importantly, we fully analyzed the low-temperature annihilation branching ratios of the 
$Z'$-mediated DM, which is crucial to derive IceCube (and Fermi-LAT) constraints.  
In order to properly understand  the expected neutrino fluxes from  DM annihilation in the Sun's core, one has to know the annihilation channels of the DM in the Sun. Simply assuming that the dominant expected source of neutrinos is the $WW$ or $b\bar b$ channel (as  is done in the IceCube papers)
is clearly insufficient. 
We found that, depending on the DM mass, one can divide the parameter 
space into three different regions. 
Below the $Zh$ mass threshold (very light DM) the annihilations are indeed 
dominated by the $b \bar b$ channel, yielding very soft neutrino fluxes, although even in this case the IceCube constraints are dominated by small branching ratio to $\tau^+\tau^-$. 
Above this threshold $Zh$ is a dominant annihilation channel and the dominant source of neutrinos almost in the entire parameter space. 
The third region is just above the $t \bar t$ mass threshold. 
If the DM mass sits in this ``island", one usually gets  comparable annihilations into $Zh$ and $t \bar t$, and both should be taken into account for the neutrino flux calculations.
The $W^+ W^-$ channel can also become important, or even the leading channel, if the $\Up$ extension is very close to being $\UBL$.
In this work we have properly recast the existing IceCube bounds including electroweak corrections, in order to derive reliable exclusion bounds on the secondary neutrinos coming from all annihilation channels.

We find that currently the strongest bounds on the SD $Z'$-mediated DM are imposed by  LHC searches (for $\mX \lesssim 400$~GeV) and by Fermi-LAT for heavier DM candidates, which are favored by  thermal 
relic considerations.
% Low m_DM implies high g_Z' and/or low m_Z' (unless we go to non perturbative g_\chi), which are in tension with direct searches of Z' at LHC.
We also find that the best direct detection bounds come mostly from LUX (PICO becomes dominant only for very light $\sim 20$~GeV DM particles). 
 These bounds are  subdominant with respect to LHC ones in the case of  light DM, but can be comparable to IC bounds for  heavy DM if $\Up$ is close to being a $\UBL$ extension. 
We have also computed the values that yield the observed DM abundance through the freeze out mechanism, and we found that experimental exclusion limits fall in the slightly underabundant region. 
%\DR{$\Omega_\text{DM}$ turns out to be around $\sim 1\%$ the observed one, for $\mX\lesssim 1\ \TeV$.}

Finally we notice that it would be interesting to see similar works for other SD DM candidates. 
It would be also useful to have bounds on the DM, annihilating into 
%$t \bar t$ \EM{$t \bar t$ is considered in \cite{Aartsen:2016exj}} and 
$Zh$ reported directly by the IC \new{and Fermi collaborations}.
We stress that this channel immediately arises when considering a consistent anomaly-free model for a $\Up$ extension which is not a pure $\UBL$. This important feature is not captured by the use of so-called simplified models if these issues are not considered.

%%%%%%%%%%%%%%%%%%%%%%%%%%%%%%%% 

%\vspace*{3em}
%\DR{I added the following literature: 
%\cite{D'Eramo:2016atc,Haisch:2013uaa, Matsumoto:2016hbs,Dreiner:2013vla,Ghorbani:2015baa,
%DeSimone:2016fbz,
%Goodman:2010ku,Goodman:2010yf,Bai:2010hh}
%}

\acknowledgments

We thank Maxim Perelstein, Stefan Pokorski, Benjamin Safdi and  Sofia Vallecorsa for discussions.
\new{We also would like to thank the anonymous referee for his/her suggestion of considering the $\gamma$-rays constraints on our model, without which our work would have been less complete.}
We are grateful to Yanou Cui and Francesco D'Eramo for useful communication and for pointing us out the importance of the $Z$ exchange. 
D.R.\ and A.R.\ are supported by the Swiss National Science Foundation (SNSF), project ``Investigating the Nature of Dark Matter'' 
(project number: 200020\textunderscore{}159223). 
A.K.\ is grateful to the organizers of KITP program ``Experimental Challenges for the LHC run~II" where this paper has been partially completed (partially supported by National Science Foundation under Grant No. NSF PHY11-25915).

\appendix 

%%%%%%%%%%%%%%%%%%%%%%%%%%%%%%%% 
\section{Details of the annihilation rate calculation}
\label{sec:annihilation}
% !TEX root = main_draft.tex
In this appendix we present the calculation of the annihilation cross sections of the DM 
into the SM  in detail. We also go in detail over the one-loop order annihilation into the $WW$. 
The results of this calculations are summarized on Fig.~\ref{fig:BR}.  

%We report in this appendix the detailed results obtained for the annihilation cross sections of DM pairs into  pairs of SM particles (see Fig.~\ref{fig:BR} for the numeric results), together with other technical considerations.

\begin{figure}[h!]
  \centering
  \includegraphics[width=\textwidth]{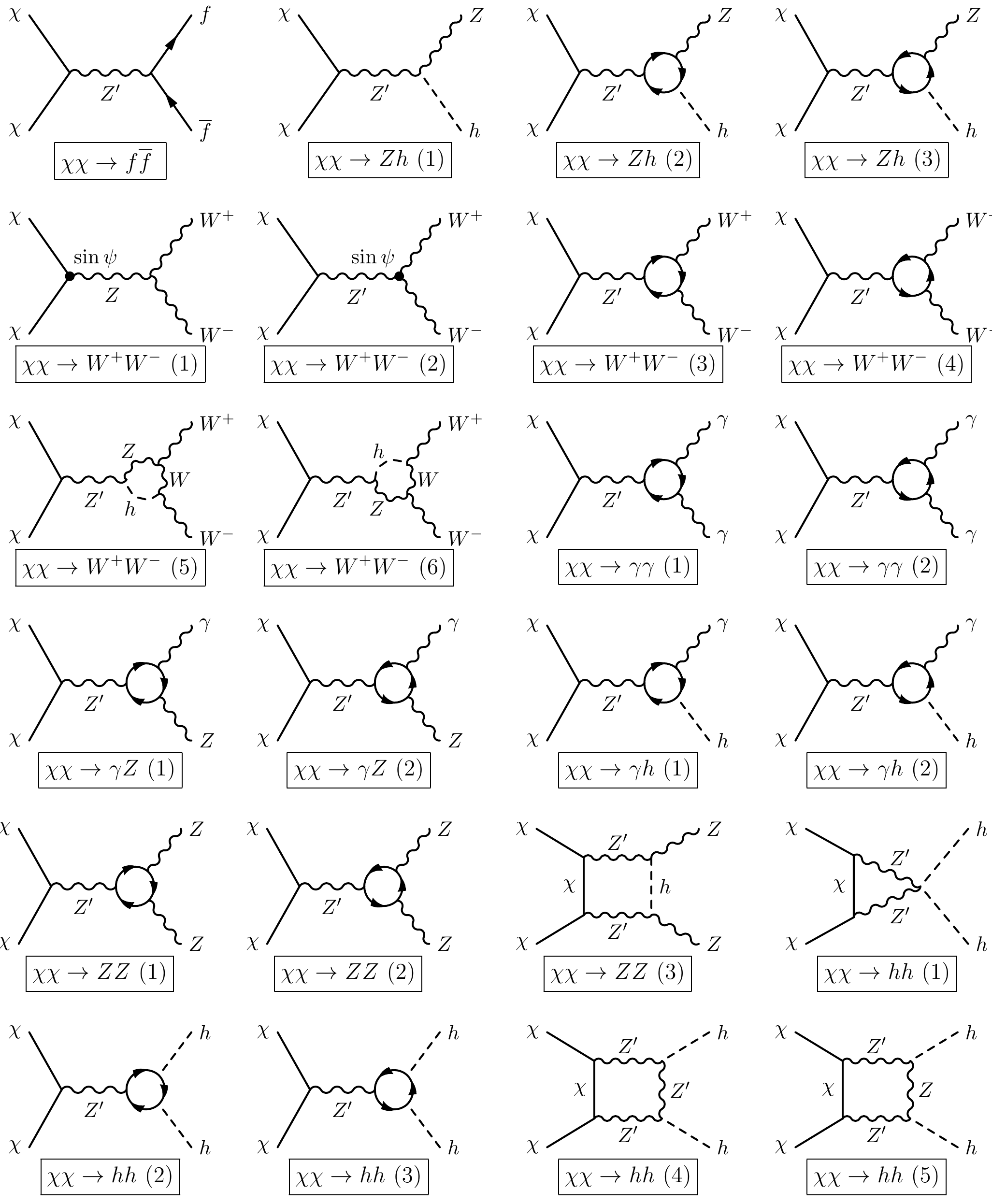}
  \caption{Feynman diagrams for the annihilation of $\chi\chi$ into pairs of SM particles that have been considered in this work.
%  Only for this figure, we denote by $Z_1\simeq Z$ the lightest mass eigenstate, and by $Z_1\simeq Z'$ the heavier mass eigenstate.
  In the fermion loops, the amplitude is summed over all the SM fermions.}
  \label{fig:feynman}
\end{figure}

The Feynman diagrams for the possible annihilation channels of $\chi\chi$ into the SM particles are shown 
on Fig.~\ref{fig:feynman}. For each channel, we consider the leading order (tree level or one-loop), 
moreover we always restrict the calculation to the leading order in the mixing angle between $Z$ and $Z'$, $\psi$.

%At tree level, the annihilation channels are $f\overline f$ and, for $\theta \neq \pi/2$, $W^+ W^-$ and $Zh$. 
At the tree level the DM annihilates into $f \bar f$, and if $\theta \neq \pi/2$ also to $W^+ W^-$ and $Zh$. 
%Recall that there is no mixing between $Z$ and $Z'$ in the pure $\UBL$ gauge theory. 
%Thus the latter two channels vanish at tree level in that case. 
%We report in 
Eqs.~\eqref{eq: xsec f f} to~\eqref{eq: xsec WT WT} summarize the annihilation cross sections into all these channels 
at the tree level, distinguishing between 
the polarization of the vector bosons in the final states through a superscript $(T)$ or $(L)$ 
for transverse or longitudinal polarization, respectively.
%\clearpage
\begin{small}
% ff
\begin{multline}
\label{eq: xsec f f}
\sigma(\chi\chi \to f \bar f) = \frac{\gX^2 \gZp^4 \cos^4 \psi\, N_c^f}{3 \pi s \big((s - \mZp^2)^2+ \GZp^2 \mZp^2\big)}
\sqrt{\frac{s - 4 \mf^2}{s - 4 \mX^2}} \Bigg(
(\cVf)^2  (s - 4 \mX^2 ) (s+ 2 \mf^2 )
 + \\
(\cAf)^2 \bigg( s (s - 4 \mX^2)  + 4 \mf^2 
\Big( \mX^2 \Big(7  - 6 \frac{s}{\mZp^2} + 3 \frac{s^2}{\mZp^4}\Big) - s \Big)
 \bigg)
\Bigg) \,,
\end{multline}
%
% Zh
\begin{multline}
\label{eq: xsec ZL h}
\sigma(\chi\chi \to Z^{ (L)} h) =
\gX^2 \gZp^4 \cos^2\theta \cos^4 \psi  
\frac{ \sqrt{ (s - (m_h^2 + m_Z^2))^2 - 4 m_h^2 m_Z^2}}
{((s-\mZp^2 )^2) + \GZp^2 \mZp^2 ) } 
\frac{1}{48 \pi s^{5/2}} \times \\
\times  \bigg( 
 \Big( m_h^4 (2 \mX^2 + s) - 2 m_h^2 (s+2\mX^2) (s  +m_Z^2) + 2\mX^2 (s^2 -10 s m_Z^2 + m_Z^4) + s (s+m_Z^2)^2\Big) \\
+ \frac{1}{\mZp^2} \Big( -6 \mX^2 s (m_h^4-2 m_h^2 (m_Z^2+s)+(m_Z^2-s)^2) \Big)\\
+ \frac{1}{\mZp^4} \Big( 3 \mX^2 s^2 (m_h^4-2 m_h^2 (s+ m_Z^2)+(s- m_Z^2)^2) \Big)
\bigg) \,,
\end{multline}
\begin{equation}
\label{eq: xsec ZT h}
\sigma(\chi\chi \to Z^{ (T)} h) =
\gX^2 \gZp^4 \cos^2\theta \cos^4 \psi  
\frac{ \sqrt{ (s - (m_h^2 + m_Z^2))^2 - 4 m_h^2 m_Z^2}}
{((s-\mZp^2 )^2) + \GZp^2 \mZp^2 ) } 
\frac{m_Z^2 \big(s - 4 \mX^2 \big)^{1/2} } {6 \pi s^{3/2} } \,,
\end{equation}
%
% WW
\begin{multline}
\label{eq: xsec WL WL}
\sigma(\chi\chi \to W^{+\, (L)}W^{-\, (L)}) = 
\gX^2 \gZp^4\aW \cos^2 \tW \cos^2 \psi \sin^2 \psi
\big( s-4 m_W^2 \big)^{3/2} \big( s-4 \mX^2 \big)^{1/2} \times \\
\times \frac{ (\mZp^2 - m_Z^2)^2 + (\GZp \mZp - \Gamma_Z m_Z)^2 }
{((s-m_Z^2)^2-\Gamma_Z^2 m_Z^2) ((s-\mZp^2)^2-\GZp^2 \mZp^2)} \frac{(2 m_W^2+s)^2}{12 m_W^4 s } \,,
\end{multline}
\begin{multline}
\label{eq: xsec WT WL}
\sigma(\chi\chi \to W^{\pm\, (T)}W^{\mp\, (L)}) = 
\gX^2 \gZp^4 \aW \cos^2 \tW \cos^2 \psi \sin^2 \psi
\big( s-4 m_W^2 \big)^{3/2} \big( s-4 \mX^2 \big)^{1/2} \times \\
\times \frac{ (\mZp^2 - m_Z^2)^2 + (\GZp \mZp - \Gamma_Z m_ Z)^2 }
{((s-m_Z^2)^2-\Gamma_Z^2 m_Z^2) ((s-\mZp^2)^2-\GZp^2 \mZp^2) }
\frac{4}{3\, m_W^2 } \,,
\end{multline}
\begin{multline}
\label{eq: xsec WT WT}
\sigma(\chi\chi \to W^{+\, (T)}W^{-\, (T)}) =
\gX^2 \gZp^4 \aW \cos^2 \tW \cos^2 \psi \sin^2 \psi
\big( s-4 m_W^2 \big)^{3/2} \big( s-4 \mX^2 \big)^{1/2} \times \\
\times \frac{ (\mZp^2 - m_Z^2)^2 + (\GZp \mZp - \Gamma_Z m_Z)^2 }
{ ((s-m_Z^2)^2-\Gamma_Z^2 m_Z^2) ((s-\mZp^2)^2-\GZp^2 \mZp^2) } \frac{2}{3 s} \,.
\end{multline}
\end{small}

A few clarifications are in order about the annihilation cross sections of the DM at tree level.
\begin{itemize}
\item[$f\overline f$:] We denote the number of colors of the fermion $f$ by $N_c^f$ , and  its vector and axial vector couplings to $Z'$ by $\cVf$, $\cAf $ respectively.
The values of $\cVf$, $\cAf $ are given in Tab.~\ref{tab:U1coefficients}
In the zero velocity limit, corresponding to $s=4\mX^2$, the cross section is proportional to $\mf^2$, 
because of the helicity suppression for pairs of annihilating fermions. In that limit, 
$\sigma \propto (\cAf)^2 \propto \cos^2 \theta$, i.e. the $a$ coefficient in the low velocity expansion 
$\sigma v\simeq a + bv^2$ comes from the $\UY$ component of the $\Up$ extension. 
%\AK{I do not understand the last sentence.}
\item[$Zh$:] The diagram on Fig.~\ref{fig:feynman} contains the tree level vertex $Z'Zh$ of Eq.~\eqref{eq:vertex Z'Zh}.
 In the zero velocity limit, the only contribution comes from the production of a longitudinally polarized $Z$, 
 since in Eq.~\eqref{eq: xsec ZT h} the factor $(s-4\mX^2)^{1/2}$ vanishes.
\item[$WW$:] We denote $\aW=g_W^2/(4\pi)$, where $g_W$ is the weak coupling constant. 
The amplitude is the sum of the two diagrams on Fig.~\ref{fig:feynman}: the annihilation occurs via the mixing 
of $Z$ and $Z'$ and the SM trilinear gauge vertex $ZWW$, see Eq.~\eqref{eq:vertex Z'WW}. For each 
of the final polarization states, the cross section is proportional to $(s-4\mX^2)^{1/2}$.
\end{itemize}

Given that $\sigma(\chi\chi \to WW)\vDM$ (where $\vDM$ is the DM velocity) is suppressed by 
$\sin^2\psi$ and by $\vDM^2$ at tree level it is worth checking whether contributions arising at one loop can 
become important. 
%The contribution to the amplitude of diagrams~5 and~6 on Fig.~\ref{fig:feynman} contains a trilinear interaction 
%vertex. Therefore their contribution to the $\langle \sigma v \rangle$ is $v^2$-suppressed by virtue of the same argument 
%as we made in Sec~\ref{sec:BR}.
%The same argument reported in Sec.~\ref{sec:BR} that explains why 
%$\sigma(\chi\chi \to WW)\vDM \simeq a+ bv^2$ has $a=0$ for the tree level channel applies also to the annihilation via 
%a bosonic loop (diagrams 5 and 6 of Fig.~\ref{fig:feynman}), since the Feynman amplitude again contains  a trilinear gauge interaction vertex. 
%However, this argument does not hold for diagrams~3 and~4% involving a fermionic loop. 
%Therefore we also computed the cross section coming from the sum of  diagrams 3 and 4.
%We ignore the interference term 
%with the other diagrams, because in the limit $\vDM\to 0$, which is relevant for IceCube, the Feynman amplitudes for the processes 1, 2, 5 and 6 vanish.
%
The contribution to the amplitude of diagrams~5 and~6 on Fig.~\ref{fig:feynman} is velocity suppressed because of the same argument reported at the end of Sec.~\ref{sec:BR}. 
Therefore we only considered the contribution to the cross section coming from the sum of  diagrams 3 and 4, also ignoring the interference terms with the other diagrams.

We also computed  the cross sections for the annihilations into $Z Z$, $\gamma \gamma$, 
$\gamma h$, $\gamma Z$ and $h h$ at one loop. It is worth computing these corrections because of the velocity 
suppression of some tree level channels, and because in the pure $\UBL$ case the tree level annihilations into 
$WW$ and $Zh$ disappear. % (as discussed at the end of this Appendix).
As for the $Z h$ channel, we computed the one loop cross section only in the pure $\UBL$ case ($\theta = \pi/2$), 
in which the tree level amplitude vanishes, and the only remaining contribution comes from 
diagrams 2 and 3 in Fig.~\ref{fig:feynman}. The results of the loop calculation are:
%%%%%%%%%%%%%%%%%%%%%%%%%%%%%
%
%% Uncomment the following line to include the cross sections for annihilation into WW
\begin{small}
\begin{multline}
\label{eq: xsec WT WT fl}
\sigma(\chi\chi \to W^{+\,(T)} W^{-\,(T)}) =
\frac{\gZp^4 \gX^2 \Nc \left(s-4 m_W^2\right)^{3/2} \alpha_W^2}
{768 \pi ^3 \left(\GZp^2 \mZp^2+\left(\mZp^2-s\right)^2\right) s \sqrt{s-4 \mX^2}} \times \\
\times\Bigg\{
%\left(
	\frac{48 \mX^2 \left(s-\mZp^2\right)^2}{\mZp^4 (s-4 m_W^2)^2}
	\Bigg| \sum_i \Nc
		\bigg\{
			(\cLd-\cRd) B_0(s,\mD^2,\mD^2) \mD^2 - \\
			-(\cLd-\cRd) (\mD^2-\mU^2-m_W^2) C_0(m_W^2,m_W^2,s,\mD^2,\mU^2,\mD^2) \mD^2
			-\frac{1}{2} (\cLd+\cLu) (s - 4 m_W^2) + \\
			+\big(
				(\cRd-\cLd) \mD^2
				+(\cRu-\cLu) \mU^2
			\big) B_0(m_W^2,\mD^2,\mU^2)
			+(\cLu-\cRu) \mU^2 B_0(s,\mU^2,\mU^2) + \\
			+(\cLu-\cRu) \mU^2 (\mD^2-\mU^2+m_W^2) C_0(m_W^2,m_W^2,s,\mU^2,\mD^2,\mU^2)
		\bigg\}
	\Bigg|^2 \\
	+ \frac{32}{81} \frac{s - 4 \mX^2}{m_W^4(s - 4 m_W^2)^4}
	\Bigg| \sum_i \Nc
		\bigg\{
			\frac{1}{2} (s - 4 m_W^2) \big(
				(3 \mD^2-3 \mU^2+7 m_W^2-5 s / 2 ) \cLd + \\
				+(3 \mD^2-3 \mU^2-7 m_W^2+5 s / 2) \cLu
			\big) m_W^2
			+\frac{3}{2} \Big[
				3 \cRd (s-4 m_W^2) \mD^2
				+\cLd \big(
					-6 \mD^4
					+2 (6 \mU^2+2 m_W^2+s) \mD^2 - \\
					-6 \mU^4+6 m_W^4+s^2-3 \mU^2 s
					-7 m_W^2 s
				\big)
			\Big] B_0(s,\mD^2,\mD^2) m_W^2
			-\frac{3}{2} \Big[
				3 \cRu (s-4 m_W^2) \mU^2 + \\
				+\cLu \big(
					-6 \mD^4
					-3 (s - 4 \mU^2) \mD^2
					-6 \mU^4
					+6 m_W^4
					+s^2
					-7 m_W^2 s
					+2 \mU^2 (s + 2 m_W^2)
				\big)
			\Big] B_0(s,\mU^2,\mU^2) m_W^2 - \\
			-\frac{9}{2} (m_W^2-\mD^2+\mU^2) \Big[
				-\cRd (s-4 m_W^2) \mD^2
				+\cLd \Big(
					2 \mD^4
					-(s+4 \mU^2) \mD^2 + \\
					+2 \big(
						\mU^4
						+(s-2 m_W^2) \mU^2
						+m_W^4
					\big)
				\Big)
			\Big] C_0(m_W^2,m_W^2,s,\mD^2,\mU^2,\mD^2) m_W^2 + \\
			+9 \Big[
				-\frac{1}{2} (s - 4 m_W^2) \cRu \mU^2
				+\big(
					\mD^4
					+(s -2 \mU^2-2 m_W^2 ) \mD^2
					+\mU^4
					+m_W^4
					-\mU^2 s/2
				\big) \cLu
			\Big] \times \\
			\times (\mD^2-\mU^2+m_W^2) C_0(m_W^2,m_W^2,s,\mU^2,\mD^2,\mU^2) m_W^2
			+\frac{3}{2} (\mD^2-\mU^2) \big(
				\cLd (\mD^2 -\mU^2 -m_W^2 ) - \\
				-\cLu ( \mD^2 -\mU^2 +m_W^2 )
			\big) (s - 4 m_W^2) B_0(0,\mD^2,\mU^2)
			+3 \Big[
				-\frac{3}{2} (s - 4 m_W^2)
				( \cRd \mD^2 -\cRu \mU^2 ) m_W^2 + \\
				+\cLu \Big(
					\frac{1}{2} \big(
						\mD^4
						-2 (\mU^2+m_W^2) \mD^2
						+\mU^4
						+m_W^4
						+\mU^2 m_W^2
					\big) s
					- m_W^2 \big(
						5 \mD^4
						+2 (m_W^2-5 \mU^2) \mD^2
						+5 \mU^4 + \\
						+5 m_W^4
						-4 \mU^2 m_W^2
					\big)
				\Big)
				+\cLd \Big(
					m_W^2 \big(
						5 \mD^4
						-2 (5 \mU^2+2 m_W^2) \mD^2
						+5 \mU^4
						+5 m_W^4
						+2 \mU^2 m_W^2
					\big) - \\
					-\frac{1}{2} \big(
						\mD^4
						+(m_W^2-2 \mU^2) \mD^2
						+(\mU^2-m_W^2)^2
					\big) s
				\Big)
			\Big] B_0(m_W^2,\mD^2,\mU^2)
		\bigg\}
	\Bigg|^2
%\right)
\Bigg\}
\end{multline}
\begin{normalsize}
We do not report the formula for $\sigma(\chi\chi \to W^{\pm\,(T)} W^{\mp\,(L)})$ because this channel turns out to have $a=0$ (therefore it is irrelevant for the annihilation process in the Sun) and the corresponding formula is too cumbersome to be reported here.
\end{normalsize}
\begin{multline}
\label{eq: xsec WL WL fl}
\sigma(\chi\chi \to W^{+\,(L)} W^{-\,(L)}) =
\frac{\gZp^4 \gX^2 \aW^2 \sqrt{s - 4 mx^2}
}{
62208\pi^3 m_W^8  [(s- \mZp^2)^2+\GZp^2 \mZp^2] s (s - 4 m_W^2)^{5/2}
}
\times \\
\times \Bigg|
%\sum_{D=\binom{u_L}{d_L}}
\sum_i
\Nc \Bigg\{
 (4 m_W^2 - s) m_W^2 \Big(\cLu (-4 m_W^4 + 12 \mU^2 m_W^2 + 2 s m_W^2 - s^2 -
       6 \mD^2 (s - 2 m_W^2 ) ) \\ 
    - \cLd (-4 m_W^4 + 12 \mD^2 m_W^2 + 2 s m_W^2 - s^2 - 
       6 \mU^2 (s - 2 m_W^2)) \Big) \\
 + 6 \Big(A_0(\mD^2) \cLd - A_0(\mU^2) \cLu\Big) (s - 4 m_W^2)^2 m_W^2  \\
 - 3 B_0(0, \mD^2, \mU^2)
    (\mD^2 - \mU^2) (4 m_W^2 - s) 
    \Big(\cLu (4 m_W^4 - 2 s m_W^2 + \mU^2 (8 m_W^2 - s) + \mD^2 (s - 8 m_W^2)) + \\
       \cLd (4 m_W^4 - 2 s m_W^2 + \mD^2 (8 m_W^2 - s) + 
       \mU^2 (s - 8 m_W^2))\Big) \\
 + 3 B_0(m_W^2, \mD^2, \mU^2)
    \Big(6 (\cRd \mD^2 - \cRu \mU^2) (4 m_W^2 - s) m_W^4 
      + \cLd [8 m_W^8 + 12 s m_W^6 - 2 s^2 m_W^4 + \\
       (\mD^2-\mU^2)^2 (32 m_W^4 - 6 s m_W^2 + s^2)  
       +\mU^2 (8 m_W^4 + 6 s m_W^2 + s^2) m_W^2 + \\
       \mD^2 (-16 m_W^6 - 12 s m_W^4 + s^2 m_W^2)] 
    - \cLu [8 m_W^8 + 12 s m_W^6 - 2 s^2 m_W^4 + \\
       (\mD^2-\mU^2)^2 (32 m_W^4 - 6 s m_W^2 + s^2)  
       + \mU^2 (-16 m_W^4 - 12 s m_W^2 + s^2) m_W^2 + 
       \mD^2 (m_W^2 (8 m_W^4 + 6 s m_W^2 + s^2))]\Big) \\
 + 3 B_0(s, \mU^2, \mU^2)
    m_W^2 \Big(6 \cRu \mU^2 (4 m_W^2 - s) m_W^2 
    + \cLu [-24 m_W^6 + 12 s m_W^4 + 4 s (s - 4 \mU^2) m_W^2 + 6 \mD^4 s \\ 
       + s (6 \mU^4 + s \mU^2 - s^2) + 
       3 \mD^2 (8 m_W^4 - 2 s m_W^2 + s (s - 4 \mU^2))] \Big)  \\
 - 3 B_0(s,\mD^2,\mD^2)
    m_W^2 \Big(6 \cRd \mD^2 (4 m_W^2 - s) m_W^2 
     + \cLd [-24 m_W^6 + 12 s m_W^4 + 4 s (s - 4 \mD^2) m_W^2 + 6 \mU^4 s \\ 
       + s (6 \mD^4 + s \mD^2 - s^2) + 
       3 \mU^2 (8 m_W^4 - 2 s m_W^2 + s (s - 4 \mD^2))] \Big)  \\
 + 18 C_0(m_W^2,m_W^2,s, \mU^2, \mD^2, \mU^2)
    m_W^2 \Big(\cRu \mU^2 (\mD^2 - \mU^2 + m_W^2) (4 m_W^2 - s) m_W^2 \\
     + \cLu [s \mD^6 + (4 m_W^4 - 2 s m_W^2 + 
          s (s - 3 \mU^2)) \mD^4 + (-8 m_W^6 + 5 s m_W^4 + 3 \mU^4 s \\
          - \mU^2 (4 m_W^4 + s m_W^2 + s^2)) \mD^2 
          - (\mU^2 - 
          m_W^2) (4 m_W^6 - 2 \mU^2 s m_W^2 + \mU^4 s)] \Big) \\
 + 18 C_0(m_W^2,m_W^2,s, \mD^2, \mU^2, \mD^2)
    m_W^2 \Big(\cRd \mD^2 (\mD^2 - \mU^2 - m_W^2) (4 m_W^2 - s) m_W^2 \\ 
    + \cLd [-4 m_W^8 + \mD^6 s - \mU^6 s - 3 \mD^4 (\mU^2 + m_W^2) s + 
       \mU^2 (8 m_W^6 - 5 m_W^4 s) \\
       - \mU^4 (4 m_W^4 - 2 s m_W^2 + s^2) + 
       \mD^2 (3 s \mU^4 + (4 m_W^4 + s m_W^2 + s^2) \mU^2 + 
          2 m_W^4 (2 m_W^2 + s))] \Big)
\Bigg\} \Bigg| ^2 \,,
\end{multline}
%
%
%% Uncomment the following line to include the cross sections for annihilation into ZZ
\begin{multline}
\label{eq: xsec ZT ZT}
\sigma(\chi\chi \to Z^{(T)} Z^{(T)}) =
\frac{4 \gX^2 \gZp^4 \mX^2 m_Z^4 (\mZp^2-s)^2}
{\pi ^5 \mZp^4 s v^4 \sqrt{s-4 \mX^2} \sqrt{s-4 m_Z^2} \left(\GZp^2 \mZp^2+\left(\mZp^2-s\right)^2\right)} \times \\
\times \Bigg| \sum_f
	N_c^f \bigg\{
		4 \cAf m_Z^2 (\cZLf)^2
		-4\cVf m_Z^2 (\cZLf)^2
		-\cAf s (\cZLf)^2
		+\cVf s (\cZLf)^2
		+4 \cAf (\cZRf)^2 m_Z^2
		+4 (\cZRf)^2\cVf m_Z^2 \\
		-\cAf (\cZRf)^2 s
		-(\cZRf)^2\cVf s
		-\bigg[
			\cAf \Big(
				(4 m_f^2-4 m_Z^2+s) (\cZLf)^2
				-8 \cZRf m_f^2 \cZLf
				+(\cZRf)^2 (4 m_f^2-4 m_Z^2+s)
			\Big) \\
			+\big((\cZLf)^2-(\cZRf)^2\big)\cVf (4 m_Z^2-s)
		\bigg] B_0(m_Z^2,m_f^2,m_f^2)
		+\bigg[
			\big((\cZLf)^2-(\cZRf)^2\big)\cVf (4 m_Z^2-s)  \\
			+\cAf \Big(
				(4 m_f^2-4 m_Z^2+s) (\cZLf)^2
				-8 \cZRf m_f^2 \cZLf
				+(\cZRf)^2 (4 m_f^2-4 m_Z^2+s)
			\Big)
		\bigg] B_0(s,m_f^2,m_f^2) \\
		+4 \cAf m_f^2 \left(
			(\cZLf)^2 m_Z^2
			+(\cZRf)^2 m_Z^2
			+\cZLf \cZRf (2 m_Z^2-s)
		\right) C_0(s,m_Z^2,m_Z^2,m_f^2,m_f^2,m_f^2)
	\bigg\}
\Bigg| ^2 \,,
\end{multline}
\begin{multline}
\label{eq: xsec ZT ZL}
\sigma(\chi\chi \to Z^{(T)} Z^{(L)}) =
\frac{4 \gX^2 \gZp^4 m_Z^2 \sqrt{s-4 \mX^2}}{3 \pi ^5 s v^4 \left(s-4 m_Z^2\right)^{3/2} \left(\GZp^2 \mZp^2+\left(\mZp^2-s\right)^2\right)} \times \\
\times \Bigg| \sum_f
	N_c^f \bigg\{
		-4 \cAf (\cZLf)^2 m_Z^4
		-4 \cAf (\cZRf)^2 m_Z^4
		+4 (\cZLf)^2\cVf m_Z^4
		-4 (\cZRf)^2\cVf m_Z^4 \\
		+\cAf (\cZLf)^2 s m_Z^2
		+\cAf (\cZRf)^2 s m_Z^2
		-(\cZLf)^2\cVf s m_Z^2
		+(\cZRf)^2\cVf s m_Z^2 \\
		+\frac{1}{2} \bigg[
			\cAf \bigg(
				-\big(
					20 m_Z^4
					-6 s m_Z^2
					+s^2
					+4 m_f^2 (s-4 m_Z^2)
				\big) (\cZLf)^2
				+8 \cZRf m_f^2 (s-4 m_Z^2) \cZLf \\
				-(\cZRf)^2 \big(
					20 m_Z^4
					-6 s m_Z^2
					+s^2
					+4 m_f^2 (s-4 m_Z^2)
				\big)
			\bigg) \\
			+\big((\cZLf)^2-(\cZRf)^2\big)\cVf \big(
				20 m_Z^4
				-6 s m_Z^2
				+s^2
			\big)
		\bigg] B_0(m_Z^2,m_f^2,m_f^2) \\
		-\frac{1}{2} \bigg[
			\cAf \bigg(
				-\big(
					20 m_Z^4
					-6 s m_Z^2
					+s^2
					+4 m_f^2 (s-4 m_Z^2)
				\big) (\cZLf)^2
				+8 \cZRf m_f^2 (s-4 m_Z^2) \cZLf \\
				-(\cZRf)^2 \big(
					20 m_Z^4
					-6 s m_Z^2
					+s^2
					+4 m_f^2 (s-4 m_Z^2)
				\big)
			\bigg) \\
			+\big((\cZLf)^2-(\cZRf)^2\big)\cVf \big(
				20 m_Z^4
				-6 s m_Z^2
				+s^2
			\big)
		\bigg] B_0(s,m_f^2,m_f^2) \\
		+2 \bigg[
			-\frac{1}{2} \big((\cZLf)^2-(\cZRf)^2\big)\cVf \Big(
				-2 m_Z^6
				+2 s m_Z^4
				+m_f^2 \big(
					8 m_Z^4
					-6 s m_Z^2
					+s^2
				\big)
			\Big) \\
			-\frac{1}{2} \cAf \bigg(
				\big(
					2 m_Z^6
					-2 s m_Z^4
					+4 m_f^2 s m_Z^2
					-m_f^2 s^2
				\big) (\cZLf)^2
				+2 \cZRf m_f^2 \big(
					8 m_Z^4
					-6 s m_Z^2
					+s^2
				\big) \cZLf \\
				+(\cZRf)^2 \left(
					2 m_Z^6
					-2 s m_Z^4
					+4 m_f^2 s m_Z^2
					-m_f^2 s^2
				\right)
			\bigg)
		\bigg] C_0(s,m_Z^2,m_Z^2,m_f^2,m_f^2,m_f^2)
	\bigg\}
\Bigg| ^2 \,,
\end{multline}
\begin{equation}
\label{eq: xsec ZL ZL}
\sigma(\chi\chi \to Z^{(L)} Z^{(L)}) = 0,
\end{equation}
\begin{multline}
\label{eq: xsec ZT h loop}
\sigma(\chi\chi \to Z^{(T)}h) \big|_{\theta=\pi/2} =
\frac{\aW  \gX^2 \gZp^4}
{24 \pi ^4  \left(\mZp^4+\GZp^2 \mZp^2-2 s \mZp^2+s^2\right)} \times \\
\times
\frac{\sqrt{s-4\mX^2}
	\sqrt{(m_H^2-m_Z^2)^2 +s^2 -2 s (m_H^2+m_Z^2)}}
	{s^{3/2} v^2 \left((m_H-m_Z)^2-s\right)^2 \left((m_H+m_Z)^2-s\right)^2}
\Bigg| \sum_f  \cVf m_f^2 N_c^f  (\cZLf+\cZRf) \times \\
\times
\bigg\{
	C_0(m_H^2,m_Z^2,s,m_f^2,m_f^2,m_f^2) s^3
	+2 (-m_H^2-m_Z^2+s) B_0(s,m_f^2,m_f^2) s \\
	-\Big[
		2 B_0(m_H^2,m_f^2,m_f^2)
		+(-4 m_f^2+3 m_H^2+m_Z^2) C_0(m_H^2,m_Z^2,s,m_f^2,m_f^2,m_f^2)
		-2
	\Big] s^2 \\
	+\Big[
		-2 B_0(m_Z^2,m_f^2,m_f^2) m_Z^2
		+2(m_H^2+2 m_Z^2) B_0(m_H^2,m_f^2,m_f^2) \\
		-\Big(
			(8 m_f^2-3 m_H^2+m_Z^2) C_0(m_H^2,m_Z^2,s,m_f^2,m_f^2,m_f^2)
			+4
		\Big) (m_H^2+m_Z^2)
%		+2(m_H^2+2 m_Z^2) B_0(m_H^2,m_f^2,m_f^2)
	\Big] s \\
%	+2 (-m_H^2-m_Z^2+s) B_0(s,m_f^2,m_f^2) s
	-(m_H^2-m_Z^2) \Big[
		-2 B_0(m_H^2,m_f^2,m_f^2) m_Z^2
		+2 B_0(m_Z^2,m_f^2,m_f^2) m_Z^2 \\
		+\Big(
			(-4 m_f^2+m_H^2-m_Z^2) C_0(m_H^2,m_Z^2,s,m_f^2,m_f^2,m_f^2)
			-2
		\Big)
		(m_H^2-m_Z^2)
	\Big]
\bigg\}
\Bigg|^2 \,,
\end{multline}
\begin{multline}
\label{eq: xsec ZL h loop}
\sigma(\chi\chi \to Z^{(L)}h) \big|_{\theta=\pi/2}=
\frac{\aW   \gX^2 \gZp^4 }{3 \pi ^4  \left(\GZp^2 \mZp^2+\mZp^4-2 \mZp^2 s+s^2\right) }\times \\
\times \frac{m_Z^2 \sqrt{s-4\mX^2}}{ \sqrt{s} v^2 \left(-2 s \left(m_H^2+m_Z^2\right)+\left(m_H^2-m_Z^2\right)^2+s^2\right)^{3/2}}
\Bigg|\sum_f \cVf m_f^2 N_c^f (\cZLf+\cZRf) \times \\
	\times \bigg\{
		m_H^2 \Big[
			(-m_H^2+m_Z^2+s) C_0(m_H^2,m_Z^2,s,m_f^2,m_f^2,m_f^2)
			-2 B_0(m_H^2,m_f^2,m_f^2)
		\Big] \\
		+(m_H^2+m_Z^2-s) B_0(m_Z^2,m_f^2,m_f^2)
		+(m_H^2-m_Z^2+s) B_0(s,m_f^2,m_f^2)
	\bigg\}
\Bigg|^2  \,.
\end{multline}
\begin{multline}
\label{eq: xsec g g}
\sigma(\chi\chi \to \gamma\gamma) =
\frac{16 \aem^2 \gX^2 \gZp^4}{\pi^3 \left(\GZp^2 \mZp^2+\left(\mZp^2-s\right)^2\right)}
\frac{\mX^2 \sqrt{s} (s-\mZp^2)^2 }{ \mZp^4 \sqrt{s-4\mX^2} } \times \\
	\times \Bigg|\sum_f \cAf N_c^f  Q_f^2
	\Big[
		2 m_f^2 C_0(0,0,s,m_f^2,m_f^2,m_f^2)
		+1
	\Big]
	\Bigg|^2 \,,
\end{multline}
\begin{multline}
\label{eq: xsec g h}
\sigma(\chi\chi \to \gamma h) =
\frac{\aem  \gX^2 \gZp^4  }{6 \pi^4 \left(\GZp^2 \mZp^2+\mZp^4-2 \mZp^2 s+s^2\right)}
\frac{\sqrt{s-4\mX^2}}{s^{3/2} v^2 \left(s-m_H^2\right) } \times \\
	\times \Bigg| \sum_f \cVf m_f^2 N_c^f  Q_f 
	\bigg\{
		-2 s B_0(m_H^2,m_f^2,m_f^2)
		+2 s B_0(s,m_f^2,m_f^2) \\
		+(s-m_H^2) \Big[
			(4 m_f^2-m_H^2+s) C_0(m_H^2,0,s,m_f^2,m_f^2,m_f^2)
			+2
		\Big]
	\bigg\}
	\Bigg|^2 \,,
\end{multline}
\begin{multline}
\label{eq: xsec g ZT}
\sigma(\chi\chi \to \gamma Z^{(T)}) =
\frac{\aem \aW  \gX^2 \gZp^4 }
{3 \pi ^3 \left([(s-\mZp^2)^2+\GZp^2 \mZp^2\right]
\mZp^4 s^{5/2} \left(s-m_Z^2\right) \sqrt{s-4\mX^2}
} \times \\
\times \Bigg\{
	\frac{\mZp^4}{2} (s-4\mX^2) \Bigg| \sum_f   N_c^f  Q_f 
	\bigg\{
		m_Z^2 s \big(\cAf (\cZLf+\cZRf)+\cVf (\cZRf-\cZLf)\big) \big(B_0(m_Z^2,m_f^2,m_f^2) - B_0(s,m_f^2,m_f^2)\big) \\
		-(s-m_Z^2) \Big[
			2 m_f^2 C_0(m_Z^2,0,s,m_f^2,m_f^2,m_f^2) \big(
				\cAf \cZLf m_Z^2+\cAf \cZRf m_Z^2-\cZLf\cVf s+\cZRf \cVf s
			\big) \\
			+m_Z^2 \big(\cAf (\cZLf+\cZRf)+\cVf (\cZRf-\cZLf)\big)
		\Big]
	\bigg\}
	\Bigg|^2
	+3 \mX^2 (s-m_Z^2)^4 (s-\mZp^2)^2  \times \\
	\times \Bigg| \sum_f    N_c^f  Q_f 
	\bigg\{
		2 \cAf m_f^2 (\cZLf+\cZRf) C_0(m_Z^2,0,s,m_f^2,m_f^2,m_f^2)
		+\cAf (\cZLf+\cZRf)+\cVf (\cZRf-\cZLf)
	\bigg\}
	\Bigg|^2 \,,
\Bigg\}
\end{multline}
\begin{multline}
\label{eq: xsec g ZL}
\sigma(\chi\chi \to \gamma Z^{(L)}) =
\frac{\aem \aW \gX^2 \gZp^4}{6 \pi ^3\left([(s-\mZp^2)^2+\GZp^2 \mZp^2\right]}
\frac{ \sqrt{s-4\mX^2}}{m_Z^2 s^{3/2} \left(s-m_Z^2\right) } \times \\
	\times\Bigg|\sum_f    N_C^f Q_f 
	\bigg\{	
		m_Z^2 s  \big(\cAf (\cZLf+\cZRf)+\cVf (\cZRf-\cZLf)\big) \big(
			B_0(m_Z^2,m_f^2,m_f^2) - B_0(s,m_f^2,m_f^2)
		\big) \\
		-(s-m_Z^2) \Big[
			2 m_f^2 C_0(m_Z^2,0,s,m_f^2,m_f^2,m_f^2) \big(
				\cAf \cZLf m_Z^2+\cAf \cZRf m_Z^2-\cZLf \cVf s+\cZRf \cVf s
			\big) \\
			+m_Z^2 \big(\cAf (\cZLf+\cZRf)+\cVf (\cZRf-\cZLf)\big)
		\Big]
	\bigg\}
	\Bigg|^2 \,,
\end{multline}
\begin{multline}
\label{eq: xsec h h}
\sigma(\chi\chi \to h h) =
\frac{\gX^4 \gZp^8 \mX^2 \cos^4\theta}{2048 \pi ^5}
\frac{\sqrt{s-4 m_h^2}}{s\big(s-4 \mX^2\big)^{3/2}} \times \\
	\times \Bigg|
		B_0(\mX^2,\mX^2,\mZp^2)
		-B_0(s,\mZp^2,\mZp^2)
		+(2 \mX^2+\mZp^2-s) C_0(\mX^2,\mX^2,s,\mZp^2,\mX^2,\mZp^2)
	\Bigg|^2 \,.
\end{multline}
\end{small}

In the cross sections involving a fermionic loop, we denoted by $\sum_f$ a sum over all the SM fermion species. In the $WW$ cross section, instead, we denoted by $\sum_i$ a sum over the six fermion families (three families of quarks and three of leptons), with $\mU,\mD$ the upper and lower component of the doublet, respectively, and with $\cLu$, $\cRu$, $\cLd$, $\cRd$ the combinations
\begin{equation}
\cLu = \cVu-\cAu,
\qquad
\cRu = \cVu+\cAu,
\qquad
\cLd = \cVd-\cAd,
\qquad
\cRd = \cVd+\cAd.
\end{equation}
The functions $A_0$, $B_0$ and $C_0$ are the standard Passarino-Veltman one loop one-, two- and three-points scalar integrals \cite{Passarino:1978jh}:
\begin{small}
\begin{equation}
A_0 \left(m_0^2\right) =  \frac{\mu^{4-D}}{i \pi^{D/2} \gamma_\Gamma}
\int{\frac{d^D k}{k^2 - m_0^2} } \,,
\end{equation}
\begin{equation}
B_0 \left(p^2,m_1^2,m_2^2\right) = \frac{\mu^{4-D}}{i \pi^{D/2} \gamma_\Gamma}
\int{\frac{d^D k}
{\left(k^2-m_1^2\right)
\left((k+p)^2-m_2^2\right)}} \,,
\end{equation}
\begin{equation}
C_0 \left(p_1^2,p_2^2,(p_1+p_2)^2,m_1^2,m_2^2,m_3^2\right) = \frac{\mu^{4-D}}{i \pi^{D/2} \gamma_\Gamma}
\int{\frac{d^D k}
{\left(k^2-m_1^2\right)
\left((k+p_1)^2-m_2^2\right)
\left((k+p_1+p_2)^2-m_3^2\right)}} \,,
\end{equation}
\end{small}
with
\begin{equation}
\gamma_\Gamma = \frac{\Gamma^2(1-\epsilon)\Gamma(1+\epsilon)}{\Gamma(1-2\epsilon)} \, , \quad D=4-2\epsilon \,,
\end{equation}
where $\gamma_\Gamma$ approaches 1 in the limit $\epsilon \rightarrow 0$.
In these equations, we denoted by $v \simeq 246\GeV$ the vacuum expectation value of the Higgs field, and by $\cZLf$, $\cZRf$ the SM coupling of the $Z$ boson to left- and right-handed fermions respectively (see Tab.~\ref{tab:Zcouplings}).
\begin{table}[h!]
\centering
\begin{tabular}{ccc}
\toprule
SM fermion $f$ & $\cZLf$ & $\cZRf$ \\
\midrule
leptons & $-\frac{1}{2} + \sin^2\tW$ & $\sin^2\tW$ \\
neutrinos & $\frac{1}{2}$ & $0$ \\
up quarks & $\frac{1}{2} -\frac{2}{3}\sin^2\tW$ & $-\frac{2}{3}\sin^2\tW $\\
down quarks & $-\frac{1}{2} +\frac{1}{3}\sin^2\tW $& $\frac{1}{3}\sin^2\tW $\\
\bottomrule
\end{tabular}
\caption{Coupling of the $Z$ boson to SM fermions.}
\label{tab:Zcouplings}
\end{table}

Let us briefly comment on the one loop cross sections:
\begin{itemize}
\item[$WW$:] The bosonic loop diagrams 5 and 6 in Fig.~\ref{fig:feynman} are velocity suppressed as expected. 
%precisely the same as the tree level contributions 1 and 2 (the justification is reported at the end of Sec.~\ref{sec:BR}). The fermionic loop diagrams 3 and 4 turn out not to be velocity suppressed. Therefore we consider only the latter ones when computing $\sigma(\chi\chi \to WW)$, because in the limit $\vDM \to 0$ the other amplitudes vanish.
\item[$Z Z$:] The box diagram (number~3 on Fig.~\ref{fig:feynman}) is suppressed at low energies by the two heavy propagators in the loop, and gives only a minor effect. Therefore we ignored it in our calculations.
\item[$\gamma\gamma$:] The cross section for annihilation into $\gamma\gamma$ vanishes on resonance, due to the factor of $(s-\mZp^2)^2$ in the numerator.
This is a consequence of the Landau-Yang theorem~\cite{Landau:1948kw,Yang:1950rg} that states that a spin-1 
particle can not decay into two photons, and is a reassuring cross-check of our results.\\
Also notice  that the $\gamma\gamma$ cross section is proportional to $\cAf$, 
the axial coupling of the fermions to the $Z'$, and vanishes in the limit of pure $B-L$.
This is due to the Dirac structure of the fermion loop in the very same way in which the cross section for annihilation 
into $\gamma h$ is proportional to the vectorial coupling $\cVf$, and can be seen as a realization of the 
Furry theorem~\cite{Weinberg:1995mt}, which states that any physical amplitude involving an odd number of photons 
vanishes (in our case one of the photons is replaced by the vectorial part of the $Z'$).
\item[$h h$:] The two diagrams with a fermionic loop (diagrams 2 and 3 in Fig.~\ref{fig:feynman}) sum  to zero, 
while the two box diagrams (numbers~4 and~5) give a contribution at most comparable to that of the 
triangular diagram (number 1).
Since the cross section for annihilation into $h h$ including only the triangular diagram is subdominant by several orders of magnitude, we can safely ignore the contribution of the two box diagrams. %\EM{Can we make this argument more robust?}
\end{itemize}

Results of our calculations show that in the low kinetic energy regime that is relevant for DM annihilation in the Sun
 loop channels are usually subdominant.
Some of the cross sections receive a velocity suppression ($\sigma v \simeq b v^2$)
 and precisely vanish  in the zero velocity limit.
Those are $W^{(T)}W^{(T)}$, $Z^{(T)}Z^{(L)}$, 
$\gamma Z^{(L)}$, $\gamma h$, $h h$, $Z^{(T)} h$ and, for $\theta=\pi/2$, $Z^{(L)} h$.
%\EM{It's not clear from the analytical expression that $h h$ has $a=0$, but numerically it seems so.}
We do not explicitly show  the analytical expansion around $v=0$ because the velocity appears as an 
argument of the Passarino-Veltman functions. The only process, which acquires a relevant contribution at the 
one-loop level is $W^{(L)} W^{(L)}$. 

%In the limit $\vDM\to 0$, 
All the cross sections we computed, except for $h h$ that has no $s$-channel exchange of a $Z'$ boson,  
vanish around $\mx = \mZp/2$ in the $\vDM \to 0$ limit. 
Again, this is a cross-check of the correctness of our calculations. 
Indeed, close to the resonance the cross section for $\chi\chi\to X X$ is proportional to the product 
$\Gamma(Z'\to\chi\chi) \cdot \Gamma(Z'\to X X)$, but $\Gamma(Z'\to\chi\chi)$ vanishes if $\mx = \mZp/2$, 
which is implied by a resonant production of $Z'$ with $\vDM=0$.
%%%%%%%%%%%%%%%%%%%%%%%%%%%%%%%% 

%%%%%%%%%%%%%%%%%%%%%%%%%%%%%%%%
% !TEX root = main_draft.tex
\section{Erratum}
In Ref.~\cite{Jacques:2016dqz-Erratum} we correct the mistakes of the original version of the present paper \cite{Jacques:2016dqz} and recalculate the relevant bounds on the $Z'$-mediated DM. 
The mistakes of the published version have to do with the calculation of the annihilation cross sections. 
In particular in this erratum we properly take into account:
\begin{itemize}
  \item the effects of the $Z$ exchange due to the mixing that are parametrically not smaller than the effects of the $Z'$ exchange;
  \item the complex mass scheme that changes the behavior on the resonances. 
\end{itemize}
This changes the dominant annihilation channels, in particular suppressing the $Zh$ channel. 
The bounds that we derive change appropriately. 

\subsection{Annihilation cross section and non-relativistic scattering}

As we have emphasized in the original paper, if the DM interactions with the SM are mediated by an anomaly free 
$Z'$, the $Z'$ necessarily mixes with the SM $Z$, inducing therefore tree level annihilations of DM into EW gauge bosons (including the Higgs) as well as SM fermions. 
There we calculated these interactions both at tree and one loop level, assuming that the dominant effect was coming from the $Z'$ exchange. 
However, due to the above-mentioned mixing between the $Z$ and the $Z'$, one should also take into account both the contributions of $Z$ and $Z'$. 
Although the former has a suppressed coupling to the DM, since it is much lighter than the $Z'$ and its coupling to the gauge bosons are unsuppressed, it is expected to be of the same order of magnitude as the contributions of the $Z'$~\cite{Cui:2017juz}.

Explicitly the relevant vertices that involve the neutral gauge bosons $Z$ and $Z'$ are:
\beqa
Z' \chi \chi & \to & 2 i \gZp \gX \gamma^\mu \gamma^5 \\
Z \chi \chi & \to & 2 i (-\sin \psi) \gZp \gX \gamma^\mu \gamma^5 \\
Z' f \bar f & \to & i \gZp \gamma^\mu (c_{V, f}^{Z'} + c_{A, f}^{Z'} \gamma^5) \\
Z f \bar f & \to & i g_Z \gamma^\mu (c_{V, f}^Z + c_{A, f}^Z \gamma^5) \\
Z' Z h & \to & i \gZp \cos \theta m_Z \eta^{\mu \nu} \\
Z Z h & \to & i g_Z m_Z \eta^{\mu \nu}
\eeqa
Hereafter we use the fact that $m_Z \ll m_{Z'}$ and keep only the terms up to order $\cO(m_Z/m_{Z'})^2$.
In this approximation $\cos \psi \approx 1$ and 
$\sin \psi \approx - \cos \theta \frac{g_{Z'}}{g_Z} \frac{m_Z^2}{m_{Z'}^2}$, so that the mixing angle $\psi$ is 
proportional to the ratio of the neutral gauge bosons squared masses. 

When we take into account all the diagrams of the same order in $\cO(m_Z/m_{Z'})^2$  we find important 
cancellations between the SM $Z$ and $Z'$ contributions.
In particular, we find that for a DM axially coupled to the $Z'$ 
\emph{there are no s-wave 
annihilation channels}.\footnote{This statement is true at any order of $m_Z/m_{Z'}$ for the fermion channels up 
to the helicity suppression, and holds at least at one loop level and at first order in $m_Z^2/m_{Z'}^2$ for 
the boson channels.}
The would-be $s$-wave contribution of the $Z'$ 
precisely cancels out against the analogous contribution of the 
SM $Z$.  
More importantly, we need to sum the contribution of the $Z$ and the $Z'$ in order to see that the process 
$ \chi \chi \to Z^{(L)}h$ vanishes at $\cO(E^2)$, such that unitarity is not violated. 

Due to these effects we find that, contrary to the claim that we make in the original paper, $t \bar t$ and $Zh$
are generically not the dominant annihilation channels both in the Galactic Center and in the Sun. 
Instead the annihilation Branching Ratios are dominated by light SM fermions, posing in this sense an additional challenge to the neutrino telescopes and indirect detection experiments. 
Moreover, the total annihilation rate is suppressed.
Among the channels that contribute to the hard neutrino signal observable at IceCube, we find that 
the bound is driven by comparable contributions of $\nu_i \overline \nu_i$, $\tau^+\tau^-$, $\mu^+\mu^-$, and a smaller contribution of $t\bar t$.
We show the relevant branching ratios in Fig~\ref{fig: BR}, which supersedes the plots on Fig.~6 in the original text.

We also include the $Z$-exchange diagrams in our calculations of the NR scattering. 
The effect on the DD is mild, but it is appreciable on the DM Solar Capture. 
In particular we find that the NR scattering operator $\cO_4$ \emph{vanishes at the leading order}, and therefore the DM scattering with nucleons is controlled by $\cO_8$ and $\cO_9$, which are velocity and momentum suppressed, respectively. 
This changes the prospects for neutrino telescopes. 
In particular, we find that due to these suppressions the amount of the DM captured by the Sun \emph{is not yet in equilibrium}, except for the resonance DM masses. We plot 
the ratio between the equilibrium time and the Sun lifetime on Fig~\ref{fig: equilibrium}. Later, whenever 
the DM is out of equilibrium we rescale the Ice Cube bound by the factor $\tanh^2 (t_\odot/\tau_\text{eq})$.

Another important point that we properly take into account in our revised calculation is the complex mass scheme, that removes unphysical effects near the resonances. 
The correct application of the mass scheme requires the replacement of all the $m^2$ factors by $m^2 - i m\Gamma$, both in the propagator and the mixing angles~\cite{Nowakowski:1993iu}. 
In particular, near the $Z$ and the $Z'$ resonances the propagators and the mixing angle have the following structure:    
\begin{align}
\frac{-i}{p^2-m_Z^2}\left( \eta^{\mu\nu} -\frac{p^\mu p^\nu}{m_Z^2} \right) 
& \quad \longrightarrow \quad 
\frac{-i}{p^2-(m_Z^2-i m_Z \GZ)}\left( \eta^{\mu\nu} -\frac{p^\mu p^\nu}{(m_Z^2-i m_Z \GZ)} \right) \,, \\
\frac{-i}{p^2-\mZp^2}\left( \eta^{\mu\nu} -\frac{p^\mu p^\nu}{\mZp^2} \right) 
& \quad \longrightarrow \quad 
\frac{-i}{p^2-(\mZp^2-i \mZp \GZp)}\left( \eta^{\mu\nu} -\frac{p^\mu p^\nu}{(m_Z^2-i \mZp \GZp)} \right) \,, \\
\spsi = - \cth \frac{\gZp}{\gZ} \frac{\mZ^2}{\mZp^2}
& \quad \longrightarrow \quad 
\spsi = - \cth \frac{\gZp}{\gZ} \frac{\mZ^2-i \mZ \GZ}{\mZp^2-i \mZp \GZp} \,.
\end{align}
Note that after applying this scheme the BRs near the $\mZp$ resonance are smooth ({\it c.f} Fig.~\ref{fig: BR}), 
in agreement with similar results obtained in~\cite{Ismail:2017ulg}.

We also notice, that we have found a bug in our calculation of the maximal allowed couplings $\gZp$ 
as a function of the angle $\theta$. 
We show the correct results on Fig.~\ref{fig: resonant Z'} that supersedes the plot on Fig.~1 in the original text.  

\subsection{Results}
\begin{figure}[h!] \centering
\includegraphics[width=.49\textwidth]{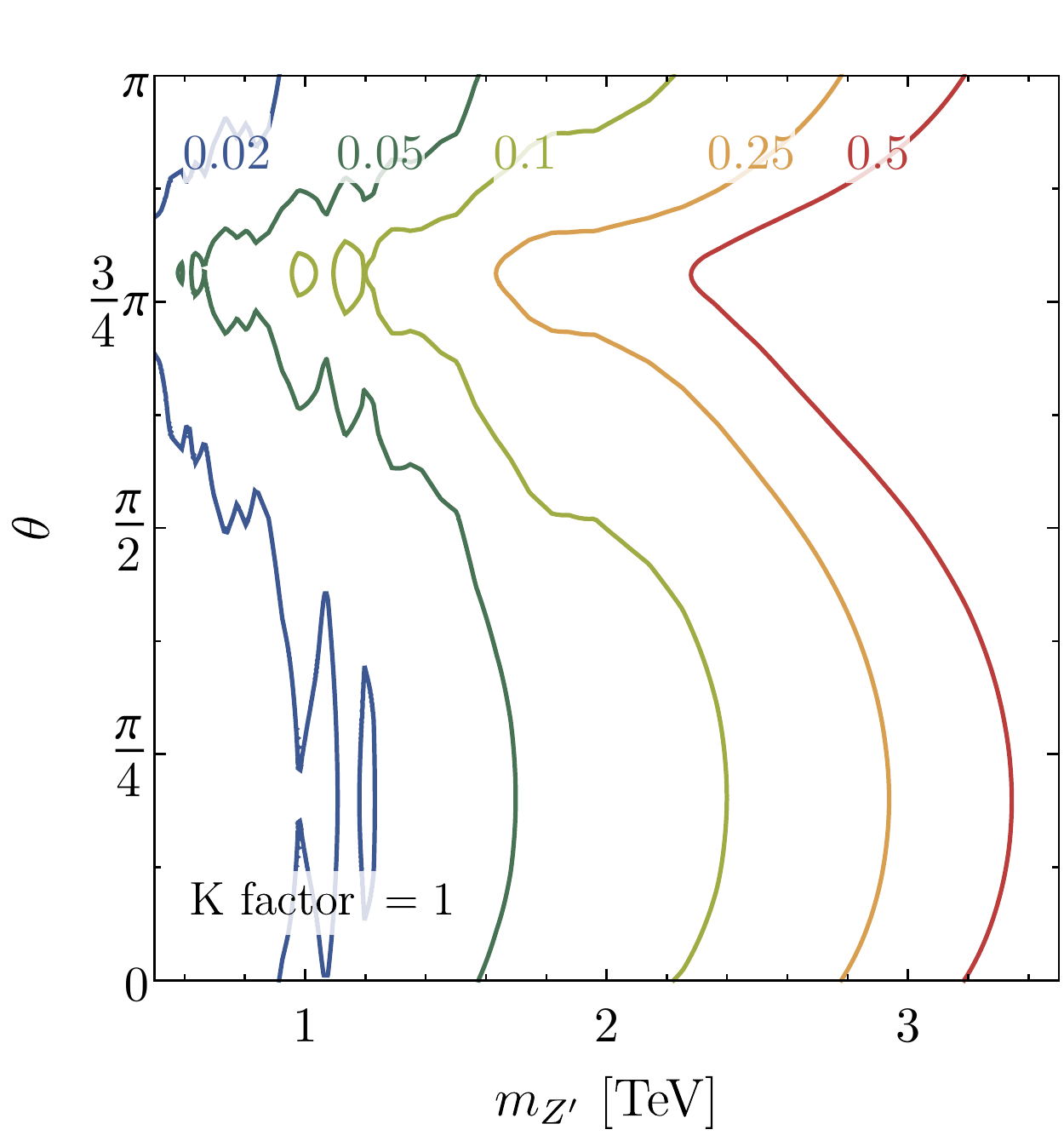}
\includegraphics[width=.49\textwidth]{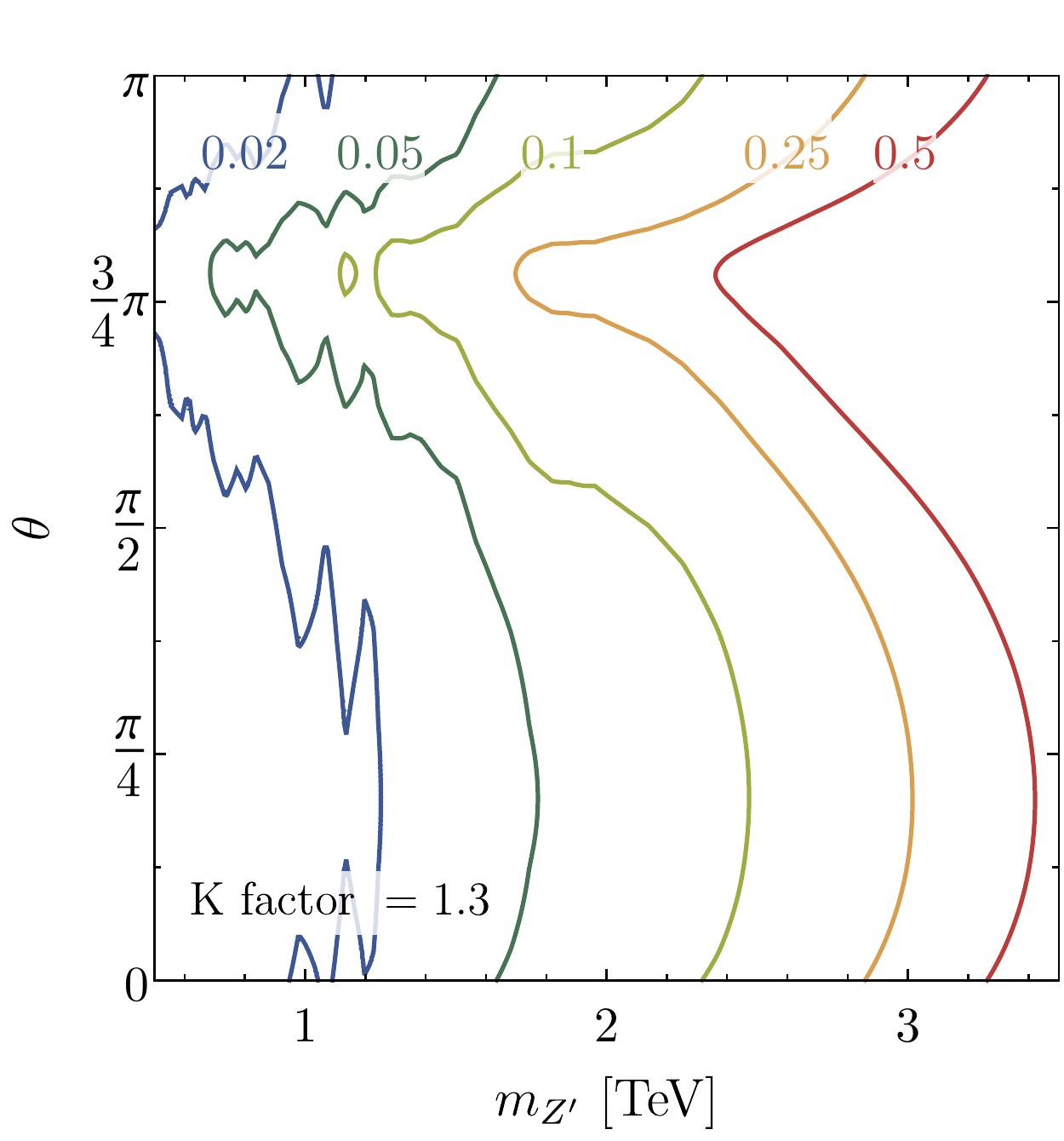}\caption{\replace{1} Contours on the maximal allowed $\gZp$ as functions of $\mZp$ and $\theta$ for K-factors of $1$ and $1.3$ (to account for non-perturbative QCD effects).}
\label{fig: resonant Z'}
\end{figure}

\begin{figure}[h!]
\centering
\includegraphics[width=.49\textwidth]{gZpRelicDensity}
\includegraphics[width=.49\textwidth]{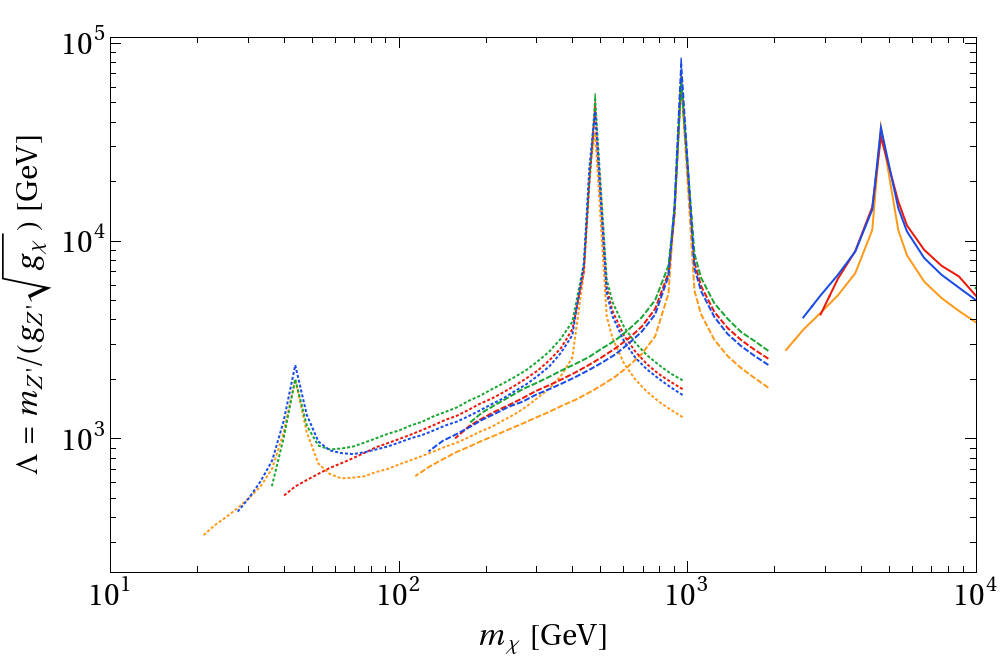}
\caption{\replace{2} Lower limits on the couplings $\gZp,\gX$ and corresponding upper limits on the effective scale $\Lambda=\mZp/(\gZp\sqrt{\gX})$ from the requirement of not overclosing the universe. The gray shaded region in the upper panel correspond to a value of the couplings such that $\GZp>\mZp$, signaling the breakdown of the perturbative description.}
\label{fig: relic}
\end{figure}

\begin{figure}[h!] \centering
\includegraphics[width=.7\textwidth]{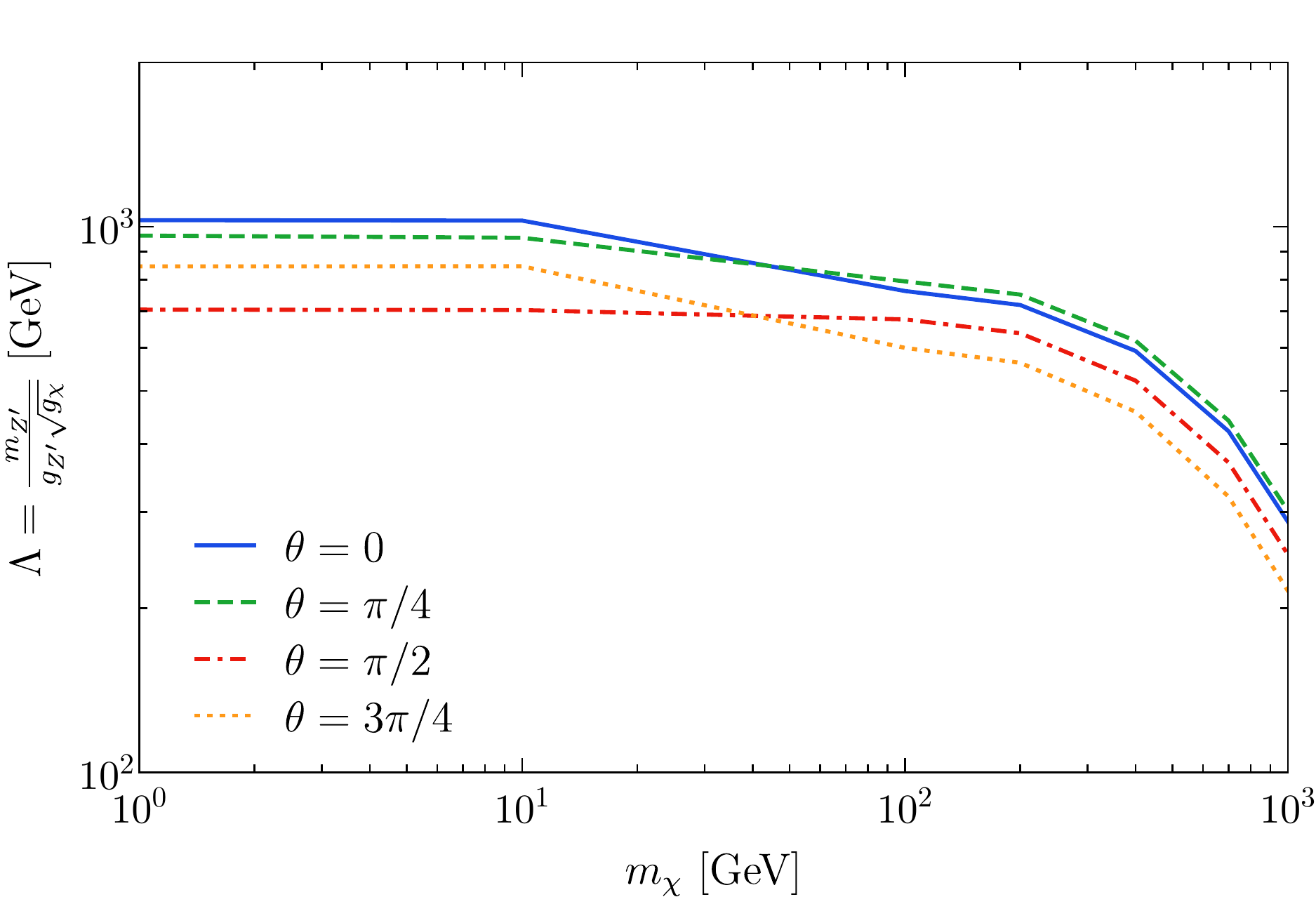}
\caption{\replace{3} Bounds on $\Lambda$ for each value of $\theta$ from the monojet search.}
\label{fig: monojet}
\end{figure}

\begin{figure}[h!] \centering
\includegraphics[width=.49\textwidth]{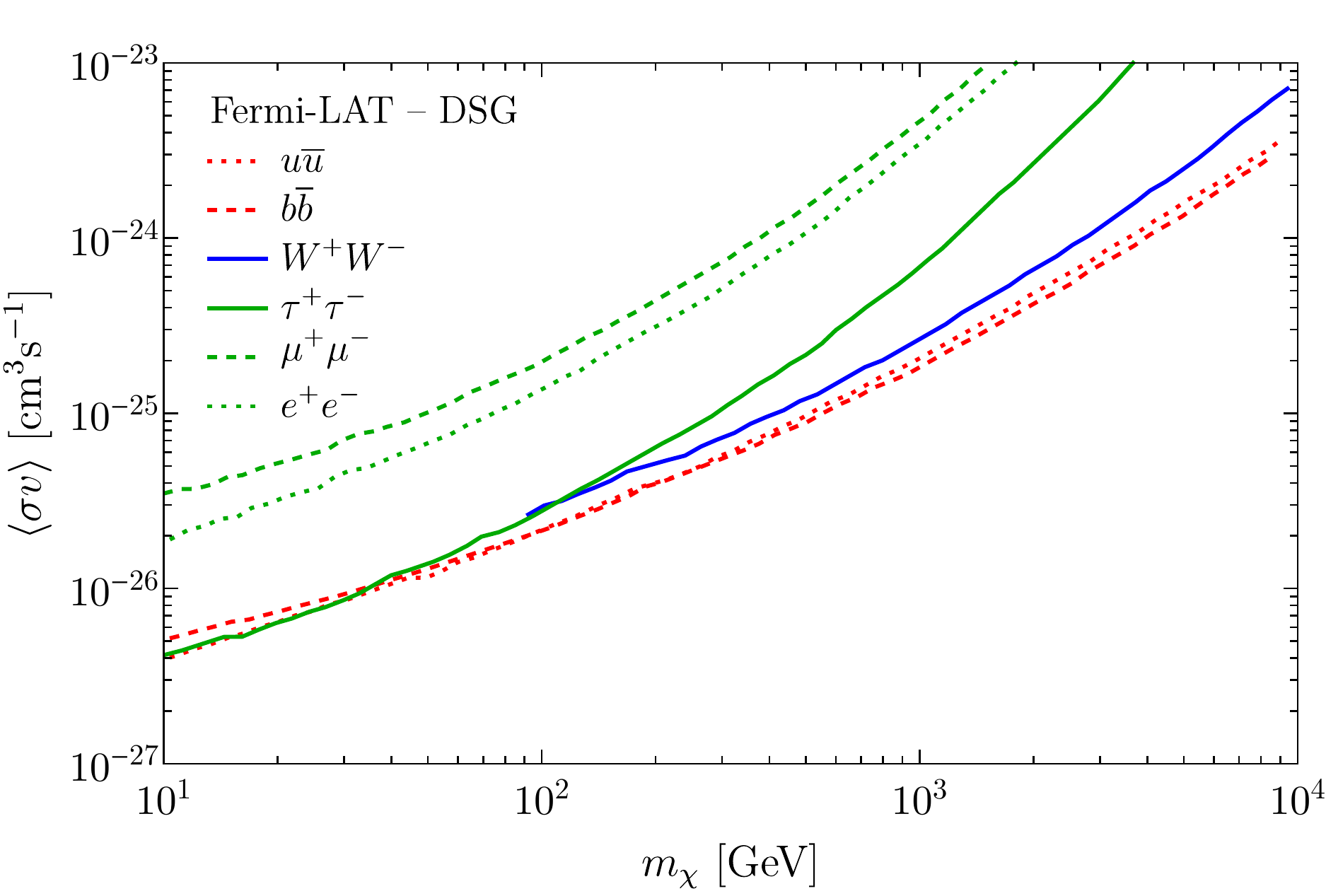}\\
\includegraphics[width=.49\textwidth]{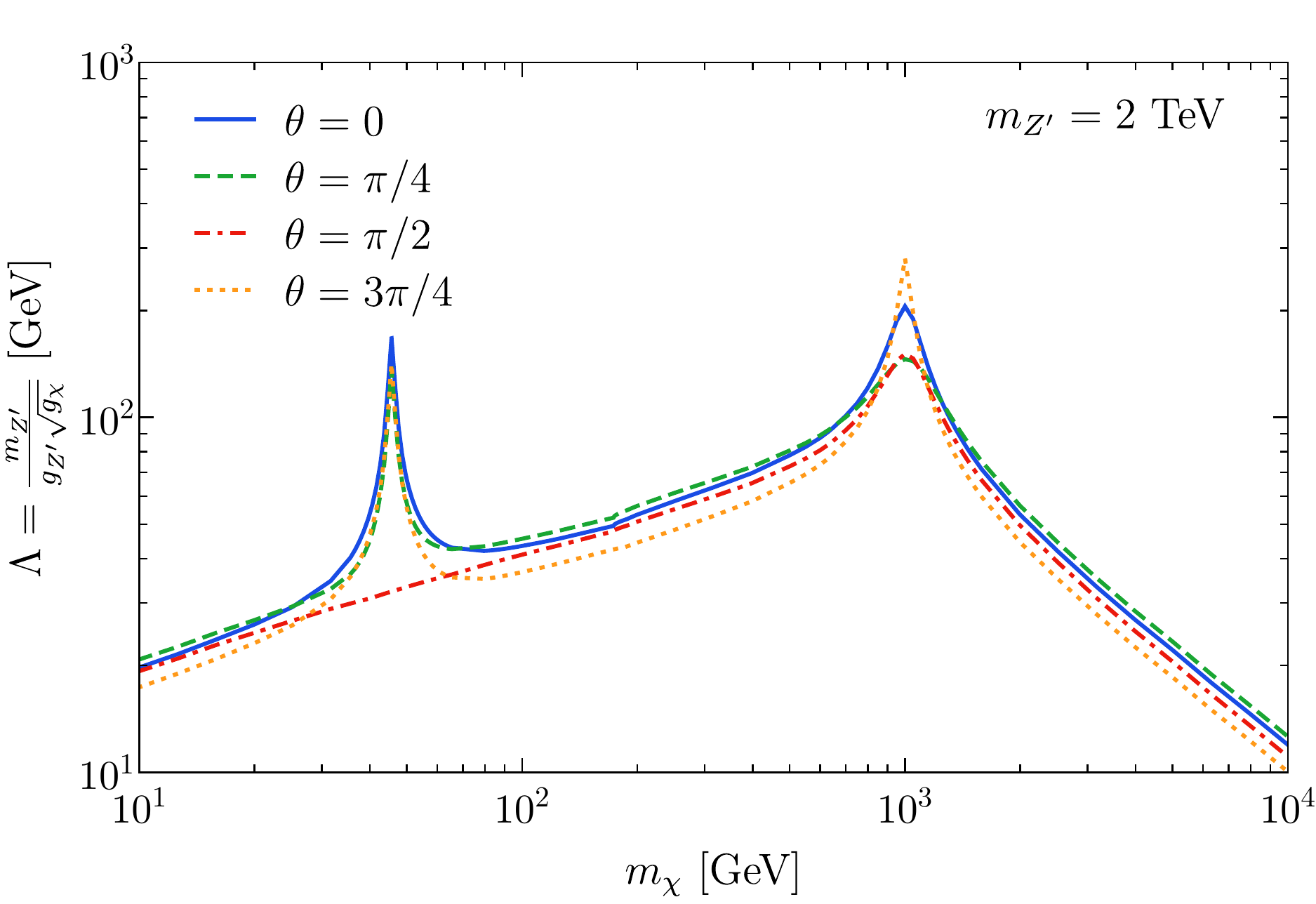}
\includegraphics[width=.49\textwidth]{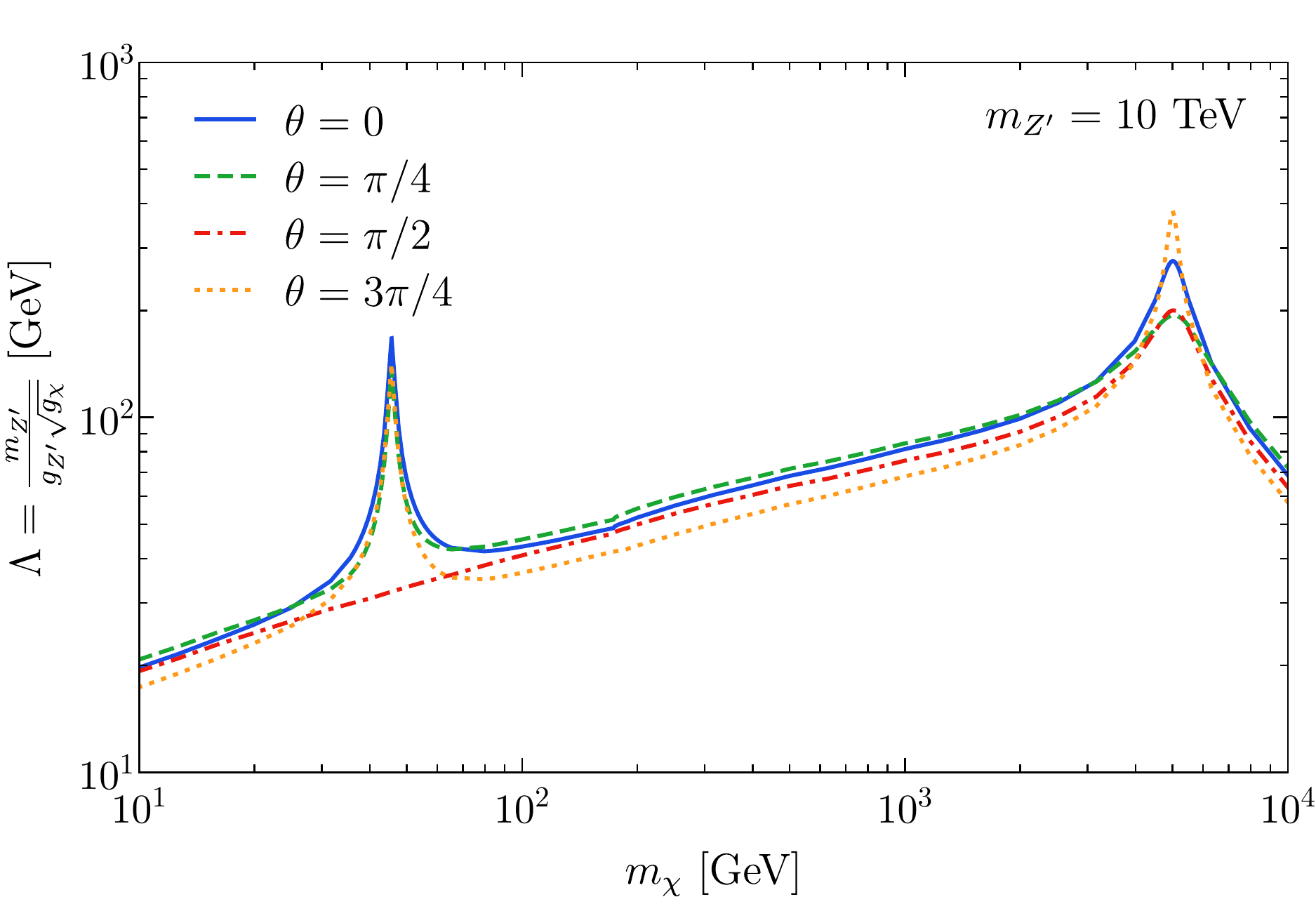}
\caption{\replace{5}
\textit{Top}: Bounds on $\sv$ from Fermi-LAT observations of dSph, 
assuming 100\% BR in the channels shown in the legend.
\textit{Bottom}: Bounds on $\sv$ from Fermi-LAT observations of dSph in our model, 
for the four values of $\theta$ we have chosen, and for $\mZp=2$ TeV (\textit{left}) and $10$ TeV (\textit{right}).
}
\label{fig: Fermi DSG}
\end{figure}

\begin{figure}[h!]
\includegraphics[width=.49\textwidth]{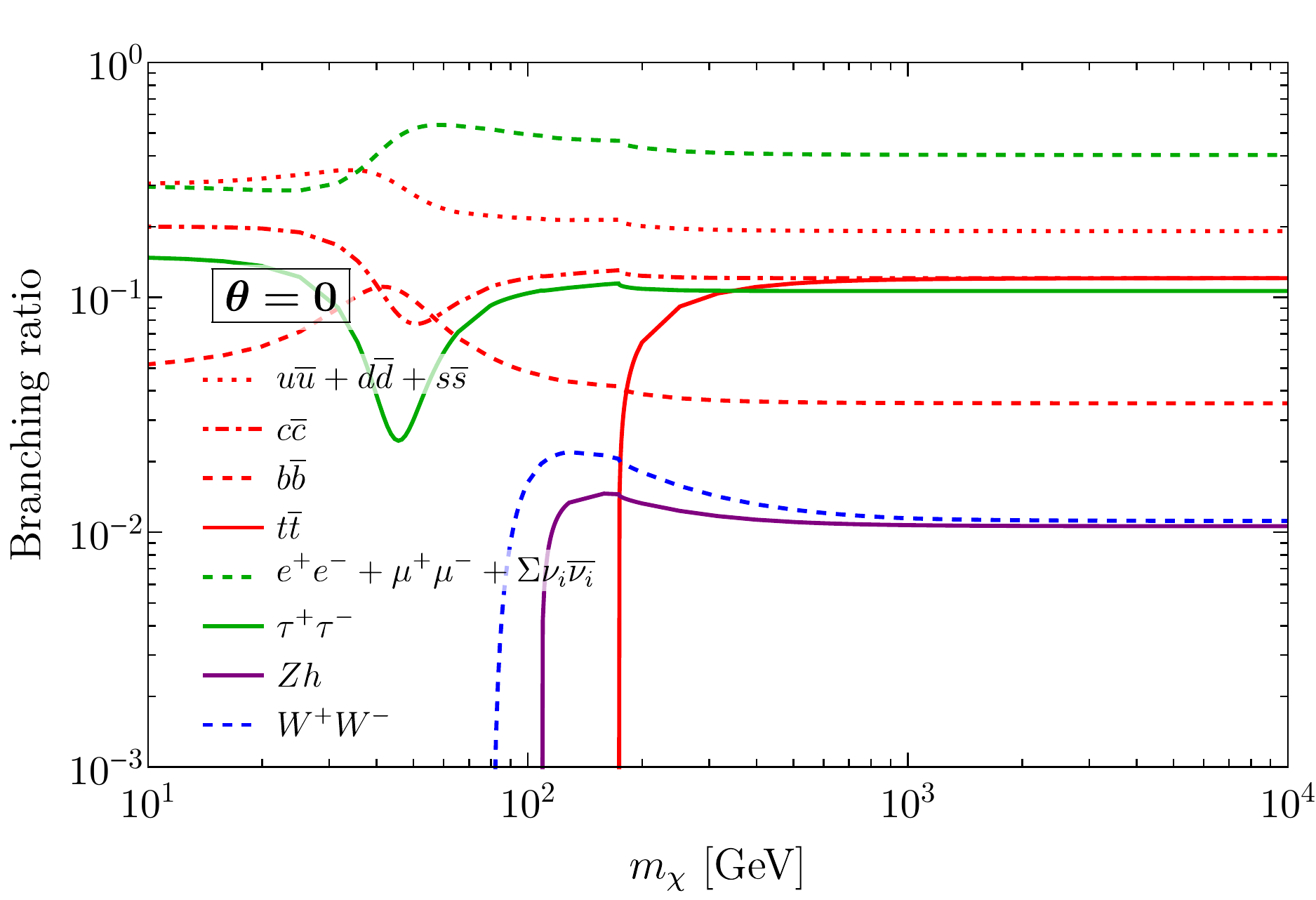} \hfill
\includegraphics[width=.49\textwidth]{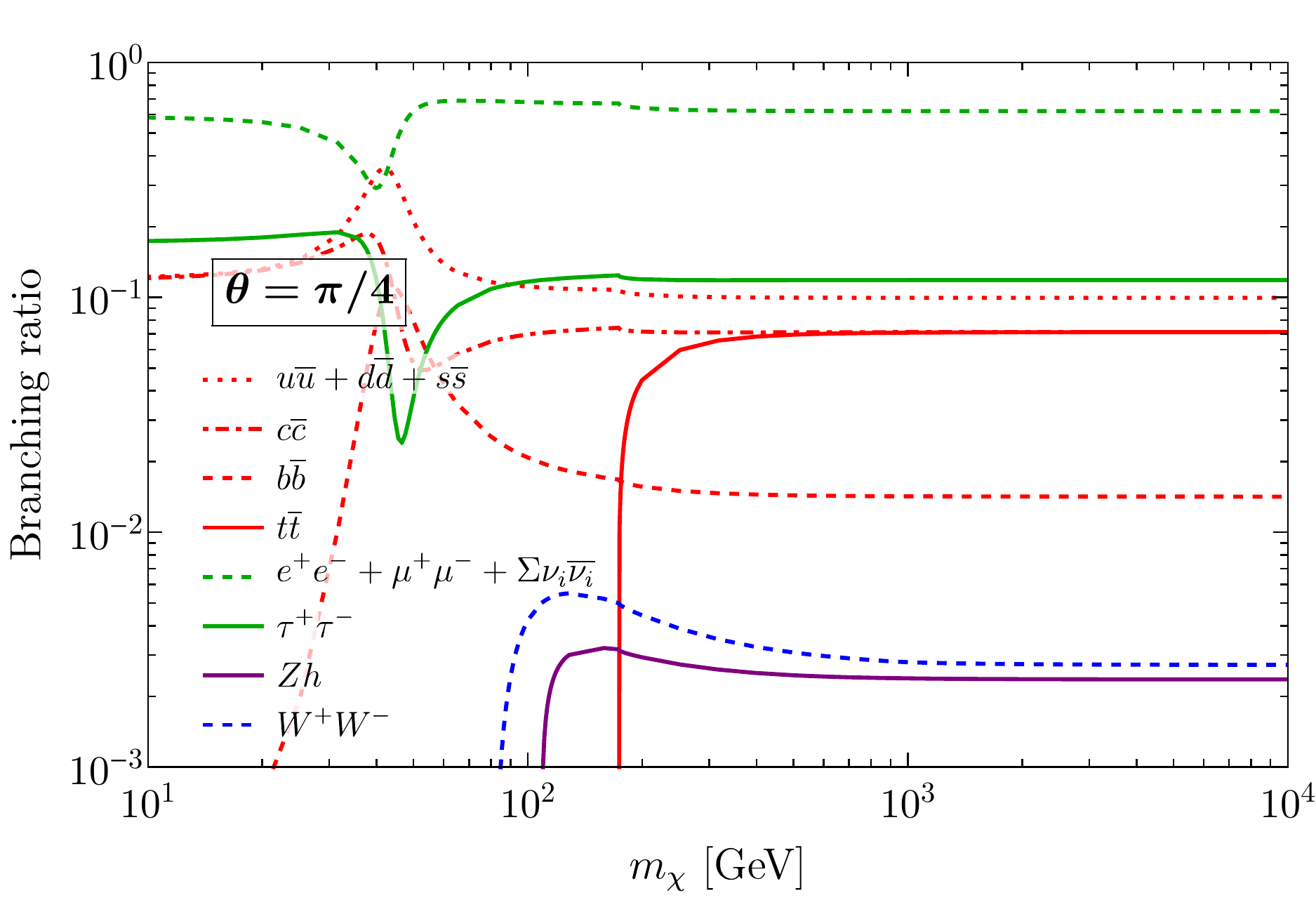} \newline
\includegraphics[width=.49\textwidth]{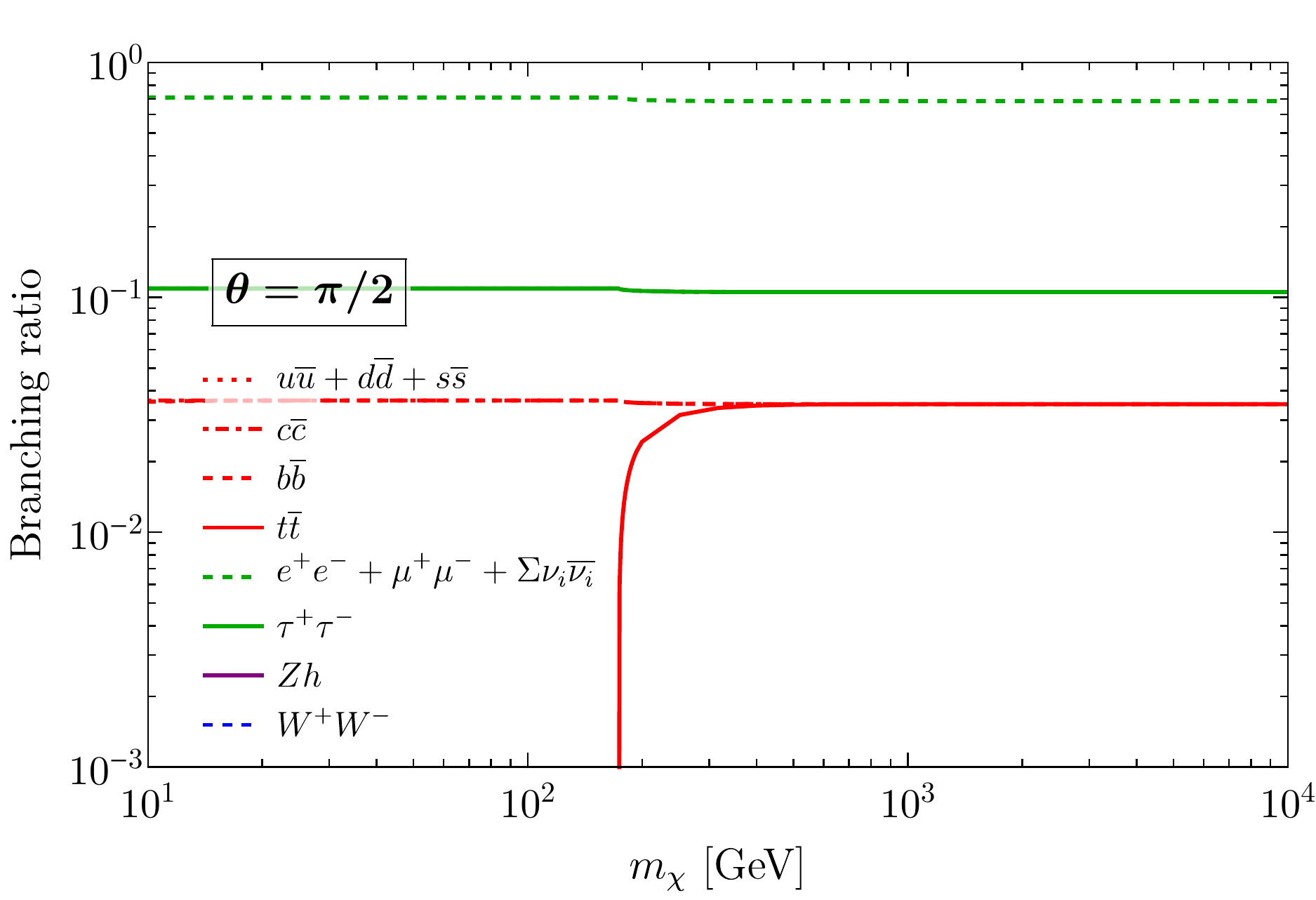} \hfill
\includegraphics[width=.49\textwidth]{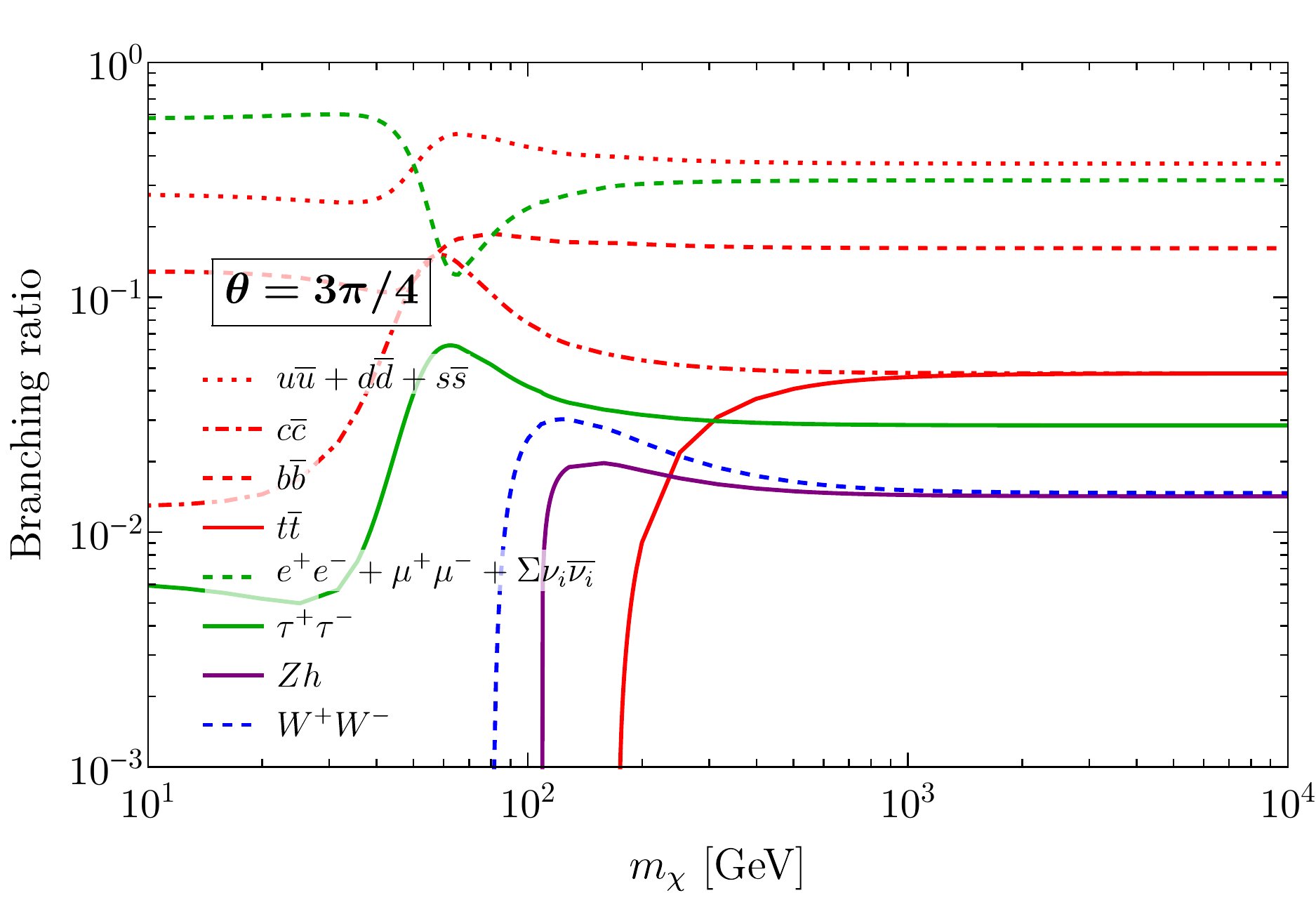}
\caption{\replace{6} Branching ratios for DM annihilations, for four different values of $\theta$. 
Annihilation at different energies have the same behavior, given that all the channels are in $p$-wave,
and the branching ratios are independent of $\mZp$.}
\label{fig: BR}
\end{figure}

\begin{figure}[h!]
\centering
\includegraphics[width=.49\textwidth]{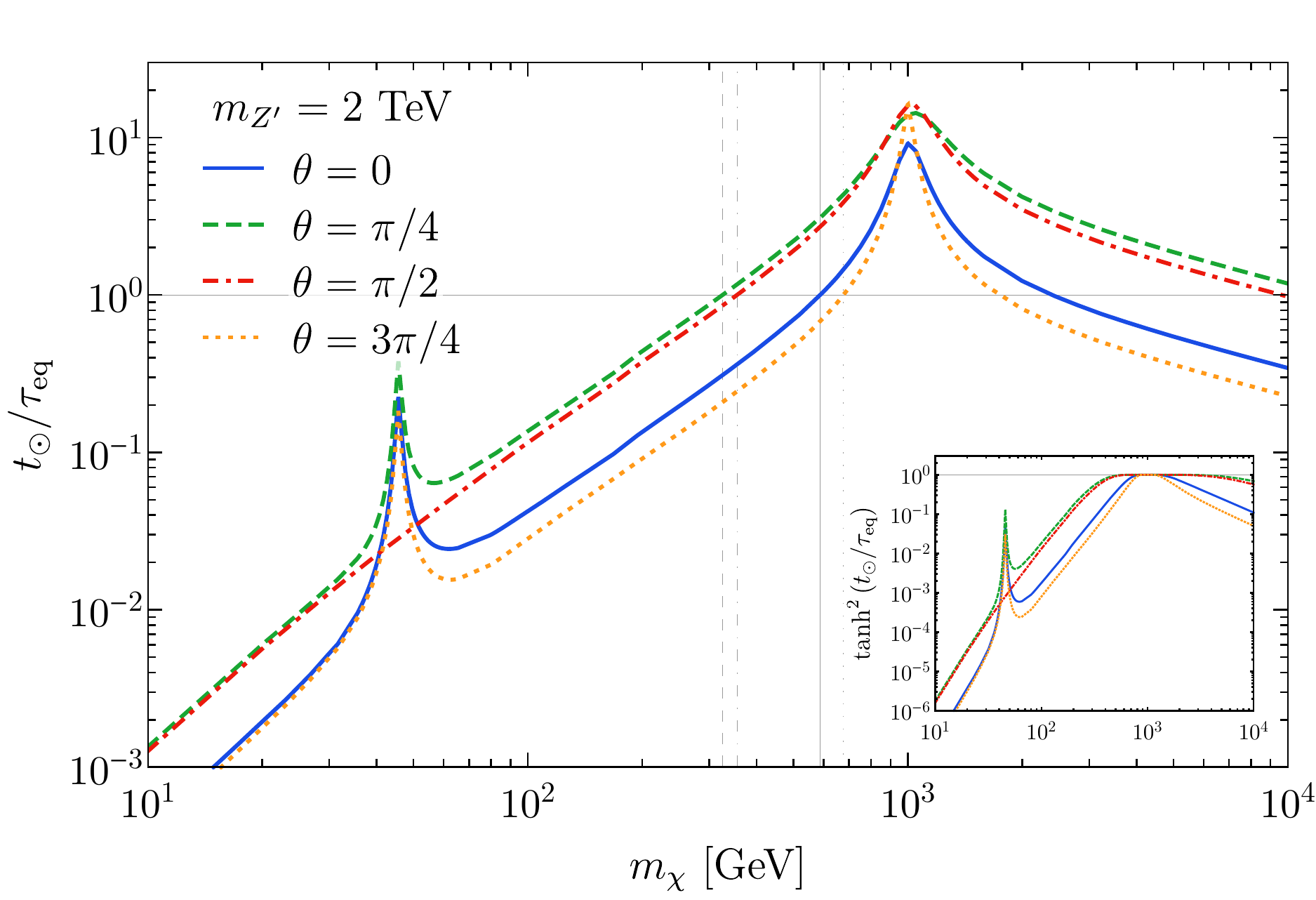}
\includegraphics[width=.49\textwidth]{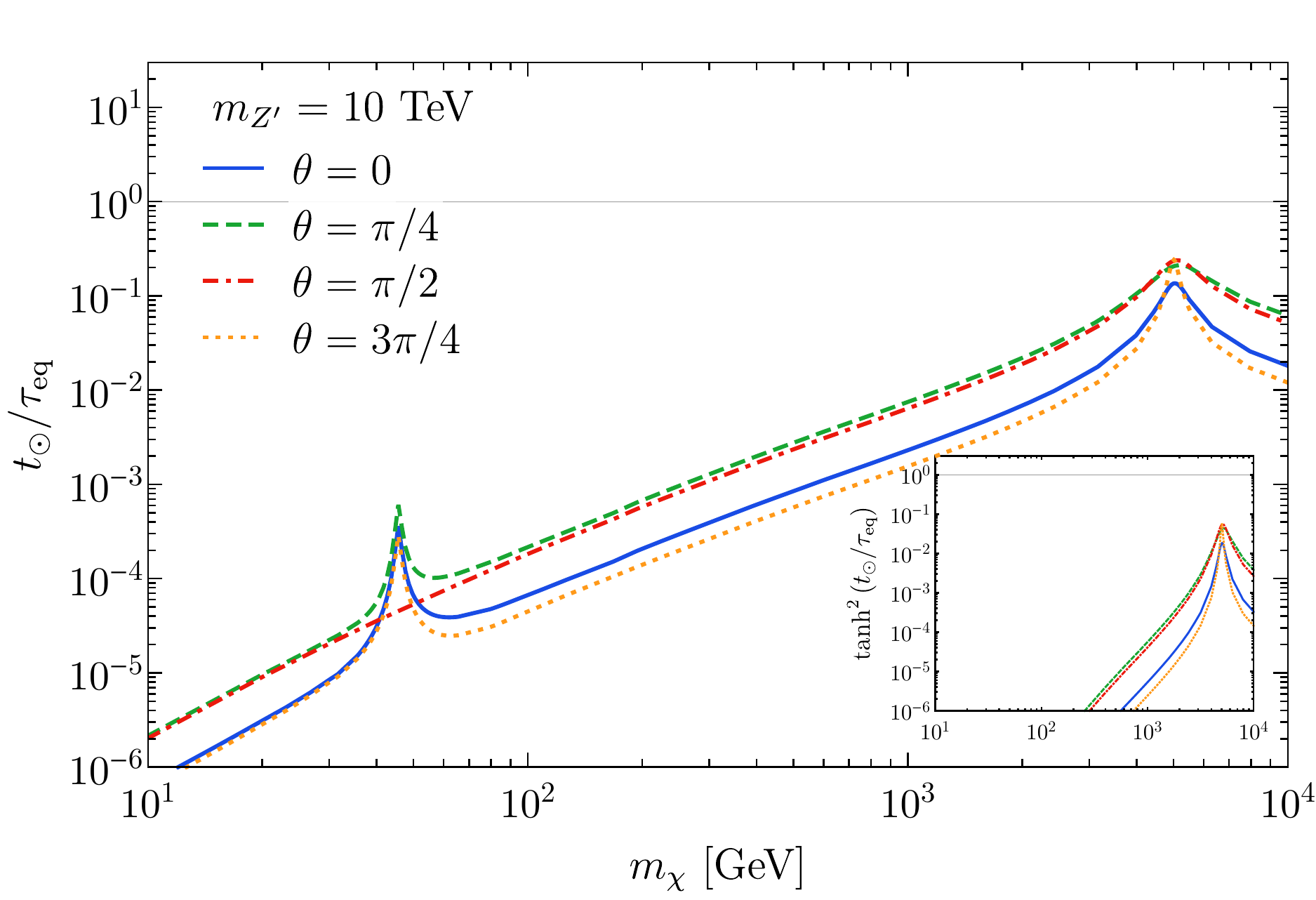}
\caption{Ratio of the age of the Sun over the timescale for the reach of equilibrium between capture and annihilation of DM, for $\mZp=2$ TeV (\textit{left}) and $10$ TeV (\textit{right}).
The ratio $t_\odot/\tau_\text{eq}$ scales as $\Lambda^{-4}$.}
\label{fig: equilibrium}
\end{figure}

\begin{figure}[h!] \centering
\includegraphics[width=.49\textwidth]{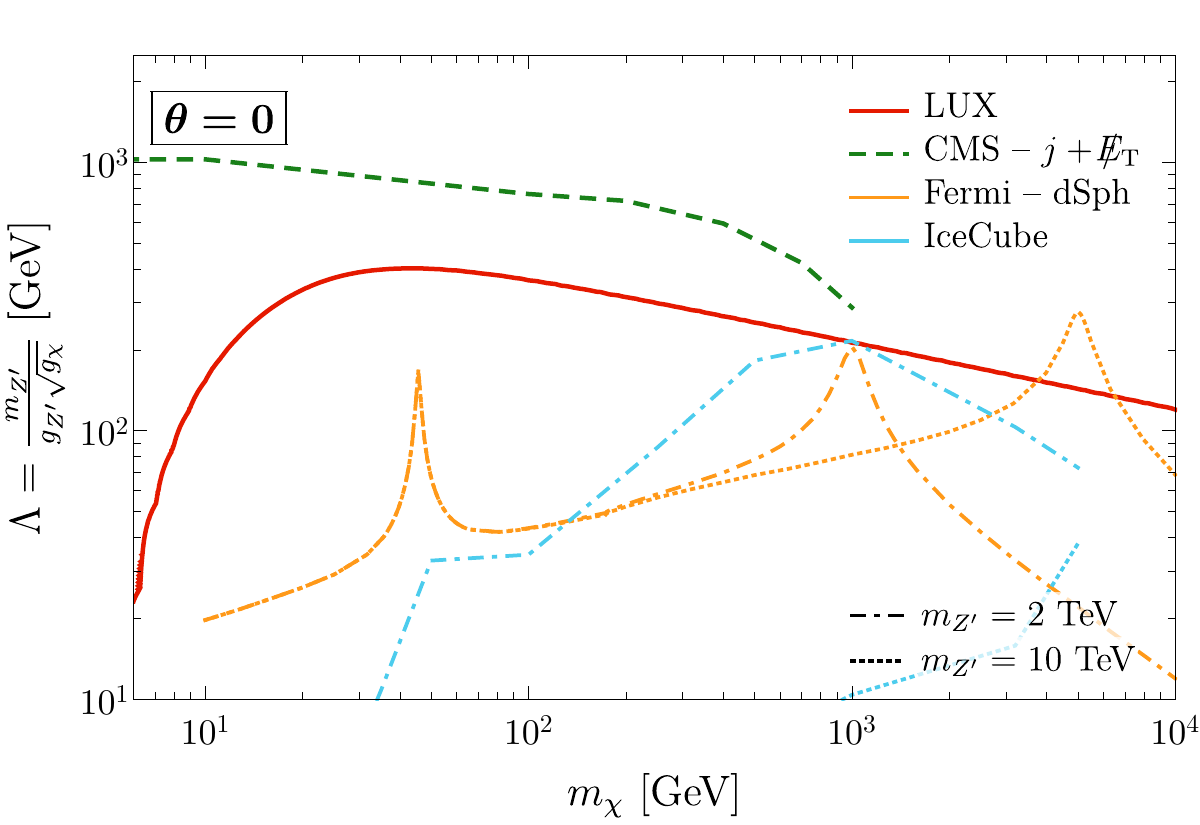} \hfill
\includegraphics[width=.49\textwidth]{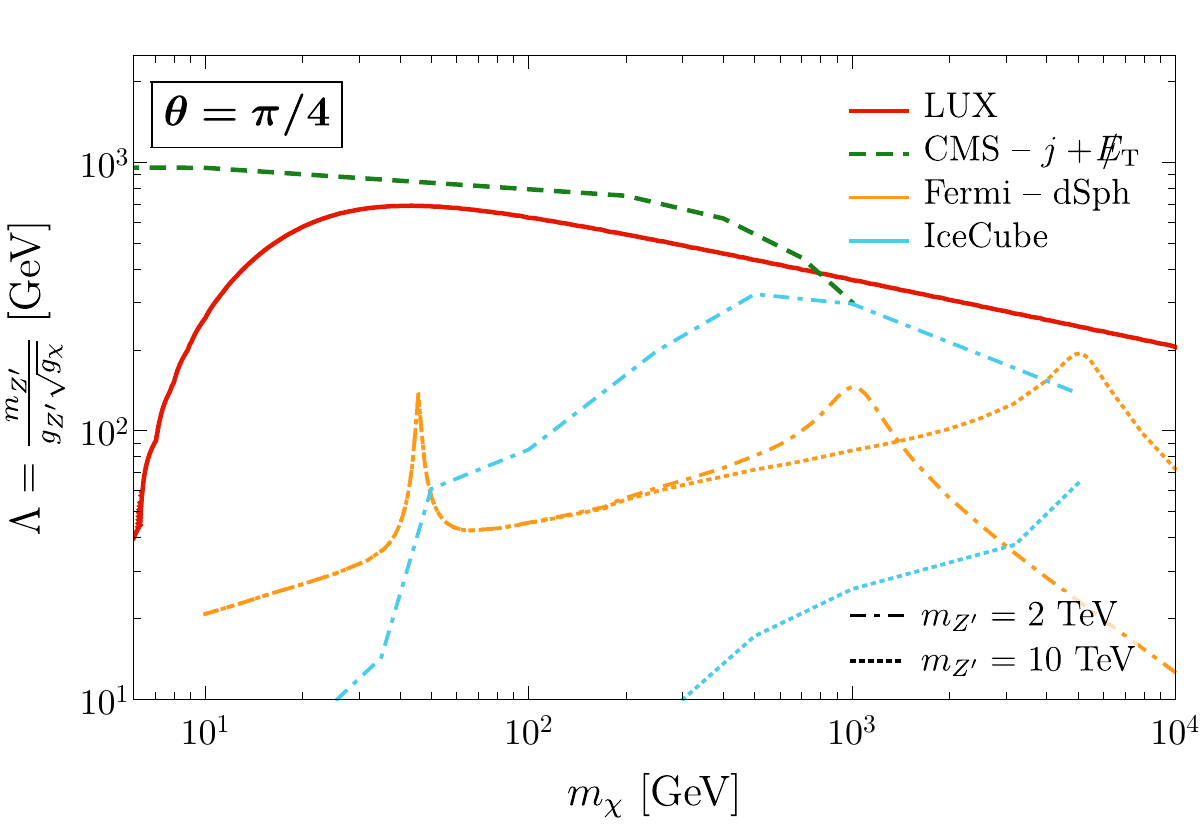} \\
\includegraphics[width=.49\textwidth]{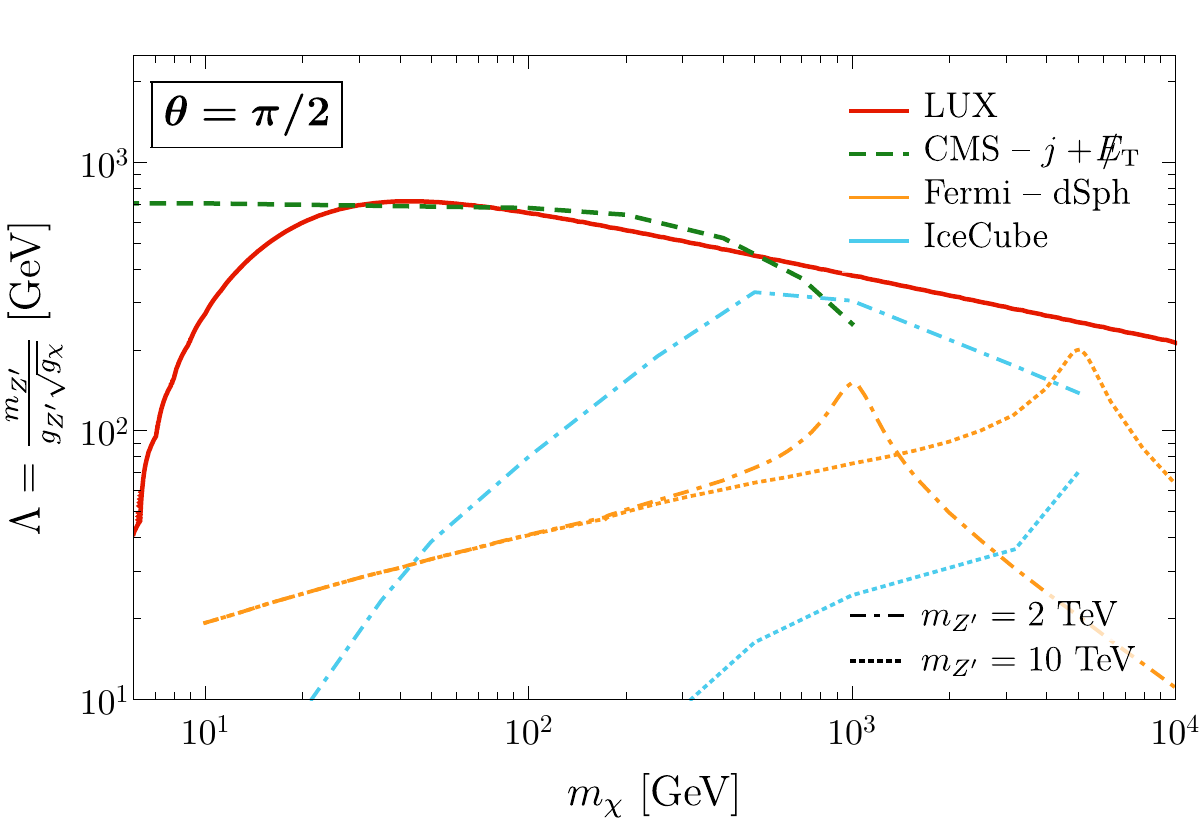} \hfill
\includegraphics[width=.49\textwidth]{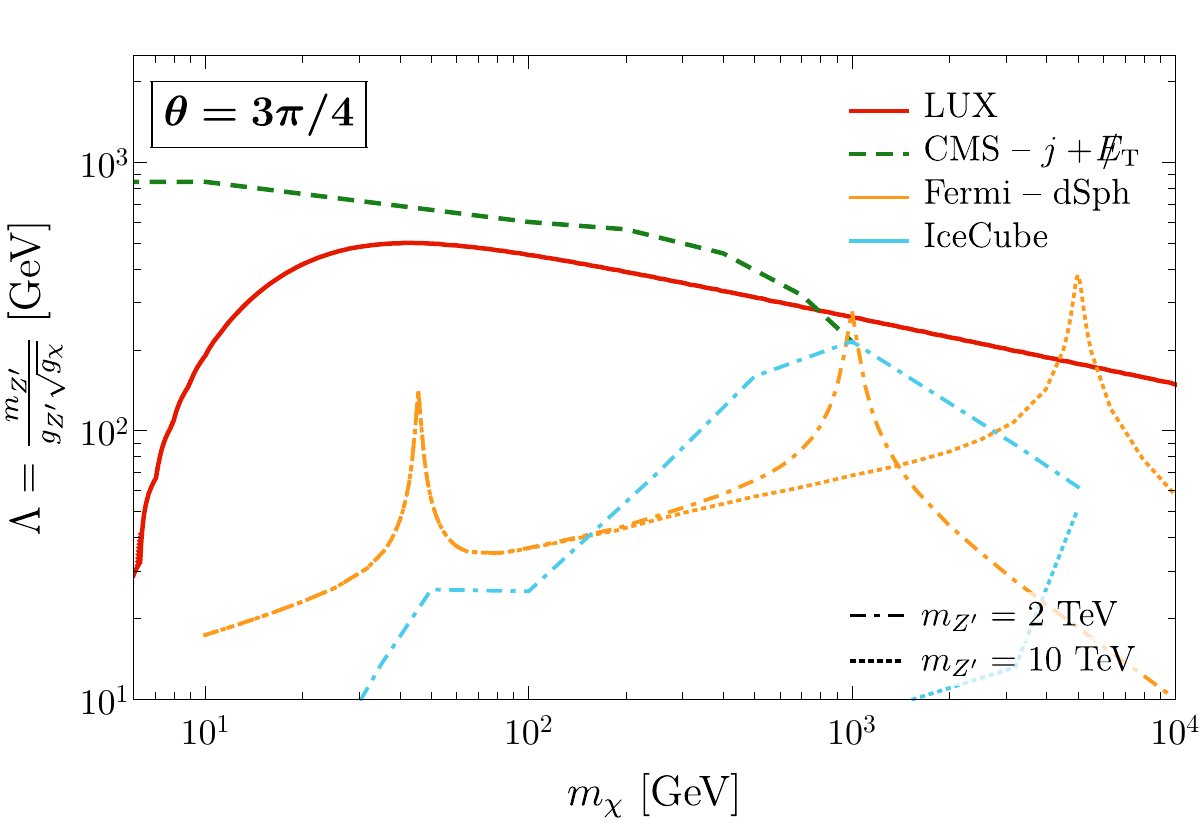}
\caption{\replace{7, 8} Exclusion limits for the four values of $\theta$ we consider.
The bound on spin dependent cross section for $\theta = 0$ (as on Fig. 7 of the original paper) is not shown, since there is no such scattering.
For the Fermi and IceCube bounds, we show two lines corresponding to $\mZp=2$ or $10$ TeV.
}
\label{fig: final plots}
\end{figure}

After fixing these errors we have replotted all the figures, since all of them are affected by the above mentioned changes in the calculations, albeit some of these corrections are truly minor. 
Hereafter in Figs.~\ref{fig: resonant Z'}, \ref{fig: relic}, \ref{fig: monojet}, \ref{fig: Fermi DSG}, \ref{fig: BR} and \ref{fig: final plots} we bring all the redone plots and indicate which of the figures they supersede in the original paper. 
At the end we also provide a full list of diagrams that we calculate, because it slightly differs from one that we present in the appendix of the original paper.
We also collect the formul\ae\ obtained for the annihilation cross sections of DM at tree level (up to corrections $\mathcal O(\mZ^4/\mZp^4)$).

% 1

% 2

% 3

% 5

% 6

% 8

\clearpage
% 9
\begin{figure}[h!]
\includegraphics[width=\textwidth]{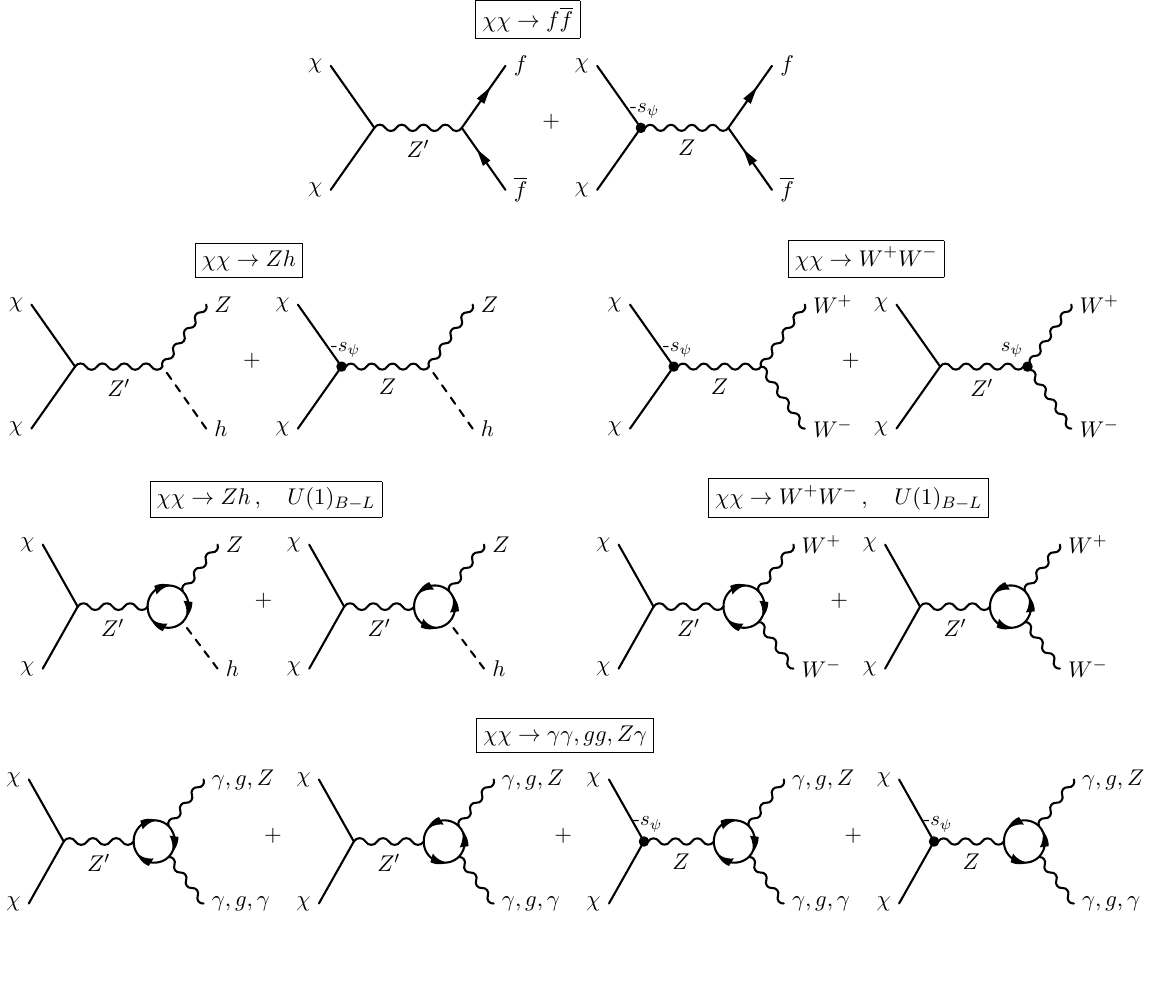}
\caption{\replace{9} Feynman diagrams for the annihilation channels $f\overline f$, $Zh$, $W^+W^-$, $\gamma\gamma$, $gg$, $Z\gamma$.}
\label{fig:feynman diagrams}
\end{figure}

We collect here the results for the annihilation cross sections of two DM particles into SM pairs of fermions or bosons, computed from the diagrams of Fig.~\ref{fig:feynman diagrams}.

The most important annihilation channels are fermions, for which
\begin{footnotesize}
% ff
\begin{align}
\label{eq: xsec ff}
\sigma (\chi\chi & \to  f \bar f) = \frac{\gX^2 \gZp^4 \, N_c^f}{3 \pi s \big((s - \mZp^2)^2+ \mZp^2 \GZp^2\big)}
  \sqrt{\frac{s - 4 \mf^2}{s - 4 \mX^2}} \ \times \\
   \times  \Bigg[
   %
   % Z' Z'
   \label{eq: ff Z'Z' exchange}
   & \Bigg(
(\cVZp)^2  (s - 4 \mX^2 ) (s+ 2 \mf^2 )
 +(\cAZp)^2 \bigg( s (s - 4 \mX^2)  - 4 \mf^2 
\Big( s - 7 \mX^2 -\frac{3s \, \mX^2 (s-2\mZp^2)}{\mZp^2 (\mZp^2 + \GZp^2)} \Big)
 \bigg)
\Bigg) \\
  %
  % Z Z 
  &+ \cos^2 \theta \frac{\mZ^2 (\mZ^2 + \GZ^2)}{\mZp^2 (\mZp^2 + \GZp^2)} \cdot \\
   &\Bigg(
(\cVZ)^2  (s - 4 \mX^2 ) (s+ 2 \mf^2 )
 +(\cAZ)^2 \bigg( s (s - 4 \mX^2)  - 4 \mf^2 
\Big( s - 7 \mX^2 -\frac{3s \, \mX^2 (s-2\mZp^2)}{\mZp^2 (\mZp^2 + \GZp^2)} \Big)
 \bigg)
\Bigg) \\
   %
   % Z Zp
  &+ \cos \theta \frac{1}{\mZp^2 (\mZp^2 + \GZp^2)\big((s - \mZ^2)^2+ \mZ^2 \GZ^2\big)}
   \Bigg(
\cVZ \cVZp (s - 4 \mX^2 ) (s+ 2 \mf^2 ) \, \mathscr C \\
 & \hspace*{5em} +\cAZ \cAZp \bigg( s (s - 4 \mX^2) \,\mathscr C  - 4 \mf^2 
\Big( s \,\mathscr C - 7 \mX^2 \, \mathscr C -3s \, \mX^2  \,\mathscr D \Big)
 \bigg)
\Bigg) 
\Bigg] \,,
\end{align}
\end{footnotesize}
where
\begin{footnotesize}
\begin{gather}
\mathscr C = (s - \mZ^2 - \GZ^2) (s - \mZp^2 - \GZp^2) \mZ^2 \mZp^2  + s^2 \mZ \mZp \GZ \GZp \,, \notag \\
\mathscr D = \mZ^2 \mZp^2 (\mZp^2 + \mZ^2 + \GZp^2 + \GZ^2)  + s^3 - s \bigg(
  \Big((s - \mZ^2)^2+ \mZ^2 \GZ^2\Big) +
  \Big((s - \mZp^2)^2+ \mZp^2 \GZp^2\Big) +\\
  + \mZ \mZp (3 \mZ \mZp - \GZ \GZp)\bigg) \,. \notag
\end{gather}
\end{footnotesize}

For a gauge group $\Up=\cos\theta\, \UY +\sin\theta \,\UBL$ the gauge boson $Z'$ has axial couplings $\cAZp$ related to the axial couplings of the $Z$ boson $\cAZ$ by $\cAZp=-\cos\theta\, \cAZ$. 
This relation implies that the term proportional to the axial coupling in $\sigma(\chi\chi \to f\overline f)$ is velocity suppressed for any value of $\mf$.
This arises as a consequence of the sum of the $Z$ and $Z'$ exchanges in the propagator (whereas the $Z'$ exchange would give just Eq.~\eqref{eq: ff Z'Z' exchange}, with an $s$-wave contribution).
Therefore the parametric behaviour in the limit $\mZp,\mX\gg \mf,\mZ$ is
\begin{equation}
\label{eq: xsec ff asymptotic}
\sigma(\chi\chi \to f\overline f) \sim \
\frac{\sqrt s \sqrt{s-4 \mX^2}}{\text{max}(s^2,\mZp^4)} \,.
\end{equation}
The annihilation cross section into $WW$ bosons reads
\begin{small}
\begin{multline}
\label{eq: xsec WW}
\sigma(\chi\chi \to W^{+}W^{-}) = 
\aW \cos^2 \tW \frac{\gX^2 \gZp^4\cos^2\theta}{\gZ^2}
\big( s-4 m_W^2 \big)^{3/2} \sqrt{s-4 \mX^2} \times \\
\times 
\frac{(\mZp^2 - \mZ^2)^2 + (\mZp \GZp - \mZ \GZ)^2}{((s-\mZp^2 )^2 + \mZp^2\GZp^2)((s-\mZ^2)^2 + \mZ^2\GZ^2)}\times 
\begin{cases}
 \frac{2}{3s} & \text{ for }W^{+\,(T)}W^{-\,(T)}\\
 \frac{4}{3 m_W^2} & \text{ for }W^{\pm\,(T)}W^{\mp\,(L)}\\
 \frac{(2 m_W^2+s)^2}{12 m_W^4 s} & \text{ for }W^{+\,(L)}W^{-\,(L)}
\end{cases}
\end{multline}
\end{small}
and the asymptotic cross section for $\mZp,\mX\gg m_W,\mZ$ reads
\begin{equation}
\label{eq: xsec WW asymptotic}
\sigma(\chi\chi \to WW) \sim \cos^2\theta \
\frac{\sqrt s \sqrt{s-4 \mX^2}}{\text{max}(s^2,\mZp^4)} \times
\begin{cases}
 \frac{m_W^4}{s^2} & \text{ for }W^{+\,(T)}W^{-\,(T)}\\
 \frac{m_W^2}{s} & \text{ for }W^{\pm\,(T)}W^{\mp\,(L)}\\
 1 & \text{ for }W^{+\,(L)}W^{-\,(L)}\\
\end{cases}
\end{equation}
Finally, the cross section for the tree level annihilation into $Zh$ turns out to be
\begin{small}
\begin{multline}
\label{eq: xsec Zh}
\sigma(\chi\chi \to Zh) =
\gX^2 \gZp^4 \cos^2\theta \frac{\sqrt{s-4\mX^2}\sqrt s }{6\pi}
\frac{ \sqrt{ (s - (m_h^2 + m_Z^2))^2 - 4 m_h^2 m_Z^2} \sqrt{\mZ^2 (\mZ^2 + \GZ^2)}}
{\mZp^2} \times\\
\times \frac{(\mZp^2 - \mZ^2)^2 + (\mZp \GZp - \mZ \GZ)^2}{((s-\mZp^2 )^2 + \mZp^2\GZp^2)((s-\mZ^2)^2 + \mZ^2\GZ^2)}
\times
\begin{cases}
 1 & \text{ for }Z^{(T)}h\\
 \frac{(s + \mZ^2 - m_h^2)^2}{8 s \mZ^2} & \text{ for }Z^{(L)}h\\
\end{cases}
\end{multline}
\end{small}
and its asymptotic behaviour in the limit $\mZp,\mX\gg \mZ, m_h$ is 
\begin{equation}
\label{eq: xsec Zh asymptotic}
\sigma(\chi\chi \to Zh) \sim \cos^2\theta \
\frac{\sqrt s \sqrt{s-4 \mX^2}}{\text{max}(s^2,\mZp^4)} \times
\begin{cases}
 \frac{\mZ^2}{s} & \text{ for }Z^{(T)}h\\
 1 & \text{ for }Z^{(L)}h\\
\end{cases}
\end{equation}
The asymptotic expansions \eqref{eq: xsec ff asymptotic}, \eqref{eq: xsec WW asymptotic}, \eqref{eq: xsec Zh asymptotic} of the cross sections show that all these annihilation channels are velocity suppressed, and explains why the branching ratios shown in Fig.~\ref{fig: BR} are basically independent of $\mZp$.

%%%%%%%%%%%%%%%%%%%%%%%%%%%%%%%%

{\footnotesize
\bibliography{lit}
}

\end{document}